\documentclass[a4paper,10pt]{article}
\pdfoutput=1
\usepackage{jheppub}
\usepackage[normalem]{ulem}
\usepackage{hyperref}
\usepackage{microtype}
\usepackage{amsmath}
\usepackage{float}
\usepackage{graphics}
\usepackage{psfrag}
\usepackage{color}
\usepackage{slashed}
\usepackage{subfigure}
\usepackage{graphicx}
\usepackage{epstopdf}
\usepackage{array,multirow}
\usepackage{amsfonts}
\usepackage{amssymb}

\newcommand*{\be}{\begin{equation}}
\newcommand*{\ee}{\end{equation}}
\newcommand*{\bea}{\begin{eqnarray}}
\newcommand*{\eea}{\end{eqnarray}}

\newcommand{\comment}[1]{}

\def\Eqn#1{Eq.\ (\ref{#1})}

\def\3Eqs#1#2#3{Eqs.\ (\ref{#1}), (\ref{#2}) and (\ref{#3})}

\def\fig#1{Fig.~(\ref{#1})}
\def\tabl#1{Table~\ref{#1}}

\newcommand{\cref}[1]{Chapter~\ref{c.#1}}

\newcommand{\lab}[1]{\label{eq:#1}}
\newcommand{\barr}{\begin{eqnarray}}
\newcommand{\earr}{\end{eqnarray}}

\def\beq{\begin{equation}}
\def\eeq{\end{equation}}
\def\bea{\begin{eqnarray}}
\def\eea{\end{eqnarray}}
\def\ba{\begin{array}}
\def\ea{\end{array}}
\def\bi{\begin{itemize}}
\def\ei{\end{itemize}}
\def\be{\begin{enumerate}}
\def\ee{\end{enumerate}}

\def\bc{\begin{center}}
\def\ec{\end{center}}
\def\bt{\begin{table}}
\def\et{\end{table}}
\def\btb{\begin{tabular}}
\def\etb{\end{tabular}}

\def\lsim{\raise0.3ex\hbox{$\;<$\kern-0.75em\raise-1.1ex\hbox{$\sim\;$}}}
\def\gsim{\raise0.3ex\hbox{$\;>$\kern-0.75em\raise-1.1ex\hbox{$\sim\;$}}}

\title{Unification with Vector-like fermions and signals at LHC}
\author[a]{Biplob Bhattacherjee,}
\author[b]{Pritibhajan Byakti,}
\author[a]{Ashwani Kushwaha,}
\author[a]{ Sudhir K  Vempati}

 \affiliation[a]{Centre for High Energy Physics, Indian Institute of Science,
Bangalore 560012}
\affiliation[b]{Department of Physics,
Pandit Deendayal Upadhyaya Adarsha Mahavidyalaya (PDUAM) Eraligool, Karimganj, 788723, Assam, India}

\emailAdd{biplob@iisc.ac.in}
\emailAdd{priti137@gmail.com}
\emailAdd{ashwanik@iisc.ac.in}
\emailAdd{vempati@iisc.ac.in}
\abstract{We look for minimal extensions of Standard Model with vector like
fermions leading to precision unification of gauge couplings. Constraints
from proton decay, Higgs stability and perturbativity are considered. 
The simplest models contain several copies of vector fermions in two 
different (incomplete) representations. Some of these models encompass
Type III seesaw mechanism for neutrino masses whereas some others have
a dark matter candidate. In all the models, at least one of the candidates
has non-trivial representation under $SU(3)_{color}$. In the limit of 
vanishing Yukawa couplings, new QCD bound states are formed, which 
can be probed at LHC. The present limits based on results from 13 TeV 
already probe these particles for masses around a TeV. Similar models
can be constructed with three or four vector representations, examples
of which are presented.}
\begin{document}
\maketitle
\section{Introduction}
For the past few decades, the path to Beyond Standard Model (BSM) physics 
has been dictated mostly by solutions to the hierarchy problem \cite{Csaki:2016kln}. However, 
with no experimental evidence to support this endeavor, from either
LEP, Tevatron or the LHC so far, one might wish to explore alternate paths
which do not contain a solution to the hierarchy problem.  Furthermore, 
there could solutions to the hierarchy problem which do not  
introduce any new particles all the way up to GUT scales.
The relaxion idea and its variant for example, propose  
a cosmological solution to the hierarchy problem without introducing any new physics at the weak scale 
\cite{Graham:2015cka,Espinosa:2015eda,Choi:2015fiu,Kaplan:2015fuy}.

One of the guiding principles for these alternate paths is the  
unification of gauge coupling constants. Popular models like split supersymmetry
\cite{ArkaniHamed:2004yi, Giudice:2004tc, ArkaniHamed:2004fb}
have been proposed which have part of the MSSM particle spectrum at the weak 
scale and rest (scalar spectrum) at an intermediate scale. The current limits
on the stable, long lived R-hadrons which are a prediction of these models
are about 1.5- 1.61 TeV \cite{Aaboud:2016uth, Khachatryan:2016sfv}. However, this framework depends
crucially on the underlying MSSM framework. Generalization without supersymmetry  are important to explore.

With this view point, we revisit extensions of the Standard Model with 
vector-like fermions which lead to precision gauge coupling unification
(for earlier works in this direction, please see
\cite{Liu:2012qua, Hartmann:2014fya, Rizzo:1991tc, Choudhury:2001hs, Morrissey:2003sc, Barger:2004sf, EmmanuelCosta:2005nh, Barger:2006fm,
Barger:2007qb, Calibbi:2009cp, Donkin:2010ta, Hall:2009nd, Dermisek:2012ke, Li:2003zh, Shrock:2008sb, Dorsner:2005fq,
Chkareuli:1994ng, Gogoladze:2010in, FileviezPerez:2007nh, Dermisek:2012as, Dorsner:2014wva, Xiao:2014kba}). There are several
virtues of these models: 

(i) They have minimal constraints from electroweak precision parameters,
especially from S and T parameters \cite{delAguila:2008pw, Aguilar-Saavedra:2013qpa, Ishiwata:2015cga,
delAguila:2000rc, Bizot:2015zaa, Lavoura:1992np, Ellis:2014dza}, as long
as the mixing between vector-like fermions and SM fermions is small.

(ii) They do not lead to any anomalies as they are vector in nature. 

(iii) They can be tested directly at the collider experiments like 
LHC. The kind of signals depend whether on the amount of  mixing 
they have with the Standard Model fields.

(iv) If they have mixing with SM quarks, it is possible that they can
be probed indirectly in flavour physics. 

To our knowledge, there has not been a recent survey of models containing vector
fermions leading to gauge coupling unification. An earlier analysis was done 
in Ref. \cite{Rizzo:1991tc} with the available LEP data at that time. We have updated
where those models stand in the Appendix, \ref{a:tom-rizzo}. In addition to improvements
in the gauge coupling measurements and theoretical threshold calculations which are 
now available at NNLO, an important role is played by the experimental discovery and the 
 (almost) precise determination of the Higgs mass. It has been shown that the  Higgs 
 potential becomes unstable from scales close to $10^{11}$ GeV \cite{Degrassi:2012ry},
 depending on the exact values of the top mass and $alpha_s$. Thus a Grand Unified
 Theory should not only lead to gauge coupling unification but also keep Higgs potential
 stable all the way up to the GUT scale. 
 
In the  models presented here as we will see the Higgs potential naturally remains stable all the way up to the
GUT scale. In the view that the primary existence of these vector particles is unification
of gauge couplings, we dub them  ``unificons". However, as we will see later, 
these models do not restrict themselves only to unification. In some 
models, we find solutions with a provision for Type III seesaw mechanism 
for neutrino masses, and in some others there is a WIMP (Weakly Interacting Massive Particle)
 dark matter candidate. Thus ``unificon" models can indeed have wide phenomenological 
 reach solving  other problems in Standard Model like neutrino masses and dark matter. 

As a search for all possible models with extra vector-like fermions would be
a herculean task, we resort to minimality. We assume unification of gauge
couplings {\'a} la SU(5). Additional vector-like particles appear as incomplete
representations of SU(5). We have looked at all possible incomplete 
decompositions emanating from SU(5) representations up to dimension 75. The number
of copies in each representation is taken to be $n$ which is an integer
between 1 and 6. The mass range of these additional vector-like fermions is 
chosen to be $m\sim k$ TeV, where $k$ is a $\mathcal{O}(1)$ number taken to
be approximately between $1/4$ to $5$.

There are no solutions with successful gauge coupling unification as long as the vector-like 
fermions come in one single representation. This holds true even if increase the number 
of copies all the way to six, the maximum we have allowed per representation\footnote{We 
have considered the Yukawa couplings of the extra vector-like fermions
and the mixing with the SM fermions to be negligible. This can be arranged by imposing discrete symmetry.}.  
The minimal set of successful models with two different representations 
each with varied number of copies is listed in \tabl{t:two-ferm}. All these
models satisfy constraints from proton decay and the stability of the 
Higgs potential. Both representations come in several copies. Some solution allows
for degeneracy between the fermions of the different representations, where as
in some cases require non-degeneracy of the 
fermions in representation 1 and fermions in representation 2. 

Interestingly all models have at least one representation with non trivial
colour quantum numbers which makes them attractive from LHC point of view. 
In the limit of negligible Yukawa couplings, these colour states in SU(3)
representations of the type 3, 6, 8  form bound states and are produced at 
LHC. The present limits on these bound states from 13 TeV run of LHC are already
touching the 1 TeV mark, depending on the decay mode and the final 
states. We provide in detail limits on the relevant SU(3) representation 
bound states. 

We also looked for solutions with three and four different representations. 
Unlike the two representation case, we considered degenerate spectrum for 
all the vector-like fermions in these two cases. Several solutions are found which
are listed in Appendix~\ref{a:three-ferm} and Appendix~\ref{a:four-ferm}. The rest of the paper is organised as follows:
In the next Section~\ref{s:recap-rg} we recap the essential RG required for gauge coupling unification and stability of the Higgs potential.
In Section~\ref{s:models}  we present the results for two fermion different representation case.
In Section~\ref{s:minmodel}  we present the properties of each successful model. In Section~\ref{s:collider-boundstate}, we discuss the bound state 
formalism of the colour vector-like fermions and limit from LHC. We close with a conclusion and outlook.
In Appendix~\ref{a:RepDyn} we have tabulated all forty representations 
of $SU(3)\times SU(2)\times U(1)$ coming from $SU(5)$ representations upto dimension 75 \cite{Slansky:1981yr} with their Dynkin index. 
In Appendix~\ref{a:mixing-SM} constraints on mixing between SM fermions with vector-like quark is summarized. Appendix~\ref{a:2lbeta}
summaries the two-loop RG equation of Standard Model. 

\section{Recap of essential RG}\label{s:recap-rg}
\subsection{One loop gauge unification}
It is well known that gauge couplings do not unify precisely in the Standard Model. 
If one insists on unification of the  guage couplings 
at the GUT scale, the required $\sin^{2}\theta_W (M_{Z}^{2})$  is 0.204 (for one loop beta functions)
instead of the current experimental value of 
$\sin^{2}\theta_W (M_{Z}^{2})$ = 0.23129 $\pm$ 0.00005 \cite{1674-1137-40-10-100001}. 
As argued in the introduction, in the present work, 
we look for additional vector-like matter fermions, close to the weak scale, 
which can compensate the deviation and lead to successful gauge coupling unification. 
At the 1-loop level, the beta functions for the three gauge couplings are given as 
\begin{eqnarray}\label{e:gcoup}
 \frac{d g_l}{d t} = -\frac{1}{16\pi^2}b_l g^3_l, \mbox{ where  } t &=& \ln\mu, 
\end{eqnarray}
where is $l = \{U(1), SU(2), SU(3)\}$ runs over all the three gauge groups. The $b_l$ functions have the
general form:
\begin{eqnarray}\label{e:one-loop}
b_l &=& \Big[ \frac{11}{3} C(V_{l}) -\frac23  T(F_{l}) - \frac13 
T(S_{l}) \Big].
\end{eqnarray}

Here $C(R)$ is quadratic Casimir and $T(R)$ is Dynkin index of representation 
R. $V, F$ and $S$ represents vector, Weyl fermion and complex scalar field 
respectively.  For U(1) group $T(R_{1})$ and $C(R_{1})$  are
\begin{eqnarray}
T(R_{1})=C(R_{1}) = \frac35 Y^2.
\end{eqnarray}
For SU(N) group $T(R)$ is defined as follows:
\begin{equation}
Tr[R^iR^j] = T(R)\delta^{ij}.
\end{equation}
The following are the list of Dynkin Indices for 
lower dimensional Representations:
\begin{center}
 \begin{tabular}{l|c}
\hline
Representation & $T(N)$\\
\hline
Fundamental & $\frac12$\\
Adjoint & $n$\\
Second Rank anti-symmetric tensor & $\frac{n-2}{2}$\\
Second Rank symmetric tensor & $\frac{n+2}{2}$\\
\hline
 \end{tabular}
\end{center}
More complete list on quadratic casimirs can be found in \cite{Aranda:2009wh}. 
Within SM, the beta functions take the value
\begin{eqnarray}
b_1^0= -\frac{41}{10},\, b_2^0 = \frac{19}{6}, \mbox{ and } b_3^0 = 7.
\end{eqnarray}

In the presence of a vector-like fermion $V_1$  at the scale $M_1$ greater than weak scale,
given the gauge coupling unification at $M_{GUT}$, the (\Eqn{e:gcoup}) take the form: 
\begin{equation}
\label{1e:msssrge}
\alpha^{-1}_l (\mu_{in})= \frac{b^{0}_l}{2 \pi} \ln\frac{\mu_{in}}{M_{GUT}} 
+ \frac{b^{{\rm V_1}}_l}{2 \pi} \ln\frac{M_{1}}{M_{GUT}} + \alpha^{-1}_l
(M_{GUT}),
\end{equation}
where $\alpha_{l}=\dfrac{g_{l}^{2}}{4{\pi}}$ and $b^{{\rm V_1}}_l$ capture effect of addition of vector-like fermions at the scale $M_{1}$.
The parameter $\bar{b}$ is an useful measure of unification of gauge
couplings. It is defined as 
\begin{eqnarray}\label{e:bbar}
\bar b (\mu_{in}) 
&=&\frac{\alpha_3^{-1}(\mu_{in}) -\alpha_2^{-1}(\mu_{in}) }{\alpha_2^{-1}(\mu_{in}) 
-\alpha_1^{-1}(\mu_{in})} \\
\label{e:bbargut}
&=& \frac {\bigtriangleup b_{32}^0 + \left(\bigtriangleup b_{32}^{{\rm V_1}}\right) 
{\ln({M_1}/{M_{GUT}})}/{\ln({\mu_{in}}/{M_{GUT}})} }{\bigtriangleup b_{21}^0 + 
\left( \bigtriangleup b_{21}^{{\rm V_1}}\right){\ln(M_1/M_{GUT})}/{\ln({\mu_{in}}/M_{GUT})} }.
\end{eqnarray}

Where the second line can be derived from \Eqn{1e:msssrge} assuming unification at $M_{GUT}$. 
The parameters $\Delta b_{lk}$ are defined as $b_l - b_k$.  In the absence of new
vector-like particles, $\bar{b}$ is independent of the running scale $\mu$. In their presence
however, there is a $\mu$ dependence but it is typically mild. For the case
where the new particles are close to weak scale $\sim $ TeV, and when $\mu_{in} = M_Z$,
the log factor, $\ln(M_1/M_{GUT})/\ln(\mu/M_{GUT})$ is close to one. In this case, 
the expression for unified theories is given by 
\begin{eqnarray}\label{e:sim-bbar}
\bar b = \frac {\bigtriangleup b_{32}^0 + \bigtriangleup b_{32}^{{\rm V_1}} 
 }{\bigtriangleup b_{21}^0 + \bigtriangleup b_{21}^{{\rm V_1}}}
\end{eqnarray}

Note that the \Eqn{e:bbar} can purely be determined from experiments at $M_Z$. 
It’s value is given by 

\begin{eqnarray}
\bar b(M_Z) = 0.718,
\end{eqnarray}

In the SM, if we insist on unified gauge couplings at $M_{GUT}$, at the weak scale, $\bar{b}$
takes the value 0.5 clearly in conflict with experiments. In MSSM, $\bar{b}$ turns out
to be $ 5/7$. Of course, these arguments are valid only at one loop. There is
 deviation in \Eqn{e:sim-bbar} when higher loops are considered.  In our analysis, 
most of the successful  models have a $\bar{b}$ of $0.67$ to $0.833$.
The above discussion can be easily generalised for more than one Vector field $V_i$ at scales $M_i$. 
It has the following general form at the 1-loop level. 
\begin{eqnarray}\label{e:bbargen}
\bar b (\mu) 
&=& \frac {\Delta b_{32}^0 + \sum\limits_{i} \left(\Delta b_{32}^{{\rm V_i}} 
{\ln({M_i}/{M_{GUT}})}\right)/{\ln({\mu}/{M_{GUT}})} }{\Delta b_{21}^0 + 
\sum\limits_{i} \left( \Delta b_{21}^{{\rm V_i}}{\ln(M_i/M_{GUT})}\right)/{\ln({\mu}/M_{GUT})} }.
\end{eqnarray}
where we assumed the hierarchy of the scales as $M_1 < M_2 < M_3 $ etc.

\subsection{Two loop RG evolution of gauge couplings} 

To improve the precision in unification of gauge couplings, we consider two loop beta functions. At the two loop level, the beta functions
involve Yukawa couplings which makes them model dependent. Vector-like fermions  which typically have ``bare'' mass terms
in the Lagrangian, can also mix with the Standard Model fermions through Yukawa interactions if allowed by the
gauge symmetry. However, this mixing is subject to strong phenomenological
constraints \cite{delAguila:2008pw, Aguilar-Saavedra:2013qpa, Ishiwata:2015cga, delAguila:2000rc, Bizot:2015zaa}.
A detailed Discussion on the mixing constraints can be found in Appendix~\ref{a:mixing-SM}. 

In the present analysis, we restrict ourselves to models with minimal or zero vector-like fermion and SM mixing through 
the Higgs mechanism. 
With this assumption, we can safely neglect the Yukawa contribution from the new sector to the gauge coupling unification. 
The RG equations at the two loop level are given by \cite{Jones:1981we, Arason:1991ic, Luo:2002ey}: 
\begin{equation}\label{e:2lrg}
 \frac{dg_{l}}{dt}=-b_l\frac{g_l^3}{16\pi^2}-\sum_k m_{lk} \frac{g_l^3g_k^2}{(16\pi^2)^2}-\frac{g_l^3}{(16\pi^2)^2}
 Tr\big\{C_{lu}Y^{\dag}_uY_u+C_{ld}Y^{\dag}_dY_d+C_{le}Y^{\dag}_eY_e\big\},
\end{equation}
where the first term in the right hand side is due to one-loop which was 
discussed in the previous subsection. The second term is purely from gauge 
interactions whereas the third terms involves the Yukawa terms $Y_{u,d,e}$ 
where the suffixes mean the up-type,down-type and lepton-type  
couplings. The expression for the coefficients appearing in the second term of 
the above equation are as follows \cite{Jones:1981we, Luo:2002ti}:
\begin{eqnarray}
 m_{lk}&=&\big(2C(F_k)d(F_k)T(F_l)d(F_m)+4C(S_k)d(S_k)T(S_k)d(S_m)\big)
 \mbox{ where } l\neq k 
 \\
 m_{ll}&=&\Big[\frac{10}{3}C(V_l)+2C(F_l)\Big]T(F_l)d(F_m)d(F_k)+
 \Big[\frac{2}{3}C(V_l)+4C(S_l)\Big]T(S_l)d(S_m)d(S_k) \nonumber
 \\&-&
  \frac{34}{3}[C(V_l)^2],
\end{eqnarray}
where $d(R)$ means dimension of the representation $R$ and other factors C(R) and T(R)
are already defined in \Eqn{e:one-loop}.

For the Standard Model, the values of $m_{lk}$ are as follows:
\begin{eqnarray}
 m^0 &=& -\left(  
 \begin{array}{ccc}
 \frac{199}{50} & \frac{27}{10} & \frac{44}{5}\\
  \frac{9}{10}  & \frac{35}{6}  & 12\\
  \frac{11}{10} & \frac{9}{2}   & -26
 \end{array}
 \right).
\end{eqnarray}

In the third term of \Eqn{e:2lrg}, we have the coefficients $C_{lf}$ and for 
the standard model particles it has the following form:
\begin{equation}
C^0=\left(\begin{array}{ccc}
       \frac{17}{10} & \frac{1}{2} & \frac{3}{2}\\
       \frac{3}{2} & \frac{3}{2} & \frac{1}{2} \\
       2  & 2 & 0
      \end{array}\right).
\end{equation}
 As we are considering the Yukawa couplings between the vector-like fermions with Higgs 
boson  to be negligible\footnote{This can be organised by imposing discrete symmetries distinguishing 
SM partners from vector-like fermions}, the contribution of vector-like particles to $C_{lf}$ coefficient can 
be taken as zero. On the other hand $\delta m_{ij}\neq 0$, where $\delta$ is used to indicate contribution 
from additional vector-like fermions. We'll give explicit values of $\delta m_{ij}$ for each of the
viable models in Section~\ref{s:minmodel}.

\begin{figure}
\includegraphics[height=3 cm,width=16 cm]{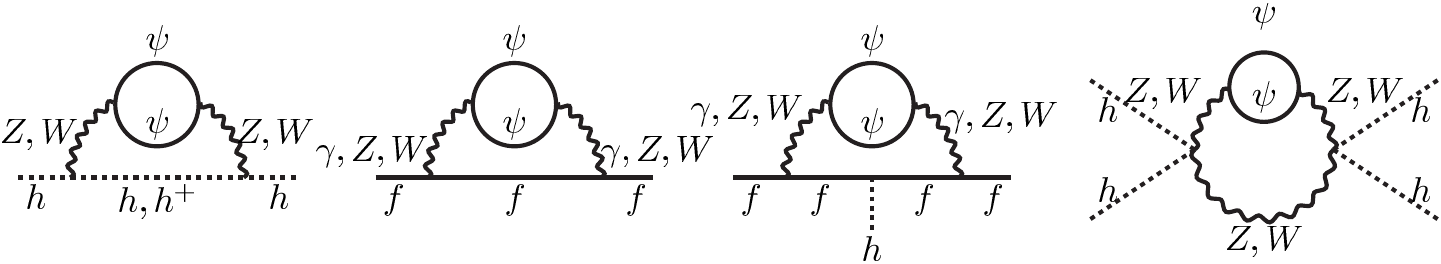}
\caption{{\sf Diagrams Contributing in Two loop RG  of Yukawa 
and Higgs quartic couplings, through new Fermion fields ($\psi$).
Here $f$ is any standard model fermion. First two diagrams 
correspond to anomalous dimension and the last two diagrams 
are giving vertex corrections.}}\label{f:two_loop_dig}
\end{figure}

Two-loop RG running for the Yukawa couplings is given as

\begin{eqnarray}\lab{e:rg-y}
Y_{u,d,e}^{-1} \frac{dY_{u,d,e}}{dt} &=& \frac{1}{16\pi^2}{\beta}^{(1)SM}_{u,d,e}+
 \frac{1}{(16\pi^2)^2}{\beta}^{(2)SM}_{u,d,e} \label{e:rg-lam}
\end{eqnarray}
The SM RG for these  Yukawa couplings are shown in Appendix~\ref{a:2lbeta}. 
Here we will address the effect of new fermion fields in RG of Yukawa  couplings 
\cite{Luo:2002ti, Machacek:1984zw, Machacek:1983fi, Machacek:1983tz}. The one 
loop beta functions of these couplings are not be affected by new 
matter(fermion) fields because we considered the Yukawa couplings 
between the vector-like fermions with Higgs boson  to be negligible. Two loop beta 
functions get contributions from the diagrams shown in~\fig{f:two_loop_dig}, which 
results in the following terms:
\begin{eqnarray}
 \delta
\beta_{u}^{(2)V} 
&=& \frac{40}{9} g_{3}^{4} T(F_{3})d(F_2)d(F_1) + \frac{29}{90} g_{1}^{4} 
T(F_{1}) d(F_3) d(F_2)\nonumber\\
&+& \frac{1}{2}g_{2}^{4}T(F_{2})d(F_3)d(F_1)\\
 \delta \beta_{d}^{(2)V} 
&=& \frac{40}{9} g_{3}^{4} T(F_{3})d(F_2)d(F_1)- \frac{1}{90} g_{1}^{4} 
T(F_{1})d(F_3)d(F_2) \nonumber\\
 &+& \frac{1}{2}g_{2}^{4}T(F_{2})d(F_3)d(F_1)\\
 \delta \beta_{e}^{(2)V} 
&=& \frac{11}{10} g_{1}^{4} T(F_{1})d(F_3)d(F_2) + \frac{1}{2}g_{2}^{4}T(F_{2})d(F_3) d(F_1)
\end{eqnarray}

\subsection{Evolution of Higgs Self coupling}

The modification of the gauge beta functions in the presence of additional vector-like particles 
can have implications on the evolution of the Higgs self coupling. At the outset, one might consider
that since there are no new large Yukawa couplings\footnote{Firstly by assumption, As we will see in the next section, 
this is automatic in most models as Yukawa couplings with new vector-like fermions are not gauge invariant.}, 
the evolution of the 
Higgs self coupling might be in the safe region. While this is true, the evolution of the SM Yukawa
couplings is itself modified in these models as seen in the previous sub-section. It is thus worthwhile
to check explicitly the stability of Higgs self coupling along with gauge coupling unification. 

To check the Higgs stability we follow \cite{EliasMiro:2011aa, 
Degrassi:2012ry, Buttazzo:2013uya} who have checked for the stability using 
three loop beta functions and NNLO matching conditions. We use the beta function of 
the $\lambda$ at the two loop and put a condition that $\lambda$ is always
positive at all scales of evolution. Two-loop RG running for the Higgs quartic coupling are 
shown below
\begin{eqnarray}
\frac{d\lambda}{dt} 
&=& \frac{1}{16\pi^{2}}\beta^{(1)SM}_{\lambda} + \frac{1}{(16\pi^{2})^{2}} 
\beta^{(2)SM}_{\lambda},
\end{eqnarray}
where beta functions for SM Higgs quartic couplings are defined in Appendix~\ref{a:2lbeta}.
The effect of new fermion fields in RG of Higgs quartic couplings are:
\begin{eqnarray}
 \delta \beta_{\lambda}^{(2)V} 
&=& -\frac{1}{25} g_{1}^{4}\left(12g_{1}^{2} + 20g_{2}^{2} - 
25{\lambda}\right)T(F_{1} )d(F_3)d(F_2) \nonumber\\
&-& \frac{1}{5}g_{2}^{4} \left(4g_{1}^{2}+20g_{2}^{2}-25{\lambda}\right)  
T(F_{2}) d(F_3) d(F_1)
\end{eqnarray}

To solve the RG equations we need boundary values  of the coupling constants 
and masses at the top mass ($M_t$) scale. The quantities of interest are Higgs 
quartic coupling ($\lambda$), Yukawa couplings and gauge coupling, which can be 
calculated in terms of physical observables W-boson mass ($M_W$), Z-boson mass 
($M_Z$), Higgs mass ($M_h$) and $\alpha_3(M_Z)$ at the two loop level. The input 
parameters are calculated in the $\overline{MS}$-scheme. More detailed  
can be found in \cite{Buttazzo:2013uya}. For the RG running we use the central value of Top mass.
The input values of SM parameters and couplings are
listed in \tabl{t:parameters}.

\begin{table}
\begin{tabular}{c c c c c}
\hline
  Parameter & Value  & Description &
  \\
\hline
  $M_W$  & 80.384 $\pm$ 0.015 GeV & Pole mass of W boson  
\cite{Agashe:2014kda}
  \\
  $M_Z$  & 91.1876 $\pm$ 0.0021 GeV & Pole mass of Z boson 
\cite{Agashe:2014kda}

  \\
  $M_h$  & 125.09 $\pm$ 0.21 $\pm$ 0.11 GeV & Pole mass of Higgs boson 
\cite{Aad:2015zhl}
  \\
  $M_t$  & 173.34 $\pm$ 0.76 $\pm$ 0.3 GeV & Pole mass of top quark 
\cite{ATLAS:2014wva}
  \\
  $\alpha_3(M_Z)$  & 0.1184 $\pm$ 0.0007  & $\overline{MS}$ gauge SU(3)$_c$ 
coupling  \cite{Bethke:2012zza}\\
\hline
\end{tabular}
  \caption{Input values of SM observales used to fix the SM fundamental 
parameters.}\label{t:parameters}
\end{table}

Values of the relevant couplings  at scale $M_t$  are as follows:

\begin{eqnarray}
\frac{\lambda(M_t)}{2} &=& 0.12604 + 0.00206 
(M_h-125.15)-0.00004(M_t-173.34)\pm 0.00030_{th},\\*
y_t(M_t)&=&0.93690 + 0.00556(M_t-173.34)
- 0.00042 \left(\alpha_{3}(M_{Z})-0.1184 \right)/0.0007,\\*
g_2(M_t) & = & 0.64779 + 0.00004(M_t 
-173.34)+0.00011\frac{M_W-80.384}{0.014},\\*
g_Y(M_t)&=& 0.35830 + 0.00011(M_t -173.34)+0.00020\frac{M_W-80.384}{0.014},\\*
g_3(M_t) &=& 1.1666+0.00314\frac{\alpha_{3}(M_{Z})-0.1184}{0.0007}
-0.00046(M_t-173.34),
\end{eqnarray}
where all the parameters with mass dimension has written in GeV. Central 
values of the above couplings are calculated upto NNLO (\cite{Degrassi:2012ry} for $\lambda$)
order for all of them except the  $y_t(M_t)$ for which we considered 
NNNLO \cite{Chetyrkin:1999ys, Chetyrkin:1999qi, Melnikov:2000qh}. 
The value of $\alpha_{3}(M_Z)$, is extracted from the global fit of 
Ref.~\cite{Bethke:2012zza} in the effective SM with 5 flavours. Including RG 
running from $M_Z$ to $M_t$ at 4 loops in QCD and at 2 loops in the electroweak 
gauge interactions, and 3 loop QCD matching at $M_t$ to the full SM with 6 
flavours, the strong gauge couplig is calculated.  The contribution of the 
bottom and tau Yukawa couplings, are computed from the $\overline{MS}$ b-quark 
mass ($M_b(M_t)=2.75$ GeV) and Tau mass ($M_\tau(M_t)=1.742$ GeV) 
\cite{Xing:2007fb}.

\paragraph{Threshold Corrections at GUT Scale}
One of the main concerns which remains now is the possible effect of threshold corrections at the GUT scale, 
which can be quite significant. These  corrections are highly model dependent.  In some GUT models, with no
extra matter at the weak scale (other than the Standard Model particle content), it is possible to achieve gauge
coupling unification through large  threshold corrections at the GUT scale \cite{Lavoura:1993su}.
 While such extreme situations are no longer valid due to the constraint on the stability of the Standard Model Higgs potential,
 it is still possible that GUT scale threshold corrections could play an important role.  To study the impact of threshold corrections
 on gauge coupling unification, we define  the following parameters:
 $\alpha_{ave.}(\mu) = (\alpha_{1}(\mu) + \alpha_{2} (\mu) + \alpha_{3}(\mu) )/3$ and 
  $\bar{\bigtriangleup_i}(\mu)= (\alpha_{i} (\mu)-\alpha_{avg}(\mu))/\alpha_{ave}(\mu) $.  Note that  $\alpha_{ave}$ coincides with $\alpha_{GUT}$  when all $\bar{\bigtriangleup_i} \to 0$, at the scale $M_{GUT}$. In the presence of threshold corrections, one could
  allow for deviations in $\alpha_{GUT}$  in terms of  $\bar{\bigtriangleup_i}$ at the GUT scale\footnote{Another model independent
  parameterisation  for the threshold corrections was presented in \cite{Ellis:2015jwa}.}
  Defining  $\bigtriangleup = max(\bar{\bigtriangleup_i)}$, we see that $\bigtriangleup$ is as large as $6\%$ in the 
  Standard Model.  In our survey of models below,  we have allowed for variations in $\bigtriangleup$ 
up to $1.2\%$.  A more conservative set of models is  tabulated in 
Appendix~\ref{a:fermdelta}  which have $\bigtriangleup$ of 3$\%$.  

   \paragraph {Proton Decay}
Models studied in this work can lead to proton decay mediated by the gauge bosons at GUT scale. 
The lifetime of proton decay is extremely sensitive to the heavy gauge bosons ($M_{(X,Y)} \sim M_{GUT}$).
For these models,  using the simple decay width formulae, $\Gamma \sim
\alpha_{Gut}\frac{m_{proton}^5}{M_{GUT}^{4}}$ we estimate the 
life time of the  proton, where the current experimental 
value is of order $ > 10^{32}-10^{34} $ years \cite{Takhistov:2016eqm}.

\section{Gauge coupling unification with vector-like fermions}\label{s:models} 

As mentioned in Introduction, in our search for successful models with gauge coupling unification,   
we focus on vector-like matter in incomplete representations of SU(5).  We have considered (incomplete) 
representations~\cite{Slansky:1981yr}  up-to 
dimension 75,  which contains a 15  of SU(3) of QCD as the largest component. The full list of incomplete representations
is presented in Appendix~\ref{a:RepDyn}.  As can be seen from the \tabl{t:rep-list}, there are 40 representations which
we have considered.
Note that representations 4, 5  in \tabl{t:rep-list}  do not come as incomplete representations of SU(5) instead they are 
singlet representations
of SU(5).    Our search strategy is start with $n_i$ copies of representation $i$ , with all  the $n_i$ copies 
degenerate in mass, $m_i$  and look for unification of the gauge couplings.  The maximum number of copies is taken
to be 10.  The number of representation types $i$ considered simultaneously  is restricted up to four.   An important
constraint comes from proton decay, which restricts the scale of unification to lie above (at least) $10^{15}$ GeV. 
As mentioned above,  in addition to unification, we also consider that the Higgs potential should be stable all 
the way up to the GUT scale.  In the computations,  we have also varied the input parameters to lie within 
their two sigma regions. The masses of new vector-like are assumed to lie between 250 GeV - 5 TeV. 

For $i=1$ we searched for the mass of the vector-like fermion from 250 GeV - 5TeV, considering number of vector-like 
fermions $n_1 = 6$. These masses for $n_1$ copies have been considered degenerate for 
simplicity and no successful model was observed.
The simplest solutions we found contain at least
two different representation content each with a different number of copies.   We call these solutions 
 ``minimal unificon models''.  These are listed in \tabl{t:two-ferm}.  We now explain the notation used 
in the Table.  The two representations considered are called Rep1 and Rep2.  The representation is 
described as $n_i ( R_{SU(3)}, R_ {SU(2)}, R_{U(1)})$, where $n_i$ introduced earlier is the number
of copies of the representation, $R_{G}$ is the representation of the field under the gauge group
G of the SM.
\begin{table}[H]
\centering
$
{\small
\begin{array}{|c|c|c||c|c||c|c|c|c|c|c|}
\hline\hline
{\rm Mod}&{\rm Rep}\, 1 & M_{Rep1}& {\rm Rep}\, 2 & M_{Rep2} & {\rm One} & 
{\rm Two} & {\rm Vaccum} & M_{\rm GUT}  & \alpha_{\rm GUT} \\
 {\rm No.}& & {\rm GeV} & & {\rm GeV} &{\rm loop} & {\rm loop} & {\rm 
Stability} & \times10^{16} {\rm GeV} & \\ \hline

 1& 6 \left( 1, 2, \frac{1}{2} \right) &\left( 250-5000\right) & 1 \left( 6, 1, 
\frac{1}{3} \right) &\left(250-5000\right)&\checkmark &\checkmark &\checkmark & \sim 0.11 
& \sim 0.038
\\ \hline

  2&6 \left( 1, 2, \frac{1}{2} \right) &\left( 250-2000\right) & 2 \left( 8, 1, 0 
\right) &\left(500-5000\right)& \checkmark&\checkmark&\checkmark& \sim 2.34 &  
\sim 0.040
\\ \hline

  3&2 \left( 1, 3, 0 \right) &\left( 250-5000\right) & 4 \left( 3, 1, \frac{1}{3} 
\right) &\left(250-5000\right)&\checkmark &\checkmark&\checkmark& \sim 2.29 
&\sim 0.030
\\ \hline

  4&2 \left( 3, 1, \frac{2}{3} \right) &\left( 250-5000\right) & 2 \left( 3, 2, 
\frac{1}{6} \right) &\left(250-4500\right)&\checkmark &\checkmark&\checkmark&\sim 4.79 
&\sim 0.040
\\ \hline

5&3 \left( 1, 3, 0 \right) &\left( 1800-5000\right) & 1 \left( 6, 1, 
\frac{2}{3} \right) &\left(250-950\right)&\checkmark & \checkmark& \checkmark&\sim 1.08 &\sim 
0.037
\\ \hline

 6&1 \left( 1, 4, \frac{1}{2} \right) &\left( 250-2000\right) & 2 \left( 6, 1, 
\frac{2}{3} \right) &\left(1000-5000\right)&\checkmark & \checkmark& \checkmark&\sim 8.58 &\sim 
0.107
\\ \hline

7&1 \left( 3, 1, \frac{1}{3} \right) &\left( 250-5000\right) & 1 \left( 3, 2, 
\frac{1}{6} \right) &\left(250-5000\right)&\checkmark &\checkmark & \checkmark&\sim 2.20 &\sim 
0.028
\\ \hline

8&4 \left( 1, 2, \frac{1}{2} \right) &\left( 300-5000\right) & 1 \left( 8, 1, 
0 \right) &\left(300-5000\right)& \checkmark &\checkmark & \checkmark&\sim 0.10 &\sim 
0.030
\\ \hline

9&3 \left( 1, 3, 0 \right) &\left( 1100 - 5000\right) & 6 \left( 3, 1, 
\frac{1}{3} \right) &\left(250-1800\right)& \checkmark &\checkmark & \checkmark&\sim 25.0 &\sim 
0.037
\\ \hline

\end{array}
}
$
\caption{Model with two vector-like fermions representation satisfying gauge coupling 
unification and vacuum stability condition, considering $\bigtriangleup$ of 1.2$\%$.}\label{t:two-ferm}
\end{table}

Furthermore,  in the above, we mentioned only one part of the representation instead 
of the complete vector multiplet for brevity. For example, $\left( 1, 2, 
\frac{1}{2} \right)$ actually means $\left( 1, 2, \frac{1}{2} \right)\oplus 
\left( 1, 2, - \frac{1}{2} \right)$. Colored representations like $\left( 3, 
1, \frac{2}{3} \right)$ may mean two possibilities: (a) $\left( 3, 
1, \frac{2}{3} \right) \oplus \left(\bar 3, 1, -\frac{2}{3} \right)$ and (b) 
$\left(\bar 3, 1, \frac{2}{3} \right) \oplus \left( 3, 1, -\frac{2}{3} 
\right)$. On the other hand, the real representations like $\left( 1, 3, 0 
\right)$ and $\left( 8, 1, 0 \right)$ are not short-hand notations. In the 
second last column, the entries are written in units of $10^{16}$ GeV.  Thus except 
the first model, all the models have unification scale larger than $10^{16}$ GeV. 
All models appeared as the solution of one loop RG equation. 
In the third and fifth columns, we show 
the mass range of the vector-like fields. One can see that if we increase the mass of 
one representation, the mass of the other field also increases (as shown in 
\fig{f:mod1}(b), \fig{f:mod2}(b), \fig{f:mod3}(b), \fig{f:mod4}(b), \fig{f:mod5}(b), \fig{f:mod6}(b), \fig{f:mod7}(b), \fig{f:mod8}(b) and \fig{f:mod9}(b)). 

Solutions with three types of representations are  also possible. These are listed in \tabl{t:three-ferm} of
Appendix~\ref{a:three-ferm}.  Here we made a restricted choice that all the representations and their copies 
are degenerate  in mass of about 1 TeV.  As can been seen from the Table,  the minimum number 
of extra vector-like fermions required is seven over the three representations, where as the maximum
number is eighteen. 
All of them have unification scale less than $10^{16}$ GeV,  which puts them
at risk with Proton decay. 
The life time of the  proton in these models is of order $\sim 10^{32}$ years which in contrast with the experimental 
value $ > 10^{32}-10^{34} $ years \cite{Takhistov:2016eqm}. 
The maximum number of representations we have chosen simultaneously is four. 
Searching for models with different masses for each copy and each 
representation is computationally very intensive.    Thus,  we have considered 
all the four representations and their copies to be degenerate in mass at 1 TeV. 
The list of successful models is given in \tabl{t:four-ferm} of Appendix~\ref{a:four-ferm}.
The minimum number of vector-like particles required over all representations is five and the
maximum is twenty.  As with the three representation case,  we find that 
the Unification scale is smaller than $10^{16}$ GeV with the exception of one
model  (Model No 17 of \tabl{t:four-ferm}).  As before from the arguments of Proton 
decay, these models can have potentially small proton life times in conflict
with experiment.   We do not address this issue here.  
\section{Minimal unificon models}\label{s:minmodel}

In this section we concentrate on the Minimal vector-like fermion Unification Models.  The list
of such of models is given in \tabl{t:two-ferm}.  Several interesting features are evident
from the \tabl{t:two-ferm}. 

(a)  Except for the first and eigth model, all the models have unification scale above $10^{16}$ 
and thus are safe with proton decay. 
(b) The minimalist model is model 7, with only two vector-like fermions one with
a mass range of 0.250-5 TeV and another within  a mass range of 250-5000 GeV.  This model
might have constraints from direct searches of vector-like quarks at LHC and elsewhere
if there is significant mixing with SM particles.  In its absence, as we assumed here, 
the bound will be different.  We will discuss it in the next section.
(c)  The maximum number of vector-like fermions needed  is nine in Model 9. 

We now discuss each of these models in detail.

\subsection{Model 1}
\begin{figure}[t]
\centering
\subfigure[]{
	\includegraphics[height=5.5 cm,width=7cm]{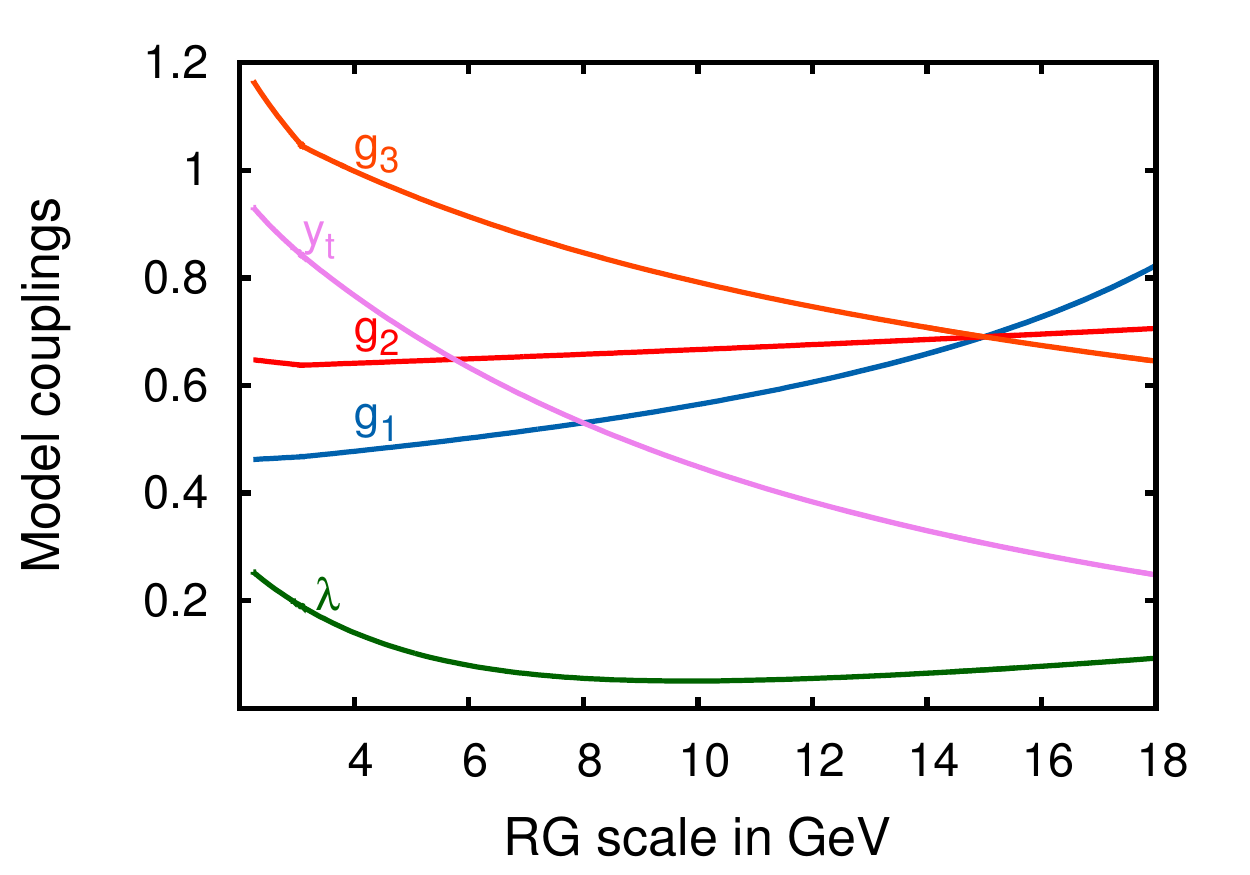}}
\subfigure[]{
	\includegraphics[height=5.4 cm,width=7.3cm]{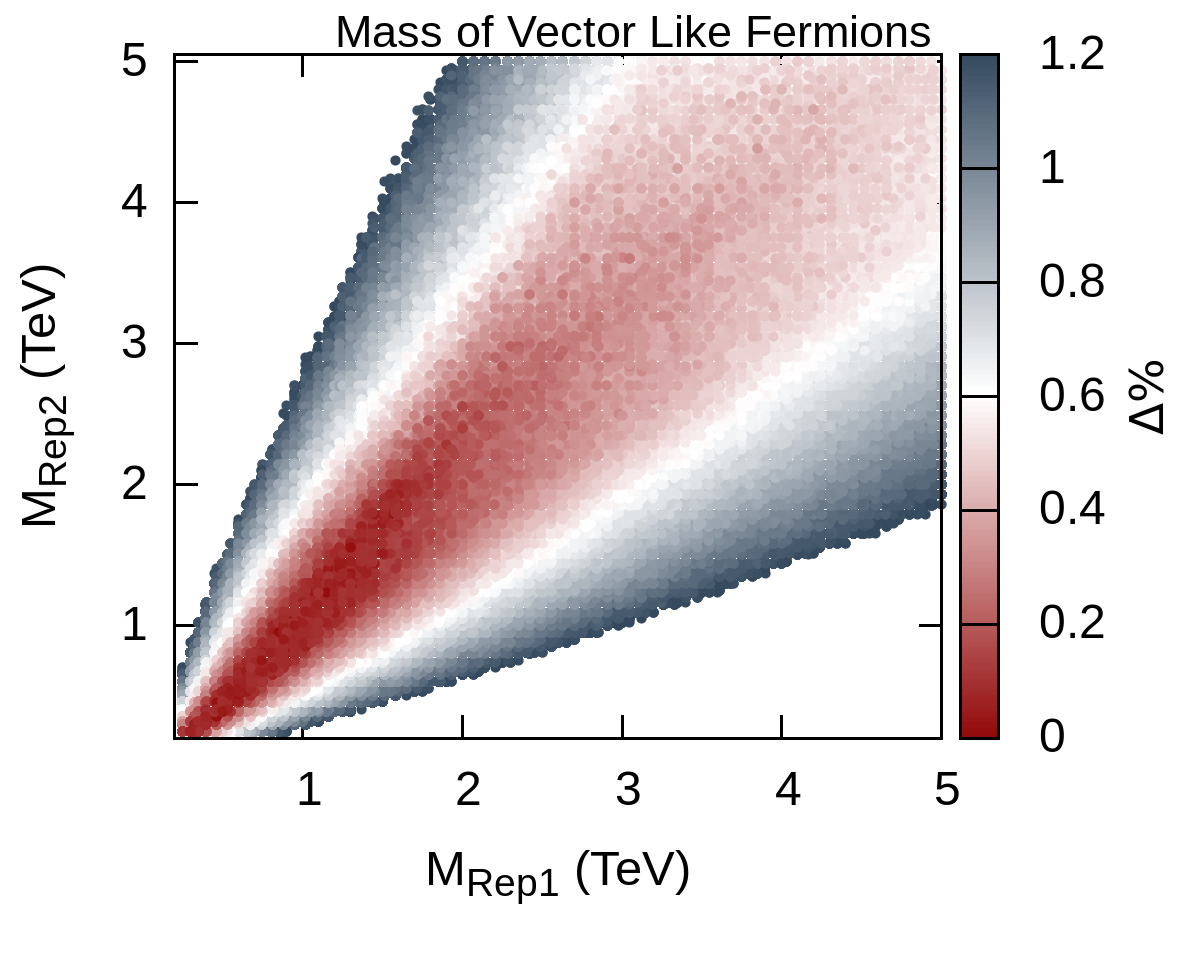}}
\caption{\sf Model 1: Fig. (a) Gauge couplings ($g_1$,$g_2$,$g_3$) 
unification and Vacuum stability ($\lambda > 0$)
plot, considering vector-like fermion in Rep.1 of mass 1210 GeV and Rep.2 of mass 1260 
GeV. Fig. (b) Mass range allowed for vector-like fermions in Rep.1 and Rep.2 
 for gauge unification and Vacuum stability.}\label{f:mod1}
\end{figure}

In this model\footnote{We have cross-checked our Two Loop RG equation of this model with the publicly 
code SARAH \cite{Staub:2008uz} for consistency.}, we have  six copies of $\left(1, 2, \frac12\right)$, which we 
called Rep1, with mass range between 250 GeV to 5000 GeV and one copy of 
$\left(6, 1, \frac13\right)$, called Rep2, with mass range from 250 GeV to 5000 
GeV. Rep1 field is lepton doublet like field and thus it can interact with right handed 
electron and the Higgs field through Yukawa interactions. This field mainly decays to gauge bosons like 
$Z$ boson and $W^\pm$. For the sake of 
simplicity of the two loop gauge coupling RG running, we impose appropriate  $Z_N$ 
symmetries to these fermion doublets. This symmetry 
cut-down all the Yukawa terms involving these fields at the 
renormalisable level and only gauge couplings are allowed.  Lightest neutral component of these fermions can be a dark 
matter candidate. This type of dark matter is called inert  
fermion doublet dark matter~\cite{Arina:2011cu, Arina:2012aj}.
Rep2 is more exotic and at the renormalisation level, 
it can interact with the gauge bosons only. It cannot decay to 
any standard model particles. Thus they form bound states. Phenomenology of 
this is studied in detail in the next Section~\ref{s:collider-boundstate}.

For most points in this model vector-like fermions in Rep1 can be degenerate
with vector-like fermions in Rep2 ($M_{Rep1}\sim M_{Rep2}$) as
shown in \fig{f:mod1}(b). However, there could be points in which 
either of the $M_{Rep1}> M_{Rep2}$ and $M_{Rep1}< M_{Rep2}$ are possible.
The change in the beta functions in these three 
possibilities are as follows:

\paragraph {(a) $M_{Rep1} = M_{Rep2}$}
\subparagraph{(I) $\mu > M_{Rep2}=M_{Rep1}$}
\begin{eqnarray}
\delta b_i(\mu > M_{Rep2} ) &=& 
\left(
\begin{array}{c}
\frac{44}{15} \\ 4 \\ \frac{10}{3} 
 \end{array}
 \right),\,\, \hphantom{< M_{Rep2}}
\delta m_{ij}(\mu > M_{Rep2} )=
\left(
\begin{array}{ccc}
  \frac{178}{150}  & \frac{54}{10}  & \frac{80}{15} \\
\frac{18}{10}  & 49 & 0 \\
 \frac{2}{3}  & 0    & \frac{250}{3} 
 \end{array}
  \right)
\end{eqnarray}
\begin{eqnarray}
 \delta \beta_{u}^{(2)}(\mu > M_{Rep2} )&=&\frac{200}{9}g_{3}^{4}+\frac{1276}{900}g_{1}^{4}\nonumber
		       + \frac{6}{2}g_{2}^{4}\\
\delta \beta_{d}^{(2)}(\mu > M_{Rep2} )&=&\frac{200}{9}g_{3}^{4}-\frac{44}{900}g_{1}^{4} \nonumber
		       + \frac{6}{2}g_{2}^{4}\\
\delta \beta_{e}^{(2)}(\mu > M_{Rep2} )&=&\frac{44}{100}g_{1}^{4}+\frac{6}{2}g_{2}^{4}
\end{eqnarray}
\begin{eqnarray}
 \delta \beta_{\lambda}^{(2)}(\mu > M_{Rep2} )&=&-\frac{44}{250}g_{1}^{4}\left(12g_{1}^{2}+20g_{2}^{2}-25{\lambda}\right) \nonumber
		       -\frac{6}{5}g_{2}^{4}\left(4g_{1}^{2}+20g_{2}^{2}-25{\lambda}\right)
\end{eqnarray}

\paragraph{(b) $M_{Rep1} > M_{Rep2}$}
\subparagraph{(I) $M_{Rep1} > \mu > M_{Rep2}$}
\begin{eqnarray}
\delta b_i(M_{Rep2}< \mu < M_{Rep1} ) &=& 
\left(
\begin{array}{c}
\frac{8}{15} \\0\\ \frac{10}{3} 
 \end{array}
 \right),\,\,
\delta m_{ij}(M_{Rep2}< \mu < M_{Rep1} )=
\left(
\begin{array}{ccc}
 \frac{16}{150}  & 0 &\frac{80}{15}   \\
 0      & 0 & 0\\
  \frac{2}{3} & 0 & \frac{250}{3} 
 \end{array}
  \right)\\ \nonumber
\end{eqnarray}

\begin{eqnarray}
 \delta \beta_{u}^{(2)}(M_{Rep2}< \mu < M_{Rep1} )&=&\frac{200}{9}g_{3}^{4}+\frac{232}{900}g_{1}^{4}\nonumber\\
 \delta \beta_{d}^{(2)}(M_{Rep2}< \mu < M_{Rep1} )&=&\frac{200}{9}g_{3}^{4}-\frac{8}{900}g_{1}^{4}\nonumber\\
 \delta \beta_{e}^{(2)}(M_{Rep2}< \mu < M_{Rep1} )&=&\frac{88}{10}g_{2}^{4}
\end{eqnarray}
\begin{eqnarray}
 \delta \beta_{\lambda}^{(2)}(M_{Rep2}< \mu < M_{Rep1} )&=&-\frac{8}{250}g_{1}^{4}\left(12g_{1}^{2}+20g_{2}^{2}-25{\lambda}\right)\nonumber
\end{eqnarray}

\subparagraph{(II) $\mu > M_{Rep1}$}
\begin{eqnarray}
\delta b_i(\mu > M_{Rep1} ) &=& 
\left(
\begin{array}{c}
\frac{44}{15} \\ 4 \\ \frac{10}{3} 
 \end{array}
 \right),\,\, \hphantom{< M_{Rep1},}
\delta m_{ij}(\mu > M_{Rep1} )=
\left(
\begin{array}{ccc}
  \frac{178}{150}  & \frac{54}{10}  & \frac{80}{15} \\
\frac{18}{10}  & 49 & 0 \\
 \frac{2}{3}  & 0    & \frac{250}{3} 
 \end{array}
  \right)
\end{eqnarray}
\begin{eqnarray}
 \delta \beta_{u}^{(2)}(\mu > M_{Rep1} )&=&\frac{200}{9}g_{3}^{4}+\frac{1276}{900}g_{1}^{4}\nonumber
		       + \frac{6}{2}g_{2}^{4}\\
 \delta \beta_{d}^{(2)}(\mu > M_{Rep1} )&=&\frac{200}{9}g_{3}^{4}-\frac{44}{900}g_{1}^{4} \nonumber
		       + \frac{6}{2}g_{2}^{4}\\
 \delta \beta_{e}^{(2)}(\mu > M_{Rep1} )&=&\frac{44}{100}g_{1}^{4}+\frac{6}{2}g_{2}^{4}
\end{eqnarray}
\begin{eqnarray}
 \delta \beta_{\lambda}^{(2)}(\mu > M_{Rep1} )&=&-\frac{44}{250}g_{1}^{4}\left(12g_{1}^{2}+20g_{2}^{2}-25{\lambda}\right) \nonumber
		       -\frac{6}{5}g_{2}^{4}\left(4g_{1}^{2}+20g_{2}^{2}-25{\lambda}\right)
\end{eqnarray}

\paragraph{(c) $M_{Rep1} < M_{Rep2}$}
\subparagraph{(I) $M_{Rep1} < \mu < M_{Rep2}$}
\begin{eqnarray}
\delta b_i(M_{Rep1}< \mu < M_{Rep2} ) &=& 
\left(
\begin{array}{c}
\frac{12}{5}\\4\\0
 \end{array}
 \right),\,\,
\delta m_{ij}(M_{Rep1}< \mu < M_{Rep2} )=
\left(
\begin{array}{ccc}
 \frac{54}{50}  & \frac{54}{10}  & 0\\
 \frac{18}{10}  & 49  & 0\\
 0    & 0   & 0
 \end{array}
  \right)
  \end{eqnarray}
 
\begin{eqnarray}
 \delta \beta_{u}^{(2)}(M_{Rep1}< \mu < M_{Rep2} )&=&\frac{1044}{900}g_{1}^{4}\nonumber
		       + \frac{6}{2}g_{2}^{4}\\
 \delta \beta_{d}^{(2)}(M_{Rep1}< \mu < M_{Rep2} )&=&\frac{36}{900}g_{1}^{4} \nonumber
		       + \frac{6}{2}g_{2}^{4}\\
 \delta \beta_{e}^{(2)}(M_{Rep1}< \mu < M_{Rep2} )&=&\frac{396}{100}g_{1}^{4}+\frac{6}{2}g_{2}^{4}
\end{eqnarray}
\begin{eqnarray}
 \delta \beta_{\lambda}^{(2)}(M_{Rep1}< \mu < M_{Rep2} )&=&-\frac{36}{250}g_{1}^{4}\left(12g_{1}^{2}+20g_{2}^{2}-25{\lambda}\right) \nonumber
		       -\frac{6}{5}g_{2}^{4}\left(4g_{1}^{2}+20g_{2}^{2}-25{\lambda}\right)
\end{eqnarray}
\subparagraph{(II) $\mu > M_{Rep2}$}
\begin{eqnarray}
\delta b_i(\mu > M_{Rep2} ) &=& 
\left(
\begin{array}{c}
\frac{44}{15} \\ 4 \\ \frac{10}{3} 
 \end{array}
 \right),\,\, \hphantom{< M_{Rep2},}
\delta m_{ij}(\mu > M_{Rep2} )=
\left(
\begin{array}{ccc}
  \frac{178}{150}  & \frac{54}{10}  & \frac{80}{15} \\
\frac{18}{10}  & 49 & 0 \\
 \frac{2}{3}  & 0    & \frac{250}{3} 
 \end{array}
  \right)
\end{eqnarray}
\begin{eqnarray}
 \delta \beta_{u}^{(2)}(\mu > M_{Rep2} )&=&\frac{200}{9}g_{3}^{4}+\frac{1276}{900}g_{1}^{4}\nonumber
		       + \frac{6}{2}g_{2}^{4}\\
 \delta \beta_{d}^{(2)}(\mu > M_{Rep2} )&=&\frac{200}{9}g_{3}^{4}-\frac{44}{900}g_{1}^{4} \nonumber
		       + \frac{6}{2}g_{2}^{4}\\
 \delta \beta_{e}^{(2)}(\mu > M_{Rep2} )&=&\frac{44}{100}g_{1}^{4}+\frac{6}{2}g_{2}^{4}
 \end{eqnarray}
 \begin{eqnarray}
 \delta \beta_{\lambda}^{(2)}(\mu > M_{Rep2} )&=&-\frac{44}{250}g_{1}^{4}\left(12g_{1}^{2}+20g_{2}^{2}-25{\lambda}\right) \nonumber
		       -\frac{6}{5}g_{2}^{4}\left(4g_{1}^{2}+20g_{2}^{2}-25{\lambda}\right)
\end{eqnarray}

A sample unification point is shown in \fig{f:mod1}(a), six copies of lepton like vector fermions with degenerate mass
of 1210 GeV and one copy of Rep2 with a mass of 1260 GeV is considered. The figure shows unification clearly.
The running of $y_t$ and $\lambda$ are also shown. The panel \fig{f:mod1}(b) has the mass distribution in Rep1-Rep2 mass plane. 
The model clearly prefers degeneracy of Rep1 and Rep2 for successful unification.

\subsection{Model 2} 

\begin{figure}
\centering
\subfigure[]{
\includegraphics[height=5.5 cm,width=7cm]{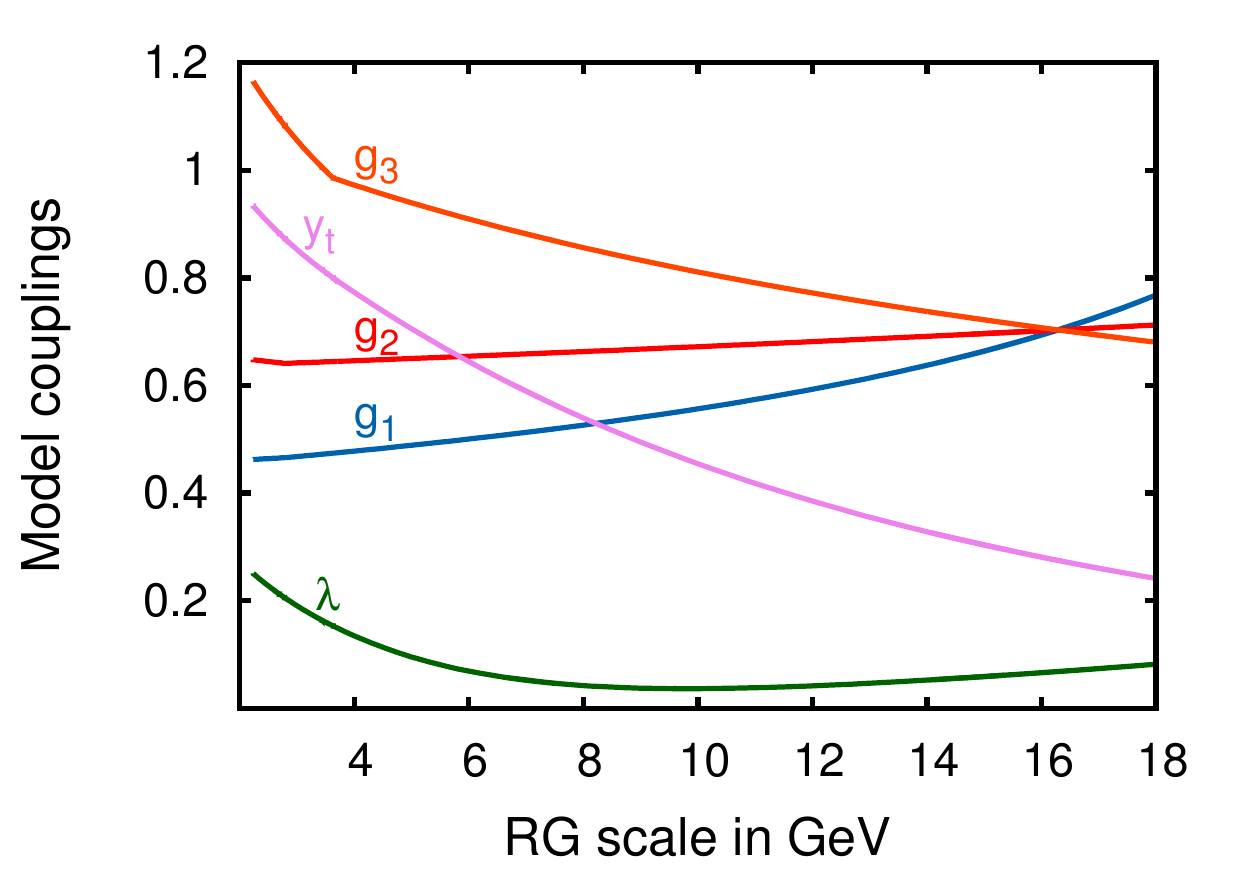}}
\subfigure[]{
\includegraphics[height=5.4 cm,width=7.3cm]{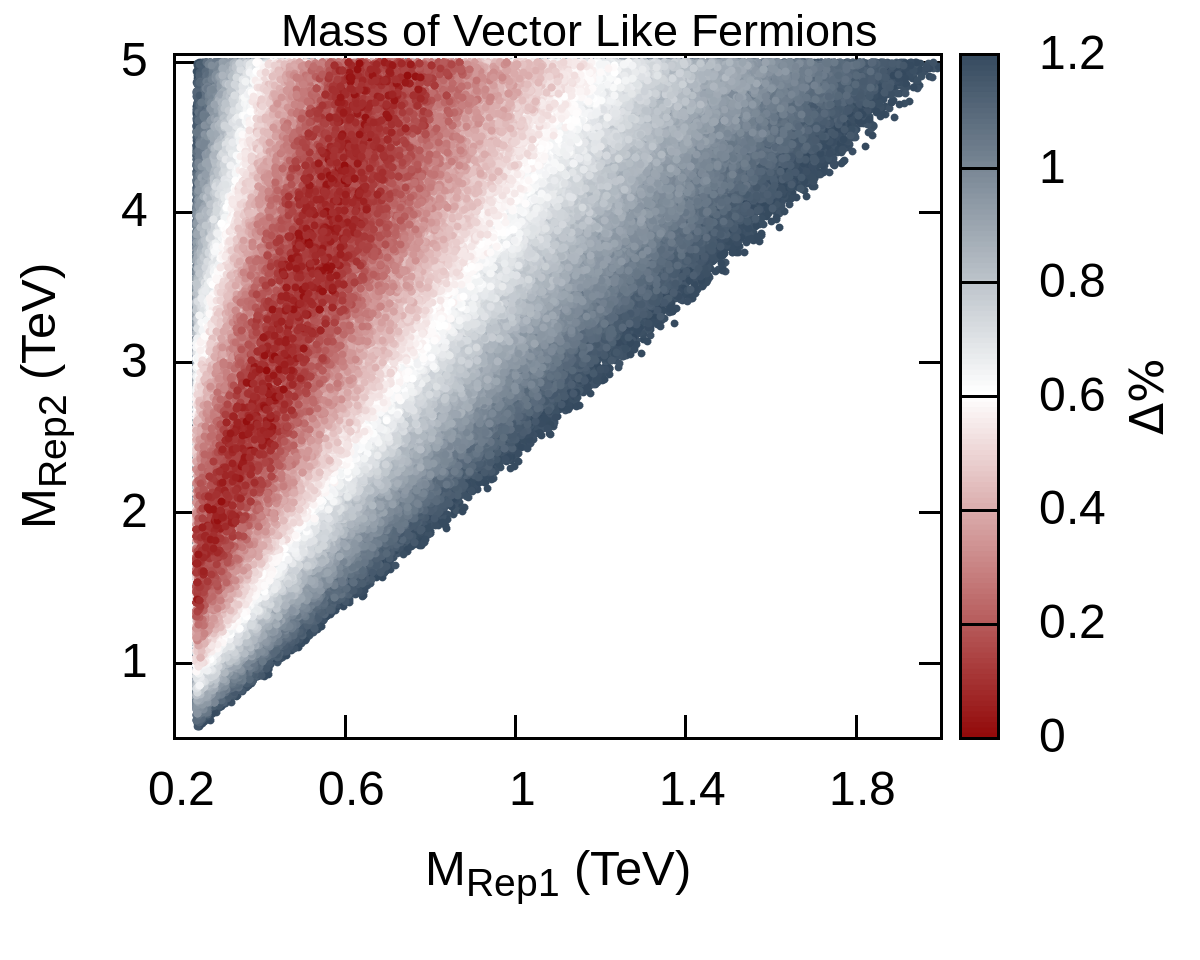}}
\caption{\sf Model 2: Fig. (a) Gauge couplings ($g_1$,$g_2$,$g_3$) 
unification and Vacuum stability ($\lambda > 0$)
plot, considering vector-like fermion in Rep.1 of mass 620 GeV and Rep.2 of mass 4310 GeV. Fig. (b) Mass range allowed for vector-like fermions in Rep.1 and Rep.2 
 for gauge unification and Vacuum stability.}\label{f:mod2}
\end{figure}

We got six copies of ${\rm Rep1}=\left(1, 2, \frac12\right)$ in mass range 
between 250 GeV to 2000 GeV and two copies of ${\rm Rep2}=\left(8, 1, 0\right)$ 
with mass range from 500 GeV to 5 TeV. Similar to the previous model, Rep1 
field is lepton like field and thus all the comments are applicable here. Rep2 
is gluino like and at the renormalisation level, it can interact with the 
gluons only and does not have any decay chain. Possibility 
of any higher dimension decaying operators and its collider phenomenology are
studied in the next Section~\ref{s:collider-boundstate}.

In the model, $M_{Rep1}$ is always less than $M_{Rep2}$. The change in the beta 
functions in the two thresholds are as follows:
\paragraph{(I) $M_{Rep1} < \mu < M_{Rep2}$}
\begin{eqnarray}
\delta b_i(M_{Rep1}< \mu < M_{Rep2} ) &=& 
\left(
\begin{array}{c}
\frac{12}{5}\\4\\0
 \end{array}
 \right),\,\,
\delta m_{ij}(M_{Rep1}< \mu < M_{Rep2} )=
\left(
\begin{array}{ccc}
 \frac{54}{50}  & \frac{54}{10} & 0\\
 \frac{18}{10}  & 49            & 0\\
       0        & 0             & 0
 \end{array}
  \right)
\end{eqnarray}
\begin{eqnarray}
 \delta \beta_{u}^{(2)}(M_{Rep1}< \mu < M_{Rep2} )&=&\frac{1044}{900}g_{1}^{4}\nonumber
		       + \frac{6}{2}g_{2}^{4}\\
 \delta \beta_{d}^{(2)}(M_{Rep1}< \mu < M_{Rep2} )&=&-\frac{36}{900}g_{1}^{4} \nonumber
		       + \frac{6}{2}g_{2}^{4}\\
 \delta \beta_{e}^{(2)}(M_{Rep1}< \mu < M_{Rep2} )&=&\frac{396}{100}g_{1}^{4}+\frac{6}{2}g_{2}^{4}
\end{eqnarray}
\begin{eqnarray}
 \delta \beta_{\lambda}^{(2)}(M_{Rep1}< \mu < M_{Rep2} )&=&-\frac{36}{250}g_{1}^{4}\left(12g_{1}^{2}+20g_{2}^{2}-25{\lambda}\right) \nonumber
		       -\frac{6}{5}g_{2}^{4}\left(4g_{1}^{2}+20g_{2}^{2}-25{\lambda}\right)
\end{eqnarray}
\paragraph{(II) $ \mu > M_{Rep2}$}
\begin{eqnarray}
\delta b_i(\mu > M_{Rep2} ) &=& 
\left(
\begin{array}{c}
\frac{12}{5}\\4\\4
 \end{array}
 \right),\,\, \hphantom{< M_{Rep2}}
\delta m_{ij}(\mu > M_{Rep2} )=
\left(
\begin{array}{ccc}
 \frac{54}{50}  & \frac{54}{10}  & 0 \\
 \frac{18}{10}  & 49             & 0 \\
            0   & 0              & 96
 \end{array}
  \right)
\end{eqnarray}
 \begin{eqnarray}
 \delta \beta_{u}^{(2)}( \mu > M_{Rep2} )&=&\frac{240}{9}g_{3}^{4}+\frac{1044}{900}g_{1}^{4}\nonumber
		       + \frac{6}{2}g_{2}^{4}\\
 \delta \beta_{d}^{(2)}( \mu > M_{Rep2} )&=&\frac{240}{9}g_{3}^{4}-\frac{36}{900}g_{1}^{4} \nonumber
		       + \frac{6}{2}g_{2}^{4}\\
 \delta \beta_{e}^{(2)}( \mu > M_{Rep2} )&=&\frac{396}{100}g_{1}^{4}+\frac{6}{2}g_{2}^{4}
\end{eqnarray}
\begin{eqnarray}
 \delta \beta_{\lambda}^{(2)}( \mu > M_{Rep2} )&=&-\frac{36}{250}g_{1}^{4}\left(12g_{1}^{2}+20g_{2}^{2}-25{\lambda}\right) \nonumber
		       - \frac{6}{5}g_{2}^{4}\left(4g_{1}^{2}+20g_{2}^{2}-25{\lambda}\right)
\end{eqnarray}
A sample unification point is shown in \fig{f:mod2}(a), six copies of lepton like vector fermions with degenerate mass
of 620 GeV and two copy of Rep2 with a mass of 4310 GeV is considered. The figure shows unification of gauge couplings as well as 
running of $y_t$ and $\lambda$. Mass distribution in Rep1-Rep2 mass plane is shown in \fig{f:mod2}(b). 

\subsection{Model 3} 

\begin{figure}
\centering
\subfigure[]{
	\includegraphics[height=5.5 cm,width=7cm]{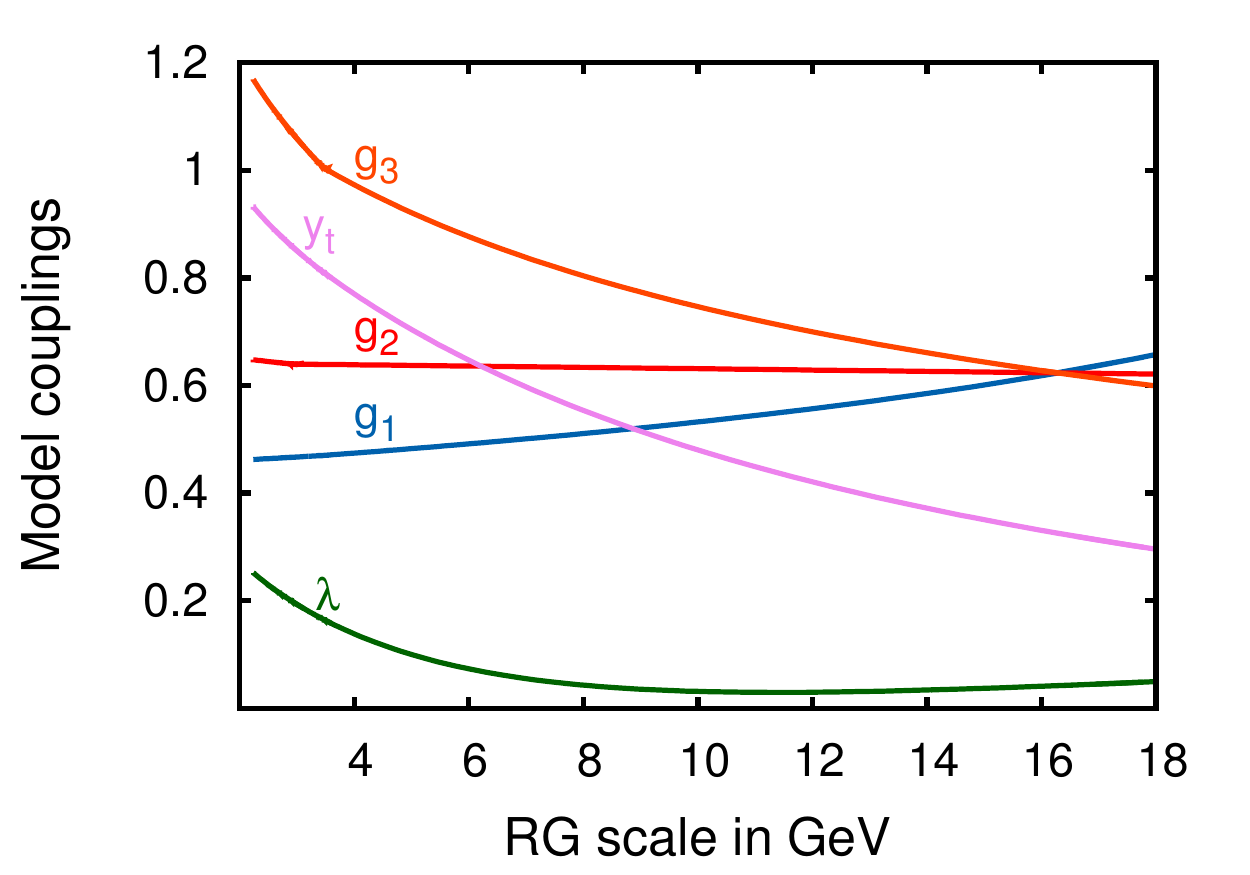}}
\subfigure[]{
	\includegraphics[height=5.4 cm,width=7.3cm]{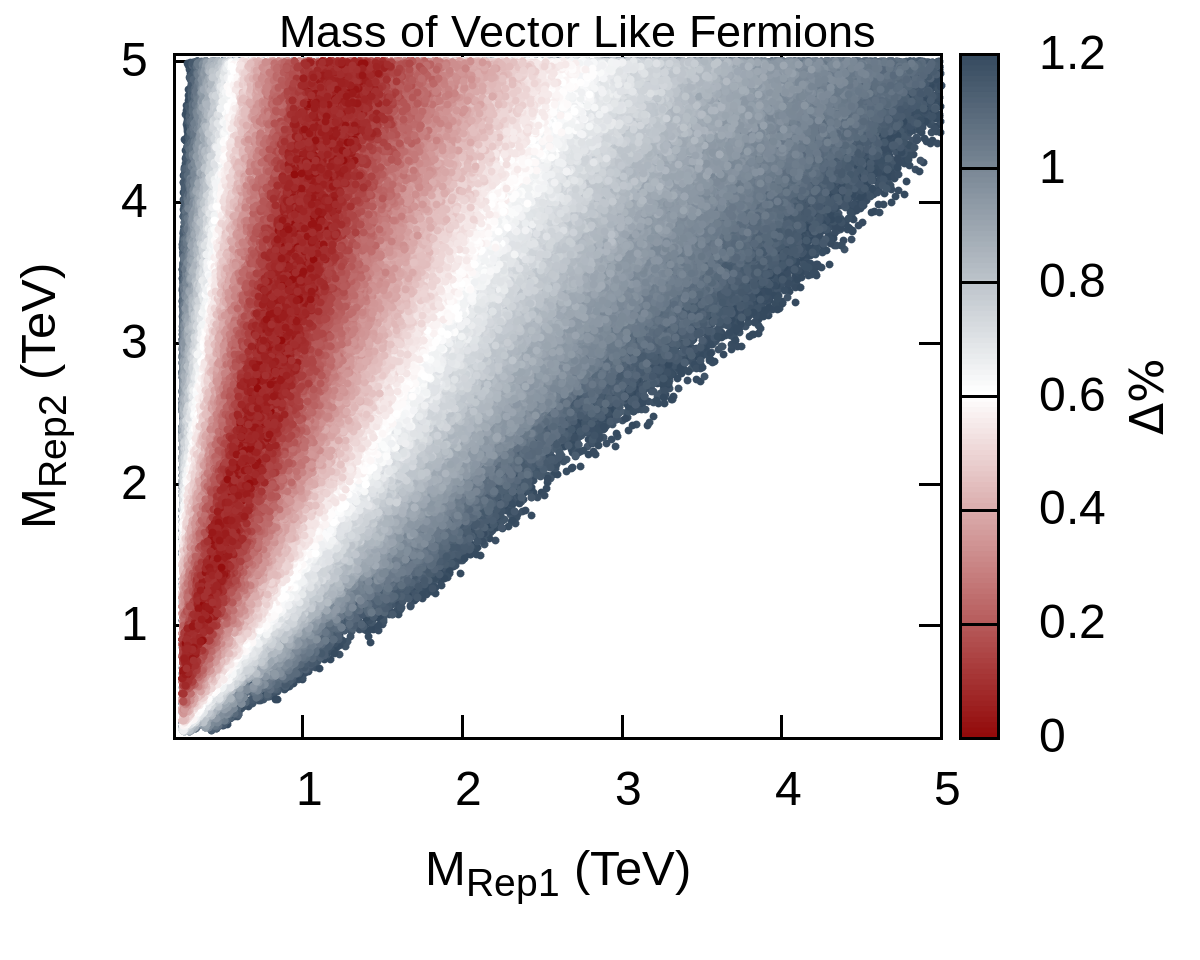}}
\caption{\sf Model 3: Fig. (a) Gauge couplings ($g_1$,$g_2$,$g_3$) 
unification and Vacuum stability ($\lambda > 0$)
plot, considering vector-like fermion in Rep.1 of mass 800 GeV and Rep.2 of mass 3030 GeV. Fig. (b) Mass range allowed for vector-like fermions in Rep.1 and Rep.2 
 for gauge unification and Vacuum stability.}\label{f:mod3}
\end{figure}

In this model, we got two copies of ${\rm Rep1}=\left(1, 3, 0\right)$ and four 
copies of ${\rm Rep2}=\left(3, 1, \frac13\right)$. The mass ranges of Rep1 and
Rep2 are (250 GeV,5 TeV) and  (250 GeV,5 TeV) respectively. 
Rep1 can be a viable candidate of type III~\cite{Foot:1988aq,Ma:1998dn} seesaw model with fermion mass 
of M.
The neutrino masses are generically given by a factor $v^{2}/M$, where $v$ is the
vacuum expectation value of the Higgs field. For large M (of the order of $10^{14}$ GeV),
small neutrino masses are generated even for Yukawa couplings of $\sim 1$. On the other hand,
either smaller Yukawa couplings $\sim 10^{-11}$ (which would not effect the RG running) or 
extended seesaw mechanisms, such as those of the inverse
seesaw models~\cite{delAguila:2008hw}, are required to obtain small neutrino masses while keeping
$M$ close to a few hundreds of GeV. However, we can also impose appropriate $Z_N$ 
symmetries. This symmetry 
removes all the Yukawa terms involving these fields at the 
renormalisable level and only gauge couplings are allowed\footnote{Seesaw 
requires Yukawa couplings, our model does not have a seesaw mechanism for neutrino masses.}.
Neutral component of these fermions is a viable dark 
matter candidate. This type of dark matter are referred as wino like dark matter 
and have been discussed in~\cite{Cirelli:2005uq, Bhattacherjee:2014dya, Cirelli:2014dsa, Moroi:2013sla}.

Rep2 has same representation like the down quark. 
This colour vector-like fermion can form a bound state and annihilate to diphoton, dijet etc. event, which we 
studied in Section~\ref{s:collider-boundstate}.

For most points in this model vector-like fermions in Rep1 can be degenerate
with vector-like fermions in Rep2 ($M_{Rep1}\sim M_{Rep2}$) as
shown in \fig{f:mod1}(b). However, there could be points in which 
either of the $M_{Rep1} = M_{Rep2}$, $M_{Rep1}> M_{Rep2}$ and $M_{Rep1}< M_{Rep2}$ are possible.
The change in the beta functions in these 
possibilities are as follows:
\paragraph {(a) $M_{Rep1} = M_{Rep2}$}
\subparagraph{(I) $\mu > M_{Rep2}=M_{Rep1}$}
\begin{eqnarray}
\delta b_i(\mu > M_{Rep2} ) &=& 
\left(
\begin{array}{c}
\frac{16}{15} \\ \frac{16}{6}\\\frac{16}{6} 
\end{array}
\right),\,\, \hphantom{< M_{Rep2}}
\delta m_{ij}(\mu > M_{Rep2} )=
\left(
\begin{array}{ccc}
\frac{16}{75}  &  0     &  \frac{64}{15} \\
0      & \frac{128}{3} & 0      \\
\frac{16}{30}  & 0     & \frac{152}{3} 
\end{array}
\right)
\end{eqnarray}
\begin{eqnarray}
\delta \beta_{u}^{(2)}( \mu > M_{Rep2} )&=&\frac{160}{9}g_{3}^{4}+\frac{464}{900}g_{1}^{4}\nonumber
+ \frac{4}{2}g_{2}^{4}\\
\delta \beta_{d}^{(2)}( \mu > M_{Rep2} )&=&\frac{160}{9}g_{3}^{4}-\frac{16}{900}g_{1}^{4} \nonumber
+ \frac{4}{2}g_{2}^{4}\\
\delta \beta_{e}^{(2)}( \mu > M_{Rep2} )&=&\frac{176}{100}g_{1}^{4}+\frac{4}{2}g_{2}^{4}
\end{eqnarray}
\begin{eqnarray}
\delta \beta_{\lambda}^{(2)}( \mu > M_{Rep2} )&=&-\frac{16}{250}g_{1}^{4}\left(12g_{1}^{2}+20g_{2}^{2}
-25{\lambda}\right) \nonumber
-\frac{4}{5}g_{2}^{4}\left(4g_{1}^{2}+20g_{2}^{2}-25{\lambda}\right)
\end{eqnarray}

\paragraph {(b) $M_{Rep1} < M_{Rep2}$}
\subparagraph{(I) $M_{Rep1} < \mu < M_{Rep2}$}
\begin{eqnarray}
\delta b_i(M_{Rep1} < \mu < M_{Rep2} ) &=& 
\left(
\begin{array}{c}
0\\ \frac{16}{6} \\0
 \end{array}
 \right),\,\,
\delta m_{ij}(M_{Rep1} < \mu < M_{Rep2} )=
\left(
\begin{array}{ccc}
 0   &  0   &  0\\
 0   &  \frac{128}{3} &  0\\
 0   & 0    &  0
 \end{array}
  \right)
\end{eqnarray}
\begin{eqnarray}
 \delta \beta_{u}^{(2)}(M_{Rep1}< \mu < M_{Rep2} )&=&\frac{4}{2}g_{2}^{4}\nonumber\\
 \delta \beta_{d}^{(2)}(M_{Rep1}< \mu < M_{Rep2} )&=&\frac{4}{2}g_{2}^{4}\nonumber\\
 \delta \beta_{e}^{(2)}(M_{Rep1}< \mu < M_{Rep2} )&=&\frac{4}{2}g_{2}^{4}
\end{eqnarray}
\begin{eqnarray}
 \delta \beta_{\lambda}^{(2)}(M_{Rep1}< \mu < M_{Rep2} )&=&-\frac{4}{5}g_{2}^{4}\left(4g_{1}^{2}+20g_{2}^{2}
 -25{\lambda}\right)\nonumber
\end{eqnarray}
\subparagraph{(II) $\mu > M_{Rep2}$}
\begin{eqnarray}
\delta b_i(\mu > M_{Rep2} ) &=& 
\left(
\begin{array}{c}
 \frac{16}{15} \\ \frac{16}{6}\\\frac{16}{6} 
 \end{array}
 \right),\,\, \hphantom{< M_{Rep2}}
\delta m_{ij}(\mu > M_{Rep2} )=
\left(
\begin{array}{ccc}
 \frac{16}{75}  &  0     &  \frac{64}{15} \\
 0      & \frac{128}{3} & 0      \\
 \frac{16}{30}  & 0     & \frac{152}{3} 
 \end{array}
  \right)
\end{eqnarray}
 \begin{eqnarray}
 \delta \beta_{u}^{(2)}( \mu > M_{Rep2} )&=&\frac{160}{9}g_{3}^{4}+\frac{464}{900}g_{1}^{4}\nonumber
		       + \frac{4}{2}g_{2}^{4}\\
 \delta \beta_{d}^{(2)}( \mu > M_{Rep2} )&=&\frac{160}{9}g_{3}^{4}-\frac{16}{900}g_{1}^{4} \nonumber
		       + \frac{4}{2}g_{2}^{4}\\
 \delta \beta_{e}^{(2)}( \mu > M_{Rep2} )&=&\frac{176}{100}g_{1}^{4}+\frac{4}{2}g_{2}^{4}
\end{eqnarray}
\begin{eqnarray}
 \delta \beta_{\lambda}^{(2)}( \mu > M_{Rep2} )&=&-\frac{16}{250}g_{1}^{4}\left(12g_{1}^{2}+20g_{2}^{2}
							  -25{\lambda}\right) \nonumber
		       -\frac{4}{5}g_{2}^{4}\left(4g_{1}^{2}+20g_{2}^{2}-25{\lambda}\right)
\end{eqnarray}

\paragraph {(c) $M_{Rep2} < M_{Rep1}$}
\subparagraph{(I) $M_{Rep2} < \mu < M_{Rep1}$}
\begin{eqnarray}
\delta b_i(M_{Rep2} < \mu < M_{Rep1} ) &=& 
\left(
\begin{array}{c}
\frac{16}{15} \\ 
0 \\
\frac{16}{6} 
\end{array}
\right),\,\, \hphantom{< M_{Rep2}}
\delta m_{ij}(M_{Rep2} < \mu < M_{Rep1} )=
\left(
\begin{array}{ccc}
\frac{16}{75}  &  0     &  \frac{64}{15} \\
0      & 0 & 0      \\
\frac{16}{30}  & 0     & \frac{152}{3} 
\end{array}
\right)
\end{eqnarray}
\begin{eqnarray}
\delta \beta_{u}^{(2)}( M_{Rep2} < \mu < M_{Rep1} )&=&\frac{160}{9}g_{3}^{4}+\frac{464}{900}g_{1}^{4}\nonumber
\\
\delta \beta_{d}^{(2)}( M_{Rep2} < \mu < M_{Rep1} )&=&\frac{160}{9}g_{3}^{4}-\frac{16}{900}g_{1}^{4} \nonumber
\\
\delta \beta_{e}^{(2)}( M_{Rep2} < \mu < M_{Rep1} )&=&\frac{176}{100}g_{1}^{4}
\end{eqnarray}
\begin{eqnarray}
\delta \beta_{\lambda}^{(2)}(M_{Rep2} < \mu < M_{Rep1} )&=&-\frac{16}{250}g_{1}^{4}\left(12g_{1}^{2}+20g_{2}^{2}
-25{\lambda}\right) 
\end{eqnarray}
\subparagraph{(II) $\mu > M_{Rep1}$}
\begin{eqnarray}
\delta b_i(\mu > M_{Rep1} ) &=& 
\left(
\begin{array}{c}
\frac{16}{15} \\ \frac{16}{6}\\\frac{16}{6} 
\end{array}
\right),\,\, \hphantom{< M_{Rep1}}
\delta m_{ij}(\mu > M_{Rep1} )=
\left(
\begin{array}{ccc}
\frac{16}{75}  &  0     &  \frac{64}{15} \\
0      & \frac{128}{3} & 0      \\
\frac{16}{30}  & 0     & \frac{152}{3} 
\end{array}
\right)
\end{eqnarray}
\begin{eqnarray}
\delta \beta_{u}^{(2)}( \mu > M_{Rep1} )&=&\frac{160}{9}g_{3}^{4}+\frac{464}{900}g_{1}^{4}\nonumber
+ \frac{4}{2}g_{2}^{4}\\
\delta \beta_{d}^{(2)}( \mu > M_{Rep1} )&=&\frac{160}{9}g_{3}^{4}-\frac{16}{900}g_{1}^{4} \nonumber
+ \frac{4}{2}g_{2}^{4}\\
\delta \beta_{e}^{(2)}( \mu > M_{Rep1} )&=&\frac{176}{100}g_{1}^{4}+\frac{4}{2}g_{2}^{4}
\end{eqnarray}
\begin{eqnarray}
\delta \beta_{\lambda}^{(2)}( \mu > M_{Rep1} )&=&-\frac{16}{250}g_{1}^{4}\left(12g_{1}^{2}+20g_{2}^{2}
-25{\lambda}\right) \nonumber
-\frac{4}{5}g_{2}^{4}\left(4g_{1}^{2}+20g_{2}^{2}-25{\lambda}\right)
\end{eqnarray}
In \fig{f:mod3}(a), a sample of gauge couplings, $y_t$ and $\lambda$ running 
is shown with two copies of weak-isospin triplet vector-like fermions with degenerate mass
of 800 GeV and four copies of bottom like vector quark with a mass of 3030 GeV. 
\fig{f:mod3}(b) shows the mass distribution in Rep1-Rep2 mass plane.

 \subsection{Model 4} 

\begin{figure}
\centering
\subfigure[]{
	\includegraphics[height=5.5 cm,width=7cm]{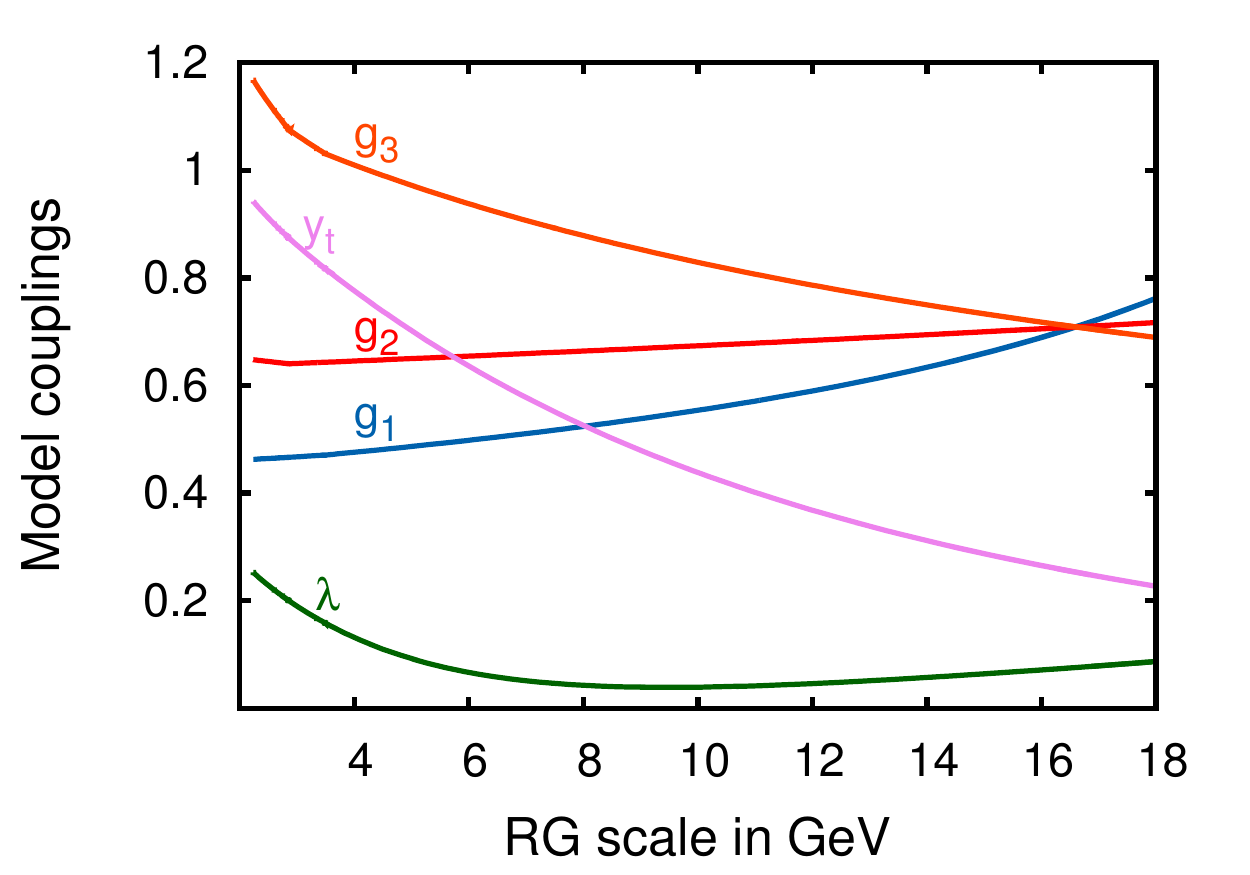}}
\subfigure[]{
	\includegraphics[height=5.4 cm,width=7.3cm]{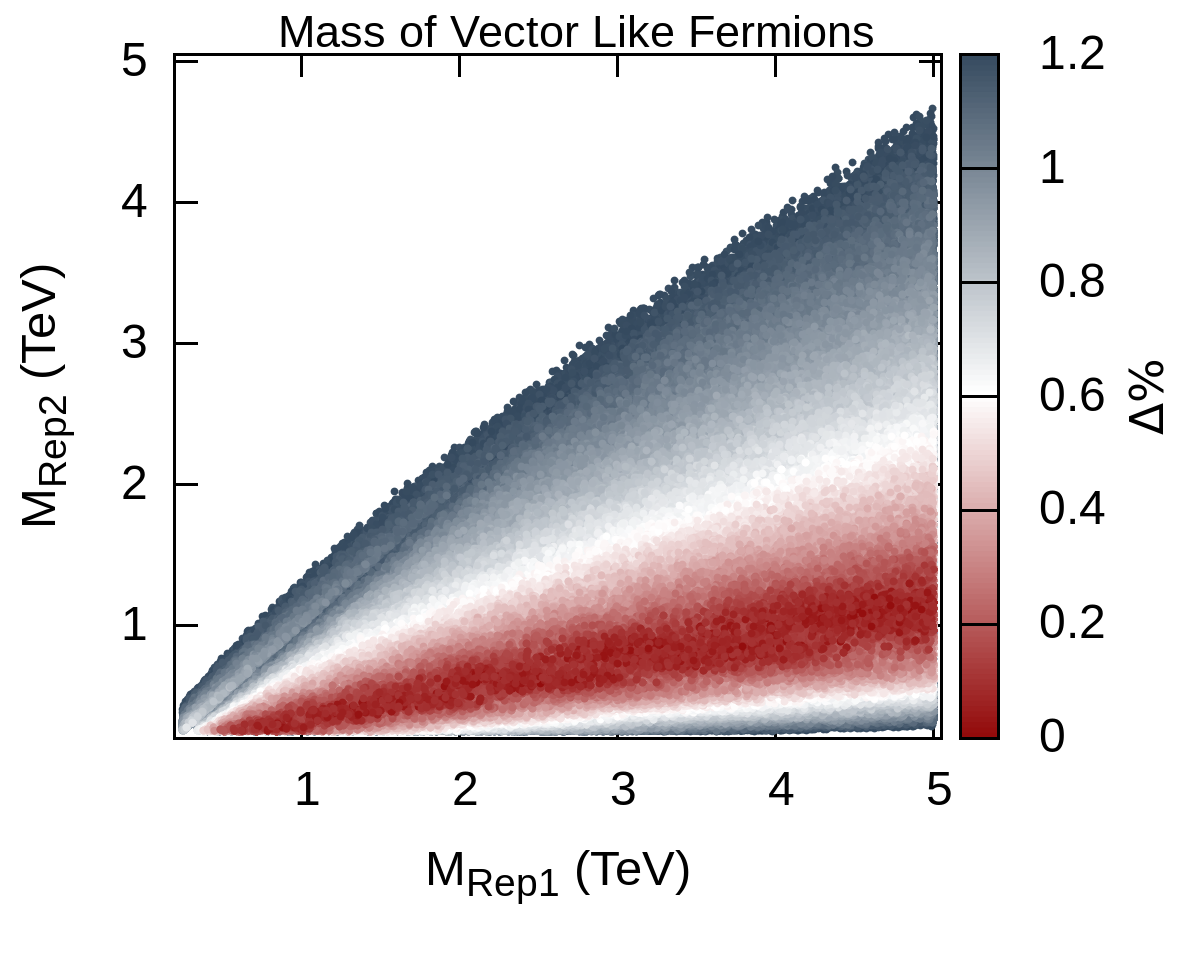}}
\caption{\sf Model 4: Fig. (a) Gauge couplings ($g_1$,$g_2$,$g_3$) 
unification and Vacuum stability ($\lambda > 0$)
plot, considering vector-like fermion in Rep.1 of mass 3175 GeV and Rep.2 of mass 730 GeV. Fig. (b) Mass range allowed for vector-like fermions in Rep.1 and Rep.2 
 for gauge unification and Vacuum stability.}\label{f:mod4}
\end{figure}

This model is interesting as representations of the vector-like matter are like up 
quarks (Rep1) and left handed quark (Rep2). They appear in two copies for each 
and their mass ranges are (500 GeV,5 TeV) and  (250 GeV,4.5 TeV) respectively. These 
vector-like quark can be probed at LHC as a bound state, which is studied in Section~\ref{s:collider-boundstate}.

In the model, $M_{Rep1}$ is greater than $M_{Rep2}$. However, some parameters of vector like fermion mass have $M_{Rep1} < M_{Rep2}$ and $M_{Rep1} = M_{Rep2}$. 
The change in the beta functions in these three possibilities are as follows:
\paragraph {(a) $M_{Rep1} = M_{Rep2}$}
\subparagraph{(I) $\mu > M_{Rep2}=M_{Rep1}$}
\begin{eqnarray}  
\delta b_i\left(\mu > M_{Rep1} \right) &=& 
\left(
\begin{array}{c}
\frac{12}{5}\\4\\4
\end{array}
\right),\,\, \hphantom{< M_{Rep1}}
\delta m_{ij}\left(\mu > M_{Rep1} \right)=
\left(
\begin{array}{ccc}
\frac{258}{150}  & \frac{3}{5}  & \frac{144}{15}\\
\frac{1}{5}       & 49   & 16\\
\frac{18}{15}    & 6    & 76
\end{array}
\right)
\end{eqnarray}
\begin{eqnarray}
\delta \beta_{u}^{(2)}( \mu > M_{Rep1} )&=&\frac{240}{9}g_{3}^{4}+\frac{1044}{900}g_{1}^{4}\nonumber
+ \frac{6}{2}g_{2}^{4}\\
\delta \beta_{d}^{(2)}( \mu > M_{Rep1} )&=&\frac{240}{9}g_{3}^{4}-\frac{36}{900}g_{1}^{4} \nonumber
+ \frac{6}{2}g_{2}^{4}\\
\delta \beta_{e}^{(2)}( \mu > M_{Rep1} )&=&\frac{396}{100}g_{1}^{4}+\frac{6}{2}g_{2}^{4}
\end{eqnarray}
\begin{eqnarray}
\delta \beta_{\lambda}^{(2)}( \mu > M_{Rep1} )&=&-\frac{36}{250}g_{1}^{4}\left(12g_{1}^{2}+20g_{2}^{2}
-25{\lambda}\right) \nonumber
-\frac{6}{5}g_{2}^{4}\left(4g_{1}^{2}+20g_{2}^{2}-25{\lambda}\right)
\end{eqnarray}
\paragraph{(b) $M_{Rep2} <  M_{Rep1}$}
\subparagraph{(I) $M_{Rep2} < \mu < M_{Rep1}$}
\begin{eqnarray}
\delta b_i\left(M_{Rep2}< \mu < M_{Rep1} \right) &=& 
\left(      
\begin{array}{c}
\frac{4}{15} \\ 4 \\ \frac{8}{3} 
 \end{array}
 \right),\,\,
\delta m_{ij}\left(M_{Rep2}< \mu < M_{Rep1} \right)=
\left(
\begin{array}{ccc}
\frac{2}{150}  &  \frac{3}{5}  & \frac{16}{15} \\
 \frac{1}{5}        & 49    & 16     \\
\frac{2}{15}     & 6     & \frac{152}{3}
 \end{array}
  \right)
  \end{eqnarray}
   
\begin{eqnarray}
 \delta \beta_{u}^{(2)}(M_{Rep2}< \mu < M_{Rep1} )&=&\frac{160}{9}g_{3}^{4}+\frac{116}{900}g_{1}^{4}\nonumber
		       + \frac{6}{2}g_{2}^{4}\\
 \delta \beta_{d}^{(2)}(M_{Rep2}< \mu < M_{Rep1} )&=&\frac{160}{9}g_{3}^{4}(4)-\frac{4}{900}g_{1}^{4}\nonumber
		       + \frac{6}{2}g_{2}^{4}\\
 \delta \beta_{e}^{(2)}(M_{Rep2}< \mu < M_{Rep1} )&=&\frac{44}{100}g_{1}^{4}+\frac{6}{2}g_{2}^{4}
\end{eqnarray}
\begin{eqnarray}
 \delta \beta_{\lambda}^{(2)}(M_{Rep2}< \mu < M_{Rep1} )&=&-\frac{4}{250}g_{1}^{4}\left(12g_{1}^{2}+20g_{2}^{2}-25{\lambda}\right) \nonumber
		       -\frac{6}{5}g_{2}^{4}\left(4g_{1}^{2}+20g_{2}^{2}-25{\lambda}\right)
\end{eqnarray}
\subparagraph{(II) $\mu > M_{Rep1}$}
\begin{eqnarray}  
\delta b_i\left(\mu > M_{Rep1} \right) &=& 
\left(
\begin{array}{c}
\frac{12}{5}\\4\\4
 \end{array}
 \right),\,\, \hphantom{< M_{Rep1}}
\delta m_{ij}\left(\mu > M_{Rep1} \right)=
\left(
\begin{array}{ccc}
 \frac{258}{150}  & \frac{3}{5}  & \frac{144}{15}\\
\frac{1}{5}       & 49   & 16\\
 \frac{18}{15}    & 6    & 76
 \end{array}
  \right)
\end{eqnarray}
 \begin{eqnarray}
 \delta \beta_{u}^{(2)}( \mu > M_{Rep1} )&=&\frac{240}{9}g_{3}^{4}+\frac{1044}{900}g_{1}^{4}\nonumber
		       + \frac{6}{2}g_{2}^{4}\\
 \delta \beta_{d}^{(2)}( \mu > M_{Rep1} )&=&\frac{240}{9}g_{3}^{4}-\frac{36}{900}g_{1}^{4} \nonumber
		       + \frac{6}{2}g_{2}^{4}\\
 \delta \beta_{e}^{(2)}( \mu > M_{Rep1} )&=&\frac{396}{100}g_{1}^{4}+\frac{6}{2}g_{2}^{4}
\end{eqnarray}
\begin{eqnarray}
 \delta \beta_{\lambda}^{(2)}( \mu > M_{Rep1} )&=&-\frac{36}{250}g_{1}^{4}\left(12g_{1}^{2}+20g_{2}^{2}
							  -25{\lambda}\right) \nonumber
		       -\frac{6}{5}g_{2}^{4}\left(4g_{1}^{2}+20g_{2}^{2}-25{\lambda}\right)
\end{eqnarray}
\paragraph{(c) $M_{Rep1} <  M_{Rep2}$}
\subparagraph{(I) $M_{Rep1} < \mu < M_{Rep2}$}
\begin{eqnarray}
\delta b_i\left(M_{Rep1}< \mu < M_{Rep2} \right) &=& 
\left(      
\begin{array}{c}
\frac{32}{15} \\ 0 \\ \frac{4}{3} 
 \end{array}
 \right),\,\,
\delta m_{ij}\left(M_{Rep1}< \mu < M_{Rep2} \right)=
\left(
\begin{array}{ccc}
\frac{256}{150}  &  0  & \frac{128}{15} \\
0        & 49    & 0    \\
\frac{16}{15}     & 0     & \frac{76}{3}
 \end{array}
  \right)
  \end{eqnarray}
   
\begin{eqnarray}
 \delta \beta_{u}^{(2)}(M_{Rep1}< \mu < M_{Rep2} )&=&\frac{80}{9}g_{3}^{4}+\frac{928}{900}g_{1}^{4}\nonumber
		      \\
 \delta \beta_{d}^{(2)}(M_{Rep1}< \mu < M_{Rep2} )&=&\frac{80}{9}g_{3}^{4}(4)-\frac{32}{900}g_{1}^{4}\nonumber
		      \\
 \delta \beta_{e}^{(2)}(M_{Rep1}< \mu < M_{Rep2} )&=&\frac{352}{100}g_{1}^{4}
\end{eqnarray}
\begin{eqnarray}
 \delta \beta_{\lambda}^{(2)}(M_{Rep1}< \mu < M_{Rep2} )&=&-\frac{32}{250}g_{1}^{4}\left(12g_{1}^{2}+20g_{2}^{2}-25{\lambda}\right) \nonumber    
\end{eqnarray}
\subparagraph{(II) $\mu > M_{Rep2}$}
\begin{eqnarray}  
\delta b_i\left(\mu > M_{Rep2} \right) &=& 
\left(
\begin{array}{c}
\frac{12}{5}\\4\\4
 \end{array}
 \right),\,\, \hphantom{< M_{Rep1}}
\delta m_{ij}\left(\mu > M_{Rep2} \right)=
\left(
\begin{array}{ccc}
 \frac{258}{150}  & \frac{3}{5}  & \frac{144}{15}\\
\frac{1}{5}       & 49   & 16\\
 \frac{18}{15}    & 6    & 76
 \end{array}
  \right)
\end{eqnarray}
 \begin{eqnarray}
 \delta \beta_{u}^{(2)}( \mu > M_{Rep2} )&=&\frac{240}{9}g_{3}^{4}+\frac{1044}{900}g_{1}^{4}\nonumber
		       + \frac{6}{2}g_{2}^{4}\\
 \delta \beta_{d}^{(2)}( \mu > M_{Rep2} )&=&\frac{240}{9}g_{3}^{4}-\frac{36}{900}g_{1}^{4} \nonumber
		       + \frac{6}{2}g_{2}^{4}\\
 \delta \beta_{e}^{(2)}( \mu > M_{Rep2} )&=&\frac{396}{100}g_{1}^{4}+\frac{6}{2}g_{2}^{4}
\end{eqnarray}
\begin{eqnarray}
 \delta \beta_{\lambda}^{(2)}( \mu > M_{Rep2} )&=&-\frac{36}{250}g_{1}^{4}\left(12g_{1}^{2}+20g_{2}^{2}
							  -25{\lambda}\right) \nonumber
		       -\frac{6}{5}g_{2}^{4}\left(4g_{1}^{2}+20g_{2}^{2}-25{\lambda}\right)
\end{eqnarray}

The running of gauge couplings, $y_t$ and $\lambda$ are shown in \fig{f:mod4}(a), considering two copies of top 
like vector fermions with degenerate mass
of 3175 GeV and two copies of left handed vector-like quark with a mass of 730 GeV. \fig{f:mod4}(b) shows the mass distribution in Rep1-Rep2 mass plane. 

\subsection{Model 5} 

\begin{figure}
\centering
\subfigure[]{
	\includegraphics[height=5.5 cm,width=7cm]{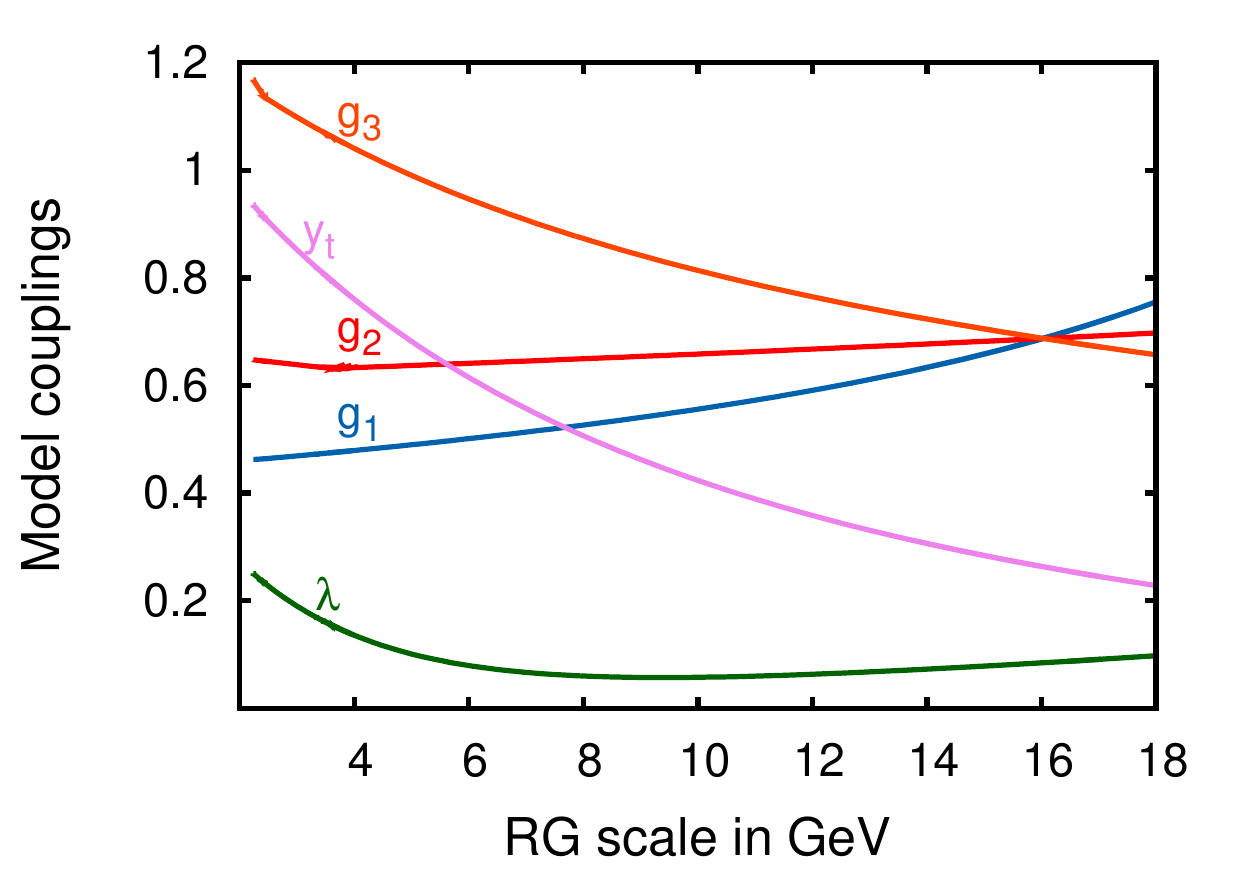}}
\subfigure[]{
	\includegraphics[height=5.4 cm,width=7.3cm]{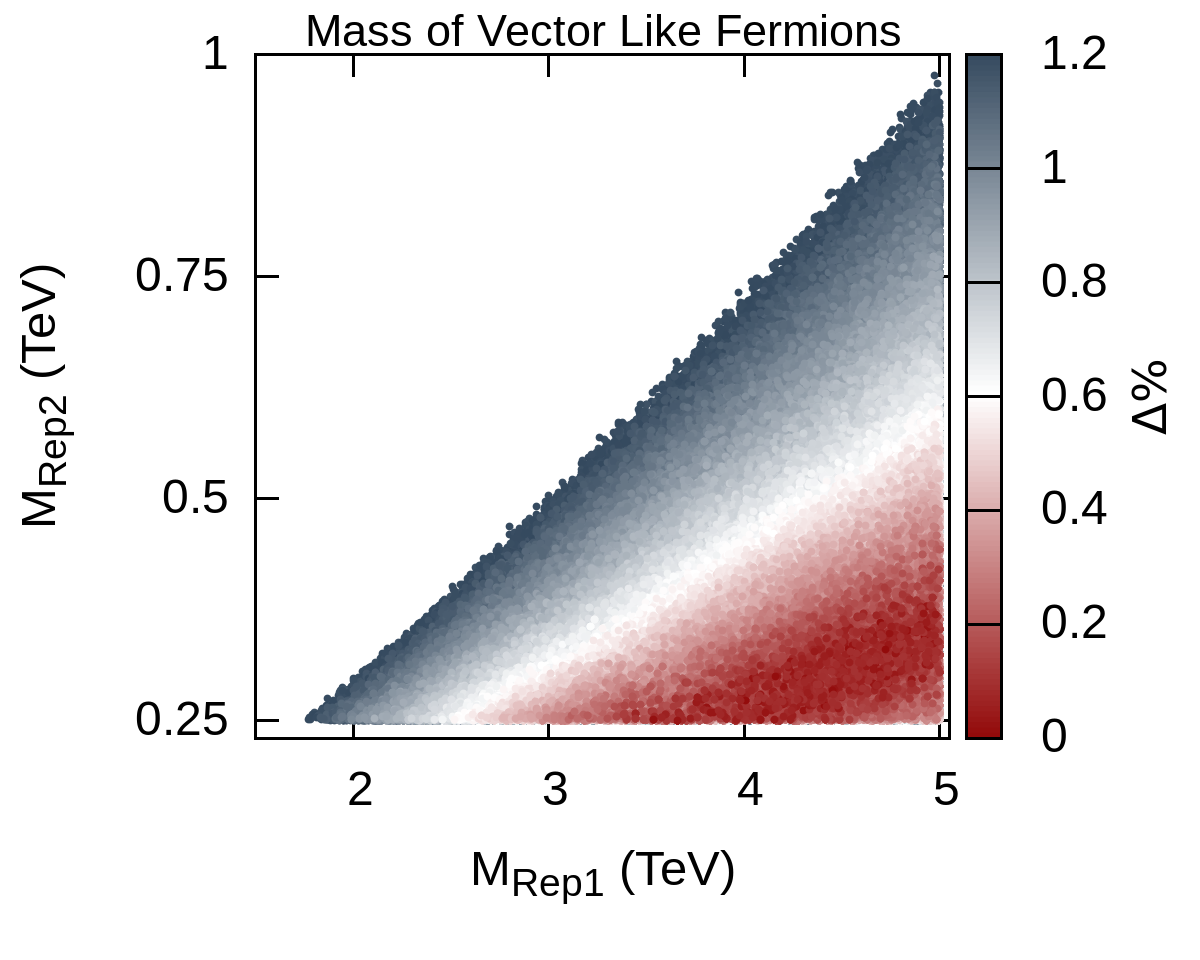}}
\caption{\sf Model 5: Fig. (a) Gauge couplings ($g_1$,$g_2$,$g_3$) 
unification and Vacuum stability ($\lambda > 0$)
plot, considering vector-like fermion in Rep.1 of mass 4.16 TeV and Rep.2 of mass 280 GeV. Fig. (b) Mass range allowed for vector-like fermions in Rep.1 and Rep.2 
 for gauge unification and Vacuum stability.}\label{f:mod5}
\end{figure}
This Model consist of 3 copies of vector-like fermion $(1,3,0)$, which is triplet under SU(2) representation (Rep1) and one copy of vector-like 
fermion $(6,1,\frac{2}{3})$ which is
sextet under SU(3) representation (Rep2). The mass range of Rep1 and Rep2 are (1.8 TeV to 5TeV) and (250 GeV to 950 GeV) respectively.
The possible scenarios of Rep1 has been discussed in Model 3 and Rep2 has been mentioned in Model 1 with  hypercharge 2/3.
In the model, $M_{Rep1}$ is greater than $M_{Rep2}$. 

The change in the beta 
functions in the two thresholds are as follows:
\paragraph{(I) $M_{Rep2} < \mu < M_{Rep1}$}
\begin{eqnarray}
\delta b_i\left(M_{Rep2}< \mu < M_{Rep1} \right) &=& 
\left(      
\begin{array}{c}
\frac{96}{45} \\ 0 \\ \frac{10}{3} 
 \end{array}
 \right),\,\,
\delta m_{ij}\left(M_{Rep2}< \mu < M_{Rep1} \right)=
\left(
\begin{array}{ccc}
\frac{128}{75} &  0  & \frac{64}{3} \\
 0      &  0  & 0    \\
 \frac{8}{3} &  0  & \frac{250}{3}
 \end{array}
  \right)
  \end{eqnarray}
\begin{eqnarray}
 \delta \beta_{u}^{(2)} (M_{Rep2}< \mu < M_{Rep1} )&=& \frac{200}{9} g_{3}^{4}  + \frac{232}{225} g_{1}^{4}\nonumber
		      \\
 \delta \beta_{d}^{(2)} (M_{Rep2}< \mu < M_{Rep1} )&=& \frac{40}{9} g_{3}^{4}(5)- \frac{32}{900} g_{1}^{4}\nonumber
		       \\
 \delta \beta_{e}^{(2)} (M_{Rep2}< \mu < M_{Rep1} )&=& \frac{352}{100} g_{1}^{4} 
 \end{eqnarray}
\begin{eqnarray}
\delta \beta_{\lambda}^{(2)}(M_{Rep2}< \mu < M_{Rep1} )&=& -\frac{32}{250} g_{1}^{4}\left(12g_{1}^{2} + 20g_{2}^{2} - 
		  25{\lambda}\right) \nonumber
\end{eqnarray}

\paragraph{(II) $\mu > M_{Rep1}$}
\begin{eqnarray}  
\delta b_i\left(\mu > M_{Rep1} \right) &=& 
\left(
\begin{array}{c}
\frac{96}{45} \\ 4 \\ \frac{10}{3} 
\end{array}
 \right),\,\, \hphantom{< M_{Rep1}}
\delta m_{ij}\left(\mu > M_{Rep1} \right)=
\left(
\begin{array}{ccc}
 \frac{128}{75} & 0  & \frac{64}{3} \\
  0      & 64   & 0\\
   \frac{8}{3} &  0  & \frac{250}{3}
 \end{array}
  \right)
\end{eqnarray}
 \begin{eqnarray}
 \delta \beta_{u}^{(2)} ( \mu > M_{Rep1} )&=& \frac{200}{9} g_{3}^{4}  + \frac{232}{225} g_{1}^{4}\nonumber
		      + \frac{6}{2}g_{2}^{4}\\
 \delta \beta_{d}^{(2)} (\mu > M_{Rep1} )&=& \frac{200}{9} g_{3}^{4}- \frac{32}{900} g_{1}^{4}\nonumber
		      + \frac{6}{2}g_{2}^{4} \\
 \delta \beta_{e}^{(2)} (\mu > M_{Rep1} )&=& \frac{352}{100} g_{1}^{4} + \frac{6}{2}g_{2}^{4}
 \end{eqnarray}
\begin{eqnarray}
\delta \beta_{\lambda}^{(2)}( \mu > M_{Rep1} )&=& -\frac{32}{250} g_{1}^{4}\left(12g_{1}^{2} + 20g_{2}^{2} - 
		  25{\lambda}\right) \nonumber
- \frac{6}{5}g_{2}^{4} \left(4g_{1}^{2}+20g_{2}^{2}-25{\lambda}\right)  
\end{eqnarray}
A sample unification point is shown in \fig{f:mod5}(a), three copies of weak isospin triplet vector-like fermions with degenerate mass
of 4.16 TeV and one copy of color sextet  vector-like fermion with a mass of 280 GeV is considered. The \fig{f:mod5}(a) shows unification clearly.
\fig{f:mod5}(b) shows the mass distribution in Rep1-Rep2 mass plane. 

\subsection{Model 6} 

\begin{figure}
\centering
\subfigure[]{
	\includegraphics[height=5.5 cm,width=7cm]{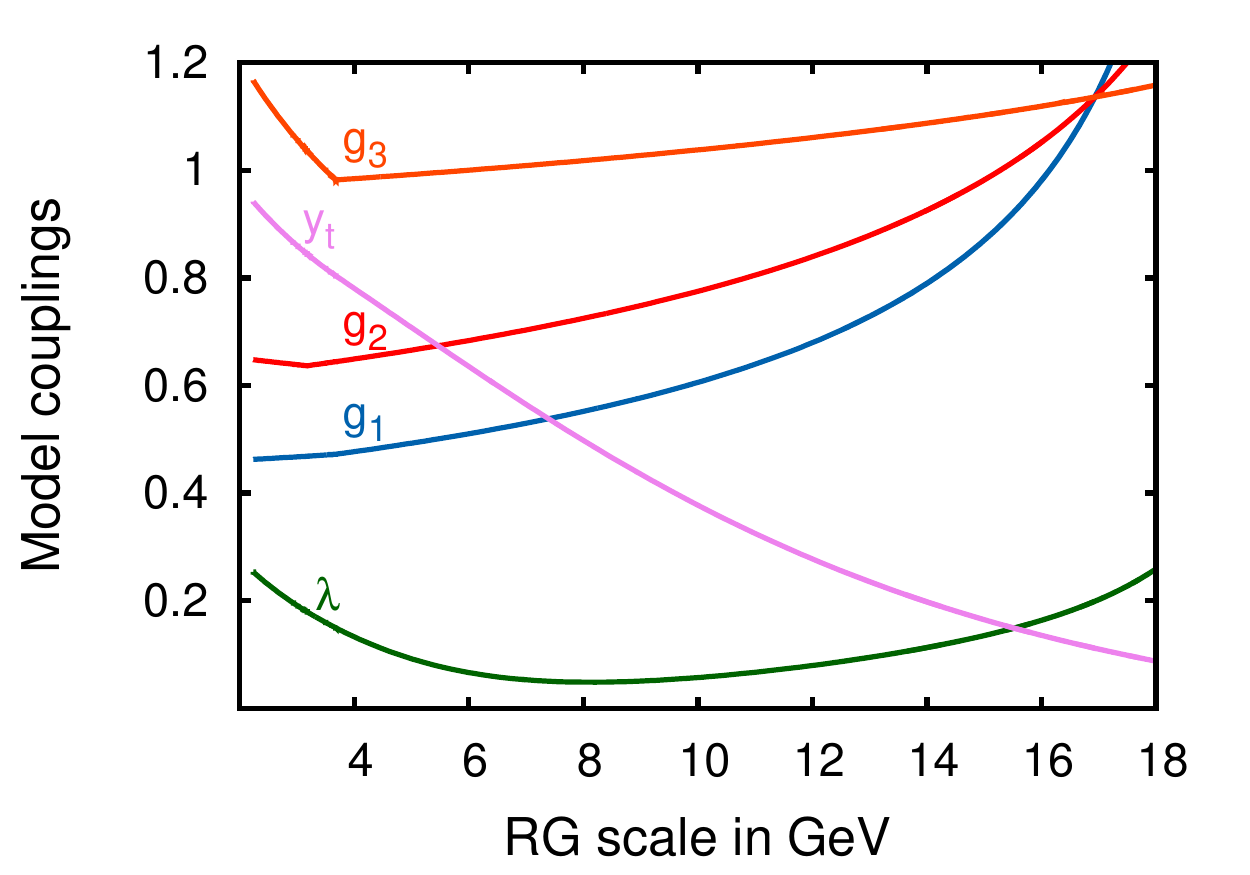}}
\subfigure[]{
	\includegraphics[height=5.4 cm,width=7.3cm]{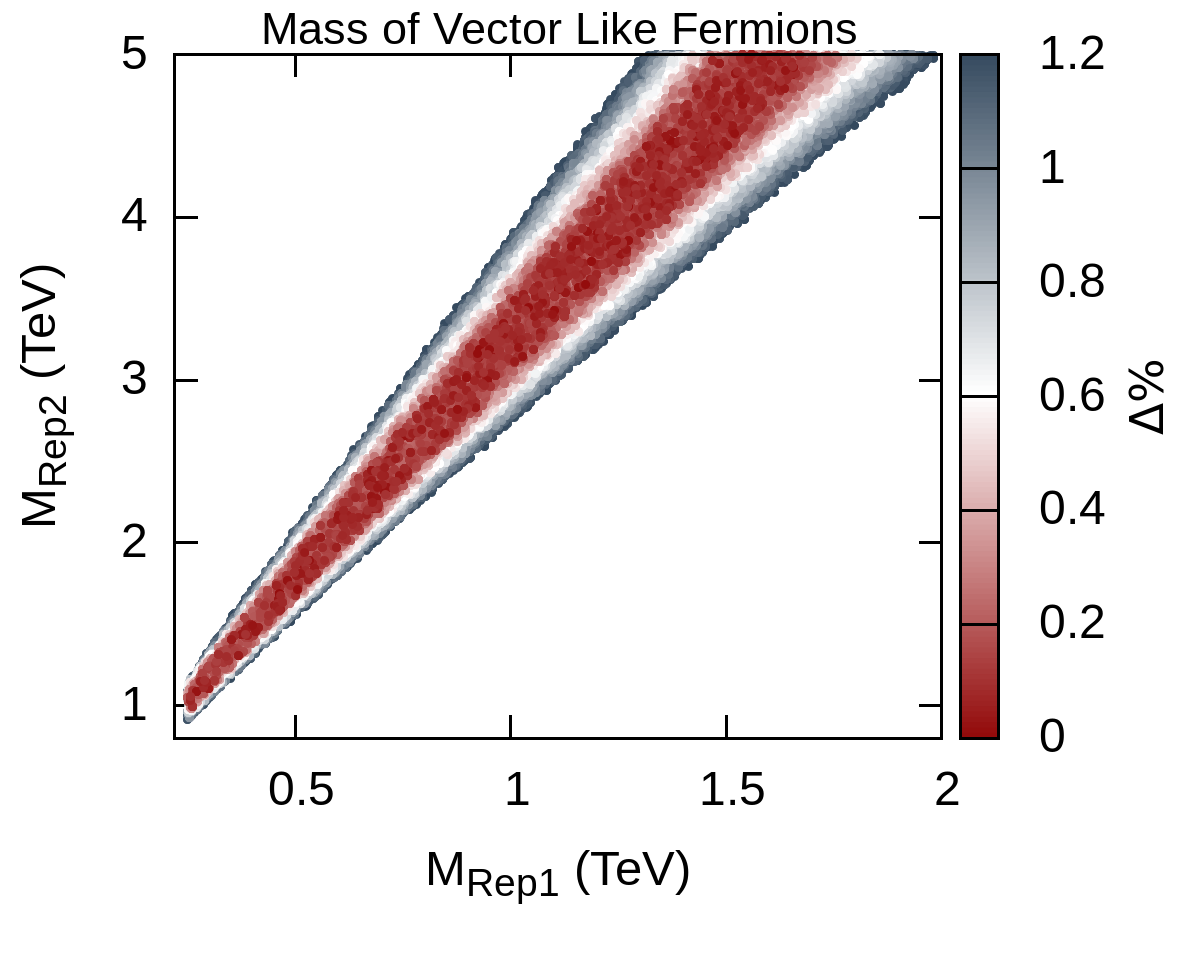}}
\caption{\sf Model 6: Fig. (a) Gauge couplings ($g_1$,$g_2$,$g_3$) 
unification and Vacuum stability ($\lambda > 0$)
plot, considering vector-like fermion in Rep.1 of mass 1.51 TeV and Rep.2 of mass 4.81 TeV. Fig. (b) Mass range allowed for vector-like fermions in Rep.1 and Rep.2 
 for gauge unification and Vacuum stability.}\label{f:mod6}
\end{figure}
This model consist of one copy of Rep1=$(1,4,\frac{1}{2})$ and two copies of Rep2=$(6,1,\frac{2}{3})$. The mass range for Rep1 and Rep2 
are (250 GeV to 2 TeV) and (1 TeV to 5 TeV)  respectively. The Rep1 is fourplet under SU(2) representation and has
been studied under minimal dark matter in Ref.~\cite{Cirelli:2005uq}. To our Knowlegde this is a first time it appeared in the unification of gauge coupling.
Rep2 is exotic sextet under SU(3), which we discussed in Model 5.
In the model, $M_{Rep2}$ is greater than $M_{Rep1}$. The change in the beta 
functions 
in the two thresholds are as follows:
\paragraph{(I) $M_{Rep1} < \mu < M_{Rep2}$}
\begin{eqnarray}
\delta b_i\left(M_{Rep1}< \mu < M_{Rep2} \right) &=& 
\left(      
\begin{array}{c}
\frac{8}{10} \\ \frac{20}{3}  \\ 0 
 \end{array}
 \right),\,\,
\delta m_{ij}\left(M_{Rep1}< \mu < M_{Rep2} \right)=
\left(
\begin{array}{ccc}
\frac{9}{25}  &  9  & 0 \\
 3            &  \frac{425}{3}  & 0 \\
 0            &  0  & 0
 \end{array}
  \right)
  \end{eqnarray}
\begin{eqnarray}
 \delta \beta_{u}^{(2)} (M_{Rep1}< \mu < M_{Rep2} )&=& \frac{348}{900} g_{1}^{4}\nonumber
		      + \frac{10}{2}g_{2}^{4}\\
 \delta \beta_{d}^{(2)} (M_{Rep1}< \mu < M_{Rep2} )&=& \frac{12}{900} g_{1}^{4}\nonumber
		      + \frac{10}{2}g_{2}^{4} \\
 \delta \beta_{e}^{(2)} (M_{Rep1}< \mu < M_{Rep2} )&=& \frac{132}{100} g_{1}^{4} + \frac{10}{2}g_{2}^{4}
 \end{eqnarray}
\begin{eqnarray}
\delta \beta_{\lambda}^{(2)}(M_{Rep1}< \mu < M_{Rep2} )&=& -\frac{12}{250} g_{1}^{4}\left(12g_{1}^{2} + 20g_{2}^{2} - 
		  25{\lambda}\right) \nonumber
- \frac{10}{5}g_{2}^{4} \left(4g_{1}^{2}+20g_{2}^{2}-25{\lambda}\right)  
\end{eqnarray}

\paragraph{(II) $\mu > M_{Rep2}$}
\begin{eqnarray}  
\delta b_i\left(\mu > M_{Rep2} \right) &=& 
\left(
\begin{array}{c}
\frac{76}{15} \\ \frac{20}{3} \\ \frac{20}{3}
\end{array}
 \right),\,\, \hphantom{< M_{Rep2}}
\delta m_{ij}\left(\mu > M_{Rep2} \right)=
\left(
\begin{array}{ccc}
 \frac{283}{75}  & 9  	          & \frac{128}{3}\\
  3              & \frac{425}{3}  & 0\\
 \frac{16}{3}    & 0              & \frac{500}{3}
 \end{array}
  \right)
\end{eqnarray}
 \begin{eqnarray}
 \delta \beta_{u}^{(2)} ( \mu > M_{Rep2} )&=& \frac{400}{9} g_{3}^{4}  + \frac{551}{225} g_{1}^{4}\nonumber
		      + \frac{10}{2}g_{2}^{4}\\
 \delta \beta_{d}^{(2)} (\mu > M_{Rep2} )&=& \frac{400}{9} g_{3}^{4}- \frac{76}{900} g_{1}^{4}\nonumber
		      + \frac{10}{2}g_{2}^{4} \\
 \delta \beta_{e}^{(2)} (\mu > M_{Rep2} )&=& \frac{836}{100} g_{1}^{4} + \frac{10}{2}g_{2}^{4}
 \end{eqnarray}
\begin{eqnarray}
\delta \beta_{\lambda}^{(2)}( \mu > M_{Rep2} )&=& -\frac{76}{250} g_{1}^{4}\left(12g_{1}^{2} + 20g_{2}^{2} - 
		  25{\lambda}\right) \nonumber
- \frac{10}{5}g_{2}^{4} \left(4g_{1}^{2}+20g_{2}^{2}-25{\lambda}\right)  
\end{eqnarray}
Gauge coupling unification and running of $y_t$ and $\lambda$ are also shown in \fig{f:mod6}(a),
with one copy of weak isospin fourplet vector-like fermions with degenerate mass
of 1.51 TeV and two copies of color sextet  vector-like fermion with a mass of 4.81 TeV. The \fig{f:mod6}(b) has the mass distribution in Rep1-Rep2 mass plane. 
\subsection{Model 7} 

\begin{figure}
\centering
\subfigure[]{
	\includegraphics[height=5.5 cm,width=7cm]{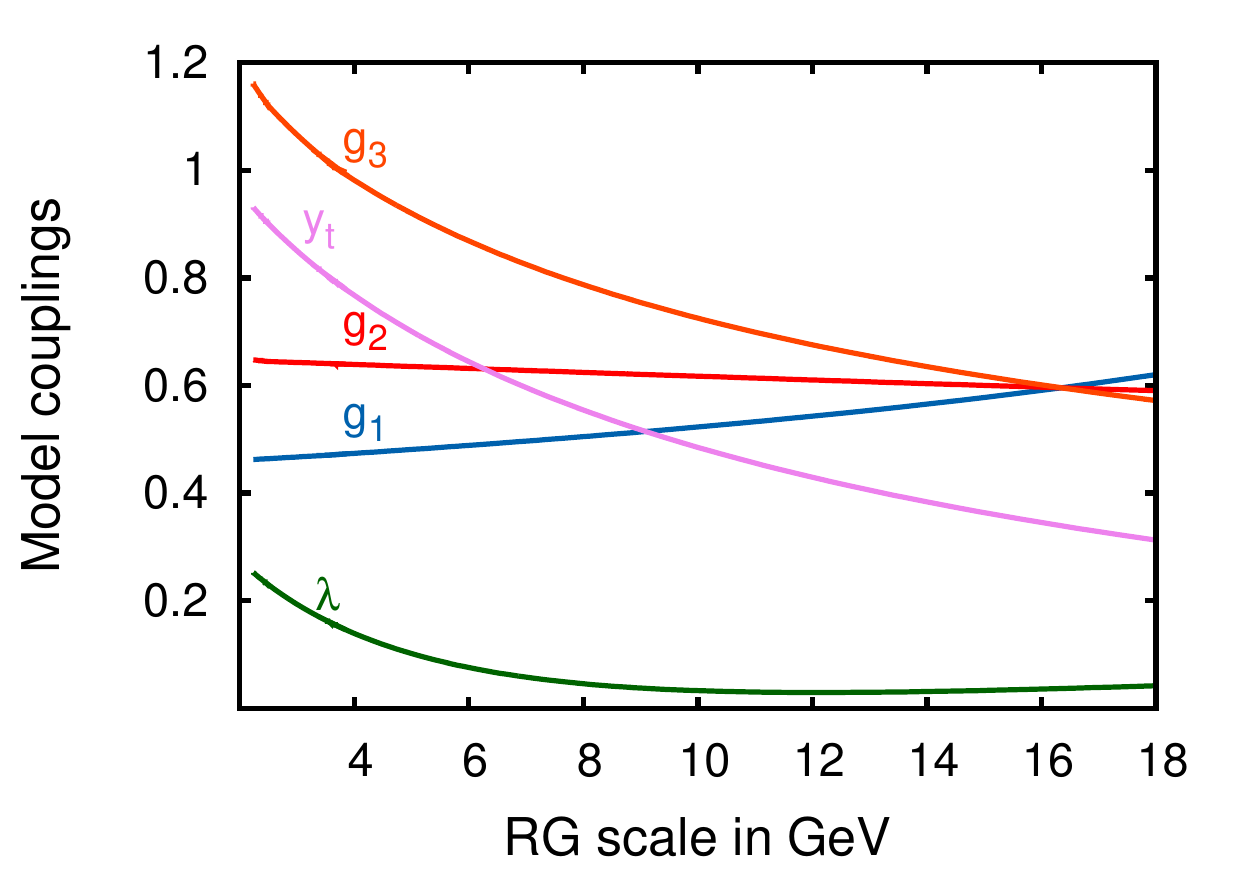}}
\subfigure[]{
	\includegraphics[height=5.4 cm,width=7.3cm]{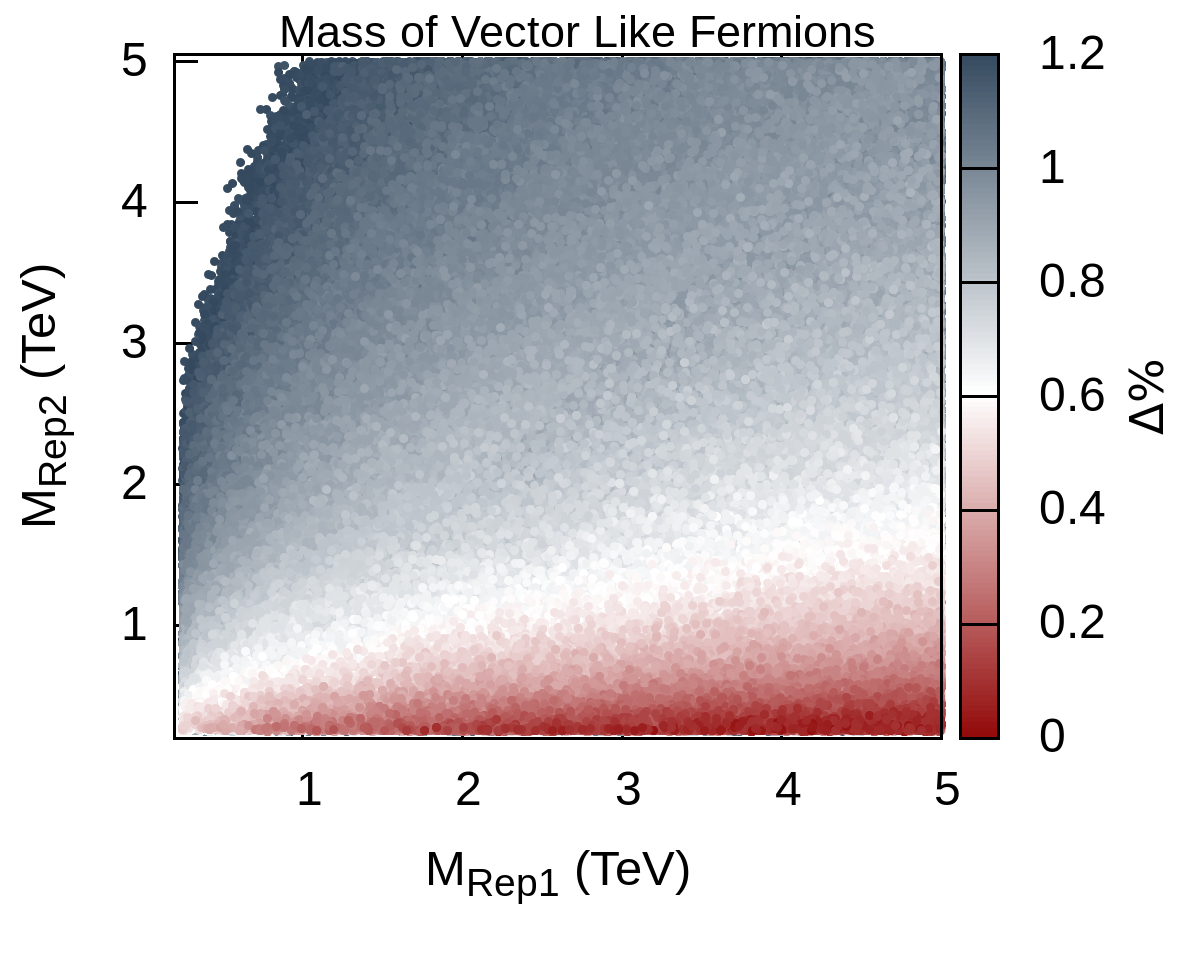}}
\caption{\sf Model 7: Fig. (a) Gauge couplings ($g_1$,$g_2$,$g_3$) 
unification and Vacuum stability ($\lambda > 0$)
plot, considering vector-like fermion in Rep.1 of mass 4.65 TeV and Rep.2 of mass 309 GeV. Fig. (b) Mass range allowed for vector-like fermions in Rep.1 and Rep.2 
 for gauge unification and Vacuum stability.}\label{f:mod7}
\end{figure}
This model consist of one copy of Rep1=$(3,1,\frac{1}{3})$ and two copies of Rep2=$(3,2,\frac{1}{6})$. The mass range for Rep1 and Rep2 
are (250 GeV to 5 TeV) and (250 GeV to 5 TeV)  respectively. Representation one has been discussed in Model 4 with Rep1 having hypercharge
2/3. The difference can been studied with their bound state decay to diphoton channel, as shown in Section~\ref{s:collider-boundstate}.
In this model, there could be points in which 
either of the $M_{Rep1} \sim M_{Rep2}$, $M_{Rep1}> M_{Rep2}$ and $M_{Rep1}< M_{Rep2}$ are possible.
The change in the beta functions in the three conditions are as follows:
\paragraph {(a) $M_{Rep1} = M_{Rep2}$}
\subparagraph{(I) $\mu > M_{Rep2}=M_{Rep1}$}
\begin{eqnarray}  
\delta b_i\left(\mu > M_{Rep1} \right) &=& 
\left(
\begin{array}{c}
\frac{4}{10} \\2\\2
\end{array}
\right),\,\, \hphantom{< M_{Rep1}}
\delta m_{ij}\left(\mu > M_{Rep1} \right)=
\left(
\begin{array}{ccc}
\frac{6}{100}   &   \frac{3}{10}    & \frac{16}{10}\\
\frac{1}{10}    & \frac{245}{10}    & 8\\
\frac{2}{10}     & 3                 & 38
\end{array}
\right)
\end{eqnarray}
\begin{eqnarray}
\delta \beta_{u}^{(2)} ( \mu > M_{Rep1} )&=& \frac{40}{3} g_{3}^{4}  + \frac{174}{900} g_{1}^{4}\nonumber
+ \frac{3}{2}g_{2}^{4}\\
\delta \beta_{d}^{(2)} (\mu > M_{Rep1} )&=& \frac{40}{3} g_{3}^{4}- \frac{6}{900} g_{1}^{4}\nonumber
+ \frac{3}{2}g_{2}^{4}\\
\delta \beta_{e}^{(2)} (\mu > M_{Rep1} )&=& \frac{66}{100} g_{1}^{4} + \frac{3}{2}g_{2}^{4}
\end{eqnarray}
\begin{eqnarray}
\delta \beta_{\lambda}^{(2)}( \mu > M_{Rep1} )&=& -\frac{6}{250} g_{1}^{4}\left(12g_{1}^{2} + 20g_{2}^{2} - 
25{\lambda}\right) \nonumber
- \frac{3}{5}g_{2}^{4} \left(4g_{1}^{2}+20g_{2}^{2}-25{\lambda}\right)  
\end{eqnarray}
\paragraph{(b)$M_{Rep1}> M_{Rep2}$}
\subparagraph{(I) $M_{Rep2} < \mu < M_{Rep1}$}
\begin{eqnarray}
\delta b_i\left(M_{Rep2}< \mu < M_{Rep1} \right) &=& 
\left(      
\begin{array}{c}
\frac{2}{15} \\ 2 \\ \frac{20}{15} 
 \end{array}
 \right),\,\,
\delta m_{ij}\left(M_{Rep2}< \mu < M_{Rep1} \right)=
\left(
\begin{array}{ccc}
\frac{67}{10000} & \frac{3}{10}   & \frac{8}{15} \\
\frac{1}{10}     & \frac{245}{10} & 8 \\
\frac{6}{90}     &  3     	  & \frac{76}{3}
 \end{array}
  \right)
  \end{eqnarray}
\begin{eqnarray}
 \delta \beta_{u}^{(2)} (M_{Rep2}< \mu < M_{Rep1} )&=& \frac{80}{9} g_{3}^{4}  + \frac{58}{900} g_{1}^{4}\nonumber
		      + \frac{3}{2}g_{2}^{4}\\
 \delta \beta_{d}^{(2)} (M_{Rep2}< \mu < M_{Rep1} )&=& \frac{80}{9} g_{3}^{4}  - \frac{2}{900} g_{1}^{4}\nonumber
		      + \frac{3}{2}g_{2}^{4} \\
 \delta \beta_{e}^{(2)} (M_{Rep2}< \mu < M_{Rep1} )&=& \frac{22}{100} g_{1}^{4} + \frac{3}{2}g_{2}^{4}
 \end{eqnarray}
\begin{eqnarray}
\delta \beta_{\lambda}^{(2)}(M_{Rep2}< \mu < M_{Rep1} )&=& -\frac{2}{250} g_{1}^{4}\left(12g_{1}^{2} + 20g_{2}^{2} - 
		  25{\lambda}\right) \nonumber
- \frac{3}{5}g_{2}^{4} \left(4g_{1}^{2}+20g_{2}^{2}-25{\lambda}\right)  
\end{eqnarray}

\subparagraph{(II) $\mu > M_{Rep1}$}
\begin{eqnarray}  
\delta b_i\left(\mu > M_{Rep1} \right) &=& 
\left(
\begin{array}{c}
\frac{4}{10} \\2\\2
 \end{array}
 \right),\,\, \hphantom{< M_{Rep1}}
\delta m_{ij}\left(\mu > M_{Rep1} \right)=
\left(
\begin{array}{ccc}
 \frac{6}{100}   &   \frac{3}{10}    & \frac{16}{10}\\
 \frac{1}{10}    & \frac{245}{10}    & 8\\
\frac{2}{10}     & 3                 & 38
 \end{array}
  \right)
\end{eqnarray}
 \begin{eqnarray}
 \delta \beta_{u}^{(2)} ( \mu > M_{Rep1} )&=& \frac{40}{3} g_{3}^{4}  + \frac{174}{900} g_{1}^{4}\nonumber
		      + \frac{3}{2}g_{2}^{4}\\
 \delta \beta_{d}^{(2)} (\mu > M_{Rep1} )&=& \frac{40}{3} g_{3}^{4}- \frac{6}{900} g_{1}^{4}\nonumber
		      + \frac{3}{2}g_{2}^{4}\\
 \delta \beta_{e}^{(2)} (\mu > M_{Rep1} )&=& \frac{66}{100} g_{1}^{4} + \frac{3}{2}g_{2}^{4}
 \end{eqnarray}
\begin{eqnarray}
\delta \beta_{\lambda}^{(2)}( \mu > M_{Rep1} )&=& -\frac{6}{250} g_{1}^{4}\left(12g_{1}^{2} + 20g_{2}^{2} - 
		  25{\lambda}\right) \nonumber
- \frac{3}{5}g_{2}^{4} \left(4g_{1}^{2}+20g_{2}^{2}-25{\lambda}\right)  
\end{eqnarray}
\paragraph{(c)$M_{Rep2} > M_{Rep1}$}
\subparagraph{(I) $M_{Rep1} < \mu < M_{Rep2}$}
\begin{eqnarray}
\delta b_i\left(M_{Rep1}< \mu < M_{Rep2} \right) &=& 
\left(      
\begin{array}{c}
\frac{4}{15} \\ 0 \\ \frac{2}{3} 
 \end{array}
 \right),\,\,
\delta m_{ij}\left(M_{Rep1}< \mu < M_{Rep2} \right)=
\left(
\begin{array}{ccc}
\frac{533}{10000} & 0   & \frac{16}{15} \\
0     & 0 & 0  \\
\frac{2}{15}     &  0   	  & \frac{38}{3}
 \end{array}
  \right)
  \end{eqnarray}
\begin{eqnarray}
 \delta \beta_{u}^{(2)} (M_{Rep1}< \mu < M_{Rep2} )&=& \frac{40}{9} g_{3}^{4}  + \frac{116}{900} g_{1}^{4}\nonumber
		  \\
 \delta \beta_{d}^{(2)} (M_{Rep1}< \mu < M_{Rep2} )&=& \frac{40}{9} g_{3}^{4}  - \frac{4}{900} g_{1}^{4}\nonumber
		       \\
 \delta \beta_{e}^{(2)} (M_{Rep1}< \mu < M_{Rep2} )&=& \frac{44}{100} g_{1}^{4} 
 \end{eqnarray}
\begin{eqnarray}
\delta \beta_{\lambda}^{(2)}(M_{Rep1}< \mu < M_{Rep2} )&=& -\frac{4}{250} g_{1}^{4}\left(12g_{1}^{2} + 20g_{2}^{2} - 
25{\lambda}\right)  \nonumber
\end{eqnarray}

\subparagraph{(II) $\mu > M_{Rep2}$}
\begin{eqnarray}  
\delta b_i\left(\mu > M_{Rep2} \right) &=& 
\left(
\begin{array}{c}
\frac{4}{10} \\2\\2
 \end{array}
 \right),\,\, \hphantom{< M_{Rep1}}
\delta m_{ij}\left(\mu > M_{Rep2} \right)=
\left(
\begin{array}{ccc}
 \frac{6}{100}   &   \frac{3}{10}    & \frac{16}{10}\\
 \frac{1}{10}    & \frac{245}{10}    & 8\\
\frac{2}{10}     & 3                 & 38
 \end{array}
  \right)
\end{eqnarray}
 \begin{eqnarray}
 \delta \beta_{u}^{(2)} ( \mu > M_{Rep2} )&=& \frac{40}{3} g_{3}^{4}  + \frac{174}{900} g_{1}^{4}\nonumber
		      + \frac{3}{2}g_{2}^{4}\\
 \delta \beta_{d}^{(2)} (\mu > M_{Rep2} )&=& \frac{40}{3} g_{3}^{4}- \frac{6}{900} g_{1}^{4}\nonumber
		      + \frac{3}{2}g_{2}^{4}\\
 \delta \beta_{e}^{(2)} (\mu > M_{Rep2} )&=& \frac{66}{100} g_{1}^{4} + \frac{3}{2}g_{2}^{4}
 \end{eqnarray}
\begin{eqnarray}
\delta \beta_{\lambda}^{(2)}( \mu > M_{Rep2} )&=& -\frac{6}{250} g_{1}^{4}\left(12g_{1}^{2} + 20g_{2}^{2} - 
		  25{\lambda}\right) \nonumber
- \frac{3}{5}g_{2}^{4} \left(4g_{1}^{2}+20g_{2}^{2}-25{\lambda}\right)  
\end{eqnarray}

A sample unification point is shown in \fig{f:mod7}(a), one copy of bottom like vector fermions with degenerate mass
of 4.65 TeV and one copy of left handed quark like  vector fermion with a mass of 309 GeV is considered. The figure shows unification clearly.
The running of $y_t$ and $\lambda$ are also shown. The panel \fig{f:mod7}(b) has the mass distibution in Rep1-Rep2 mass plane. 
\subsection{Model 8} 

\begin{figure}
\centering
\subfigure[]{
	\includegraphics[height=5.5 cm,width=7cm]{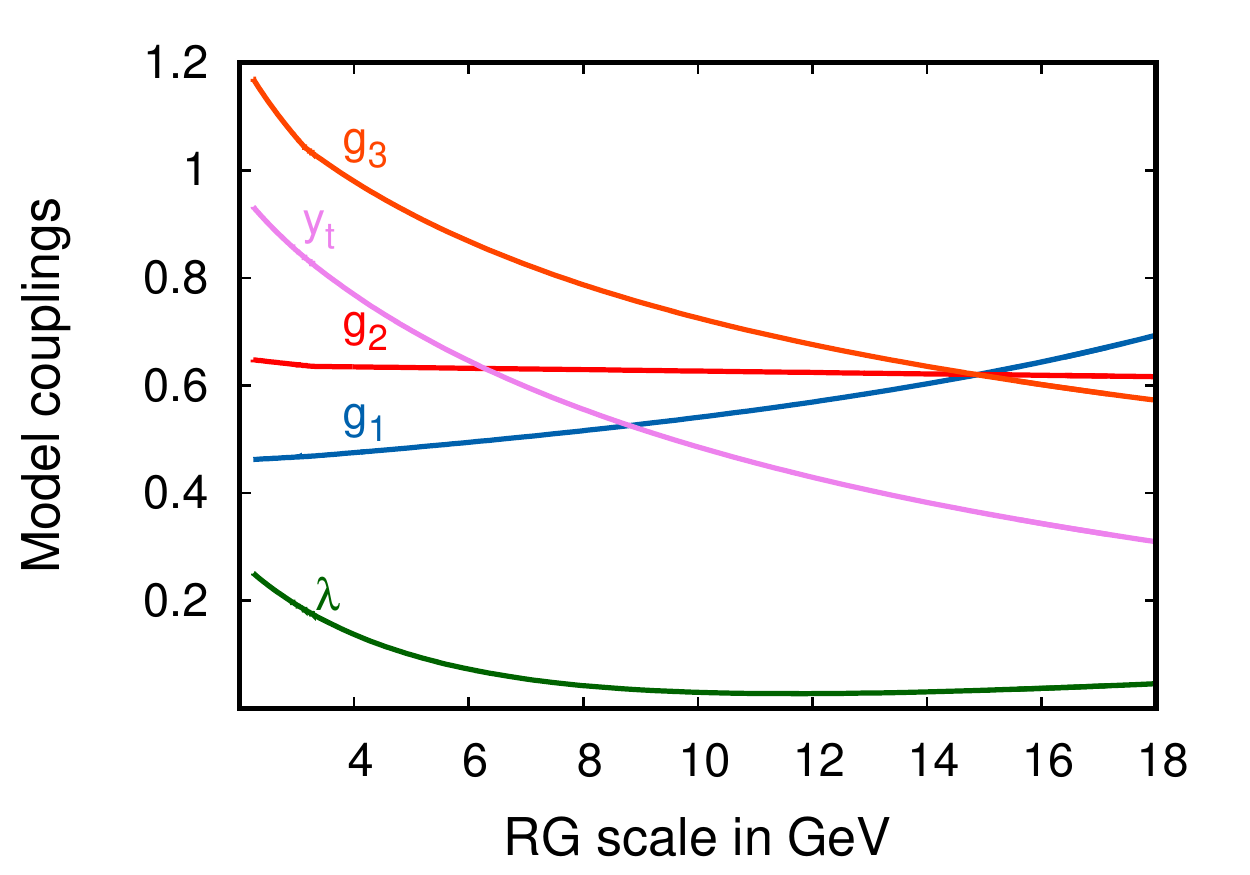}}
\subfigure[]{
	\includegraphics[height=5.4 cm,width=7.3cm]{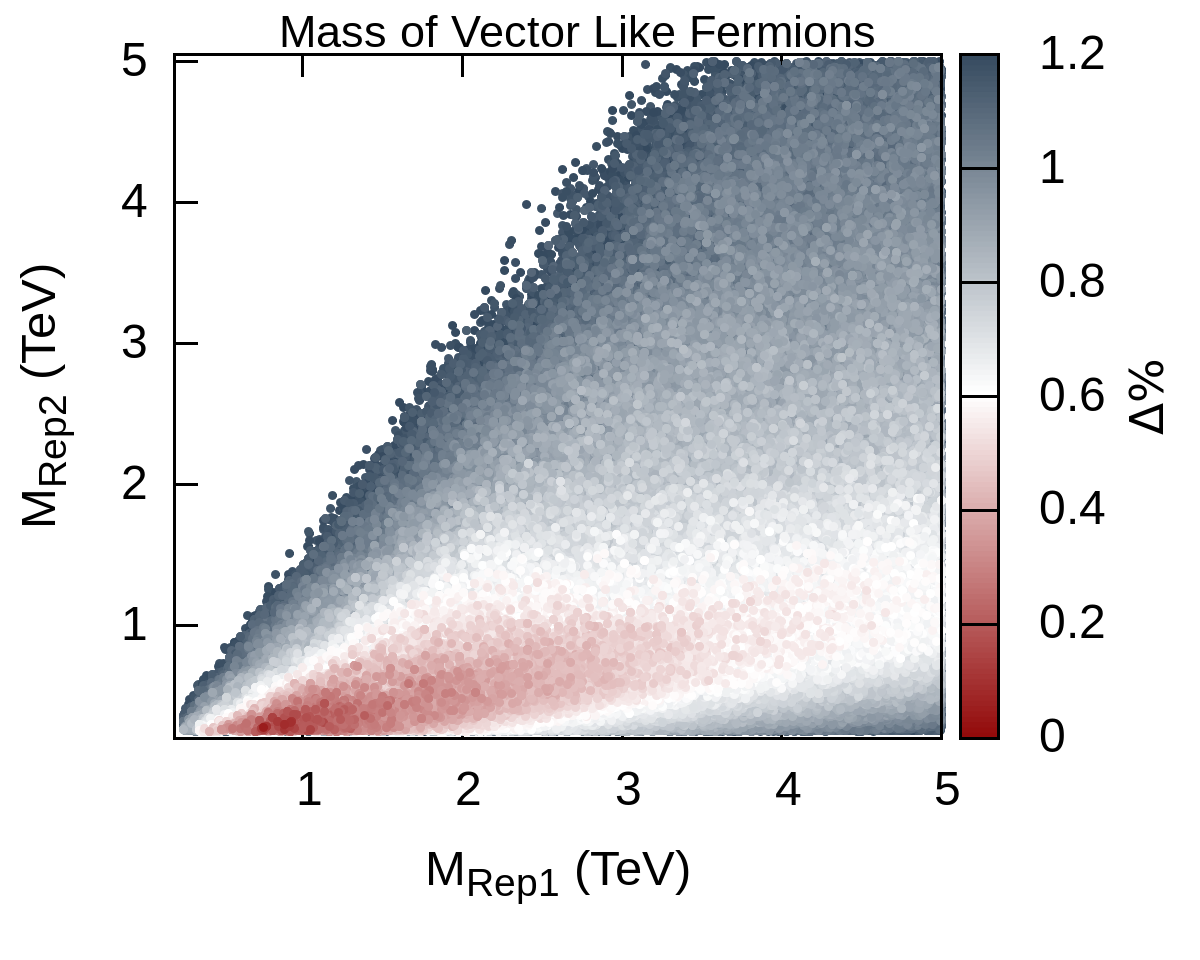}}
\caption{\sf Model 8: Fig. (a) Gauge couplings ($g_1$,$g_2$,$g_3$) 
unification and Vacuum stability ($\lambda > 0$)
plot, considering vector-like fermion in Rep.1 of mass 1.86 TeV and Rep.2 of mass 1.38 TeV. Fig. (b) Mass range allowed for vector-like fermions in Rep.1 and Rep.2 
 for gauge unification and Vacuum stability.}\label{f:mod8}
\end{figure}
This model consist of four copies of Rep1=$(1,2,\frac{1}{2})$ and one copy of Rep2=$(8,1,0)$. The mass range for Rep1 and Rep2 
are (300 GeV to 5 TeV) and (300 GeV to 5 TeV)  respectively. This representation has been discussed in Model 2 with different 
number of particles for each represenatation. The difference can been studied with their bound state decay to diphoton channel and dijet, as shown in Section~\ref{s:collider-boundstate}.

For most points in this model vector-like fermions in Rep1 can be degenerate
with vector-like fermions in Rep2 ($M_{Rep1}\sim M_{Rep2}$) as
shown in \fig{f:mod8}(b). However, there could be points in which 
either of the $M_{Rep1}> M_{Rep2}$ and $M_{Rep1}< M_{Rep2}$ are possible.

The change in the beta functions in the two thresholds are as follows:
\paragraph {(a) $M_{Rep1} = M_{Rep2}$}
\subparagraph{(I) $\mu > M_{Rep2}=M_{Rep1}$}
\begin{eqnarray}  
\delta b_i\left(\mu > M_{Rep1} \right) &=& 
\left(
\begin{array}{c}
\frac{16}{10} \\\frac{24}{9}\\2
\end{array}
\right),\,\, \hphantom{< M_{Rep1}}
\delta m_{ij}\left(\mu > M_{Rep1} \right)=
\left(
\begin{array}{ccc}
\frac{72}{100}   &   \frac{36}{10}    & 0\\
\frac{12}{10}    & \frac{294}{9}    & 0\\
0                & 0                 & 48
\end{array}
\right)
\end{eqnarray}
\begin{eqnarray}
\delta \beta_{u}^{(2)} ( \mu > M_{Rep1} )&=& \frac{40}{3} g_{3}^{4}  + \frac{696}{900} g_{1}^{4}\nonumber
+ \frac{4}{2}g_{2}^{4}\\
\delta \beta_{d}^{(2)} (\mu > M_{Rep1} )&=& \frac{40}{3} g_{3}^{4}- \frac{24}{900} g_{1}^{4}\nonumber
+ \frac{4}{2}g_{2}^{4}\\
\delta \beta_{e}^{(2)} (\mu > M_{Rep1} )&=& \frac{264}{100} g_{1}^{4} + \frac{4}{2}g_{2}^{4}
\end{eqnarray}
\begin{eqnarray}
\delta \beta_{\lambda}^{(2)}( \mu > M_{Rep1} )&=& -\frac{24}{250} g_{1}^{4}\left(12g_{1}^{2} + 20g_{2}^{2} - 
25{\lambda}\right) \nonumber
- \frac{4}{5}g_{2}^{4} \left(4g_{1}^{2}+20g_{2}^{2}-25{\lambda}\right)  
\end{eqnarray}
\paragraph {(b) $M_{Rep1} > M_{Rep2}$}
\subparagraph{(I) $M_{Rep2} < \mu < M_{Rep1}$}
\begin{eqnarray}
\delta b_i\left(M_{Rep2}< \mu < M_{Rep1} \right) &=& 
\left(      
\begin{array}{c}
0 \\ 0 \\ 2 
 \end{array}
 \right),\,\,
\delta m_{ij}\left(M_{Rep2}< \mu < M_{Rep1} \right)=
\left(
\begin{array}{ccc}
0 & 0   & 0 \\
0 & 0 & 0 \\
0 &  0& 48
 \end{array}
  \right)
  \end{eqnarray}
\begin{eqnarray}
 \delta \beta_{u}^{(2)} (M_{Rep2}< \mu < M_{Rep1} )&=& \frac{40}{3} g_{3}^{4} \\
 \delta \beta_{d}^{(2)} (M_{Rep2}< \mu < M_{Rep1} )&=& \frac{40}{3} g_{3}^{4} \\
 \delta \beta_{e}^{(2)} (M_{Rep2}< \mu < M_{Rep1} )&=& 0
 \end{eqnarray}
\begin{eqnarray}
\delta \beta_{\lambda}^{(2)}(M_{Rep2}< \mu < M_{Rep1} )&=& 0
\end{eqnarray}

\subparagraph{(II) $\mu > M_{Rep1}$}
\begin{eqnarray}  
\delta b_i\left(\mu > M_{Rep1} \right) &=& 
\left(
\begin{array}{c}
\frac{16}{10} \\\frac{24}{9}\\2
 \end{array}
 \right),\,\, \hphantom{< M_{Rep1}}
\delta m_{ij}\left(\mu > M_{Rep1} \right)=
\left(
\begin{array}{ccc}
 \frac{72}{100}   &   \frac{36}{10}    & 0\\
 \frac{12}{10}    & \frac{294}{9}    & 0\\
0                & 0                 & 48
 \end{array}
  \right)
\end{eqnarray}
 \begin{eqnarray}
 \delta \beta_{u}^{(2)} ( \mu > M_{Rep1} )&=& \frac{40}{3} g_{3}^{4}  + \frac{696}{900} g_{1}^{4}\nonumber
		      + \frac{4}{2}g_{2}^{4}\\
 \delta \beta_{d}^{(2)} (\mu > M_{Rep1} )&=& \frac{40}{3} g_{3}^{4}- \frac{24}{900} g_{1}^{4}\nonumber
		      + \frac{4}{2}g_{2}^{4}\\
 \delta \beta_{e}^{(2)} (\mu > M_{Rep1} )&=& \frac{264}{100} g_{1}^{4} + \frac{4}{2}g_{2}^{4}
 \end{eqnarray}
\begin{eqnarray}
\delta \beta_{\lambda}^{(2)}( \mu > M_{Rep1} )&=& -\frac{24}{250} g_{1}^{4}\left(12g_{1}^{2} + 20g_{2}^{2} - 
		  25{\lambda}\right) \nonumber
- \frac{4}{5}g_{2}^{4} \left(4g_{1}^{2}+20g_{2}^{2}-25{\lambda}\right)  
\end{eqnarray}
\paragraph {(c) $M_{Rep2} > M_{Rep1}$}
\subparagraph{(I) $M_{Rep1}< \mu < M_{Rep2}$}
\begin{eqnarray}
\delta b_i\left(M_{Rep1}< \mu < M_{Rep2} \right) &=& 
\left(      
\begin{array}{c}
\frac{16}{10} \\ \frac{24}{9} \\ 0 
\end{array}
\right),\,\,
\delta m_{ij}\left(M_{Rep1}< \mu < M_{Rep2} \right)=
\left(
\begin{array}{ccc}
\frac{72}{100}   &   \frac{36}{10}    & 0\\
\frac{12}{10}    & \frac{294}{9}    & 0\\
0                & 0                 & 0
\end{array}
\right)
\end{eqnarray}
\begin{eqnarray}
 \delta \beta_{u}^{(2)} ( M_{Rep1}< \mu < M_{Rep2} )&=& \frac{696}{900} g_{1}^{4}\nonumber
 + \frac{4}{2}g_{2}^{4}\\
 \delta \beta_{d}^{(2)} (M_{Rep1}< \mu < M_{Rep2} )&=&- \frac{24}{900} g_{1}^{4}\nonumber
 + \frac{4}{2}g_{2}^{4}\\
 \delta \beta_{e}^{(2)} (M_{Rep1}< \mu < M_{Rep2} )&=& \frac{264}{100} g_{1}^{4} + \frac{4}{2}g_{2}^{4}
 \end{eqnarray}
 \begin{eqnarray}
 \delta \beta_{\lambda}^{(2)}(M_{Rep1}< \mu < M_{Rep2} )&=& -\frac{24}{250} g_{1}^{4}\left(12g_{1}^{2} + 20g_{2}^{2} - 
 25{\lambda}\right) \nonumber
 - \frac{4}{5}g_{2}^{4} \left(4g_{1}^{2}+20g_{2}^{2}-25{\lambda}\right)  
\end{eqnarray}

\subparagraph{(II) $\mu > M_{Rep2}$}
\begin{eqnarray}  
\delta b_i\left(\mu > M_{Rep2} \right) &=& 
\left(
\begin{array}{c}
\frac{16}{10} \\\frac{24}{9}\\2
\end{array}
\right),\,\, \hphantom{< M_{Rep2}}
\delta m_{ij}\left(\mu > M_{Rep2} \right)=
\left(
\begin{array}{ccc}
\frac{72}{100}   &   \frac{36}{10}    & 0\\
\frac{12}{10}    & \frac{294}{9}    & 0\\
0                & 0                 & 48
\end{array}
\right)
\end{eqnarray}
\begin{eqnarray}
\delta \beta_{u}^{(2)} ( \mu > M_{Rep2} )&=& \frac{40}{3} g_{3}^{4}  + \frac{696}{900} g_{1}^{4}\nonumber
+ \frac{4}{2}g_{2}^{4}\\
\delta \beta_{d}^{(2)} (\mu > M_{Rep2} )&=& \frac{40}{3} g_{3}^{4}- \frac{24}{900} g_{1}^{4}\nonumber
+ \frac{4}{2}g_{2}^{4}\\
\delta \beta_{e}^{(2)} (\mu > M_{Rep2} )&=& \frac{264}{100} g_{1}^{4} + \frac{4}{2}g_{2}^{4}
\end{eqnarray}
\begin{eqnarray}
\delta \beta_{\lambda}^{(2)}( \mu > M_{Rep2} )&=& -\frac{24}{250} g_{1}^{4}\left(12g_{1}^{2} + 20g_{2}^{2} - 
25{\lambda}\right) \nonumber
- \frac{4}{5}g_{2}^{4} \left(4g_{1}^{2}+20g_{2}^{2}-25{\lambda}\right)  
\end{eqnarray}

A sample unification point is shown in \fig{f:mod8}(a), four copies of  lepton like vector fermions with degenerate mass
of 1.86 TeV and one copy of gluion  like  vector fermion with a mass of 1.38 TeV is considered. The figure shows unification clearly.
The running of $y_t$ and $\lambda$ are also shown. The panel \fig{f:mod8}(b) has the mass distibution in Rep1-Rep2 mass plane. 
\subsection{Model 9} 

\begin{figure}
\centering
\subfigure[]{
	\includegraphics[height=5.5 cm,width=7cm]{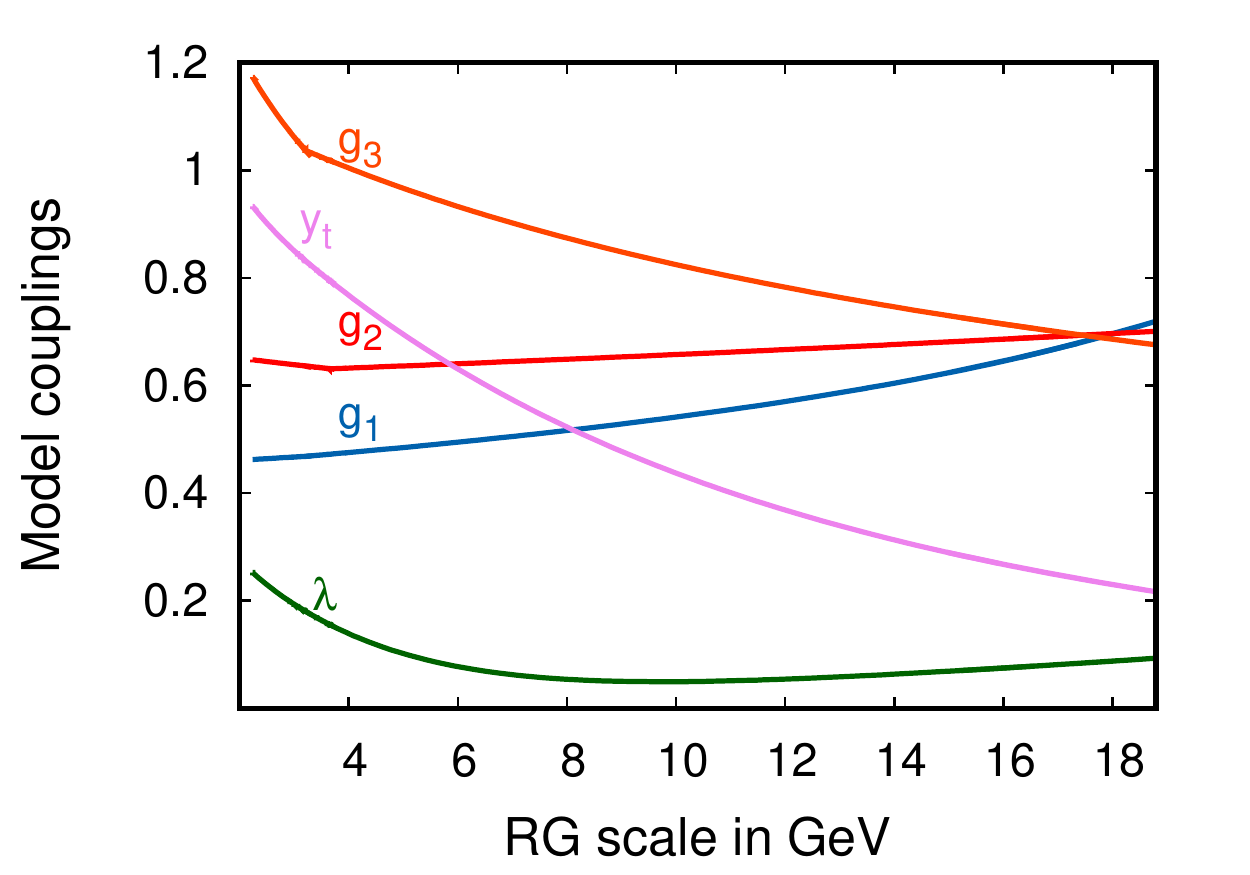}}
\subfigure[]{
	\includegraphics[height=5.4 cm,width=7.3cm]{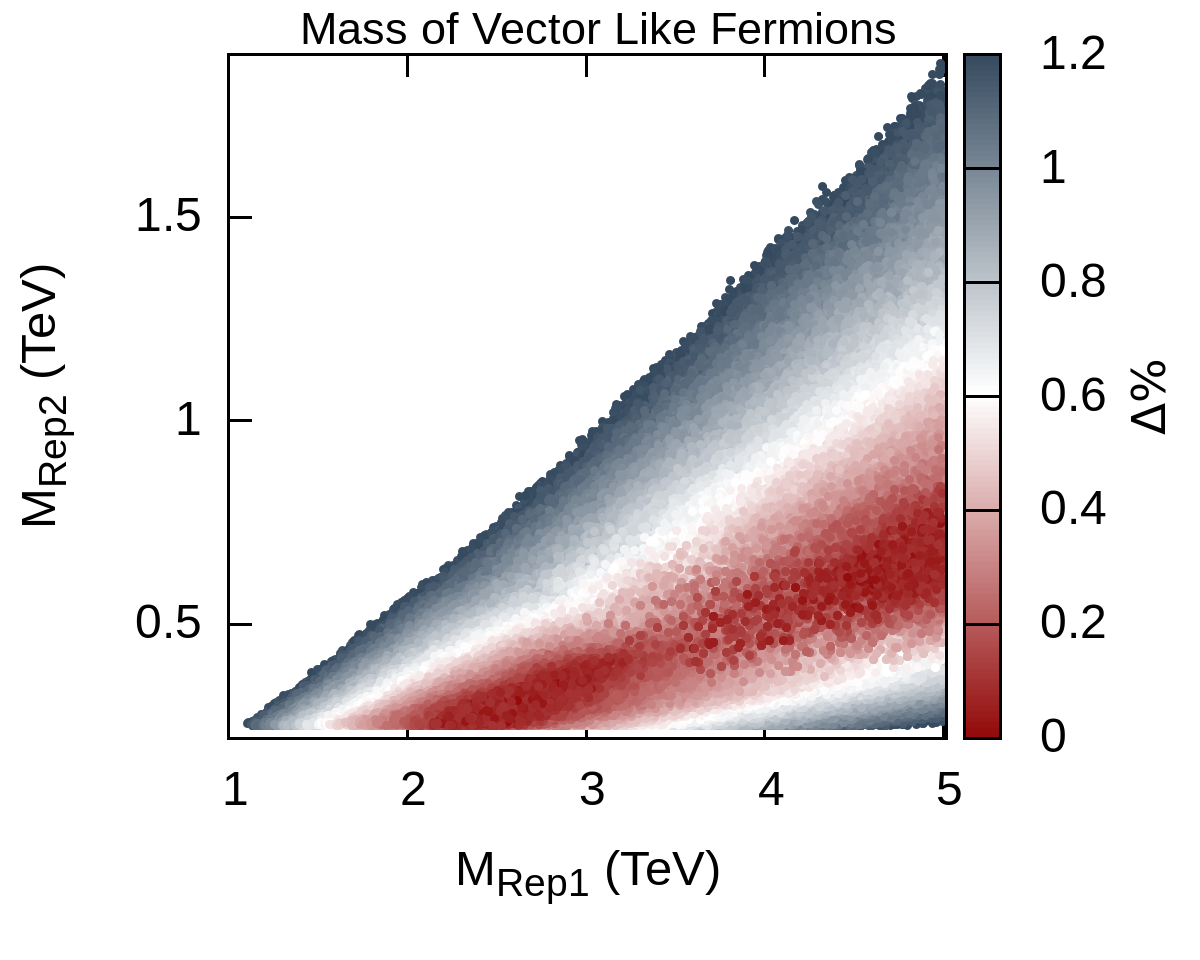}}
\caption{\sf Model 9: Fig. (a) Gauge couplings ($g_1$,$g_2$,$g_3$) 
unification and Vacuum stability ($\lambda > 0$)
plot, considering vector-like fermion in Rep.1 of mass 4.6 TeV and Rep.2 of mass 1.6 TeV. Fig. (b) Mass range allowed for vector-like fermions in Rep.1 and Rep.2 
 for gauge unification and Vacuum stability.}\label{f:mod9}
\end{figure}
This model consist of three copies of Rep1=$(1,3,0)$ and six copies of Rep2=$(3,1,\frac{1}{3})$. The mass range for Rep1 and Rep2 
are (1.1 TeV to 5 TeV) and (250 GeV to 1.8 TeV)  respectively. This representation has been discussed in Model 3 with different 
number of particle for each represenatation. The difference can been studied with their bound state decay to diphoton channel and dijet, as shown in Section~\ref{s:collider-boundstate}. In the model, $M_{Rep1}$ is greater than $M_{Rep2}$. 
The change in the beta functions in the two thresholds are as follows:
\paragraph{(I) $M_{Rep2} < \mu < M_{Rep1}$}
\begin{eqnarray}
\delta b_i\left(M_{Rep2}< \mu < M_{Rep1} \right) &=& 
\left(      
\begin{array}{c}
\frac{16}{10} \\ 0 \\ 4 
 \end{array}
 \right),\,\,
\delta m_{ij}\left(M_{Rep2}< \mu < M_{Rep1} \right)=
\left(
\begin{array}{ccc}
\frac{32}{100} & 0   & \frac{64}{10} \\
0              & 0  & 0 \\
\frac{8}{10}     &  0     	  & 76
 \end{array}
  \right)
  \end{eqnarray}
\begin{eqnarray}
 \delta \beta_{u}^{(2)} (M_{Rep2}< \mu < M_{Rep1} )&=& \frac{240}{9} g_{3}^{4}  + \frac{696}{900} g_{1}^{4}\nonumber\\
 \delta \beta_{d}^{(2)} (M_{Rep2}< \mu < M_{Rep1} )&=& \frac{240}{9} g_{3}^{4}  - \frac{24}{900} g_{1}^{4}\nonumber \\
 \delta \beta_{e}^{(2)} (M_{Rep2}< \mu < M_{Rep1} )&=& \frac{264}{100} g_{1}^{4}
 \end{eqnarray}
\begin{eqnarray}
\delta \beta_{\lambda}^{(2)}(M_{Rep2}< \mu < M_{Rep1} )&=& -\frac{24}{250} g_{1}^{4}\left(12g_{1}^{2} + 20g_{2}^{2} - 
		  25{\lambda}\right) \nonumber
\end{eqnarray}

\paragraph{(II) $\mu > M_{Rep1}$}
\begin{eqnarray}
\delta b_i\left(\mu > M_{Rep1}\right) &=& 
\left(      
\begin{array}{c}
\frac{16}{10} \\ 4 \\ 4 
\end{array}
\right),\,\,
\delta m_{ij}\left(\mu > M_{Rep1} \right)=
\left(
\begin{array}{ccc}
\frac{32}{100} & 0   & \frac{64}{10} \\
0              & 64  & 0 \\
\frac{8}{10}     &  0     	  & 76
\end{array}
\right)
\end{eqnarray}
\begin{eqnarray}
\delta \beta_{u}^{(2)} \mu > M_{Rep1} )&=& \frac{240}{9} g_{3}^{4}  + \frac{696}{900} g_{1}^{4}\nonumber + \frac{6}{2}g_{1}^{4}\\
\delta \beta_{d}^{(2)} (\mu > M_{Rep1} )&=& \frac{240}{9} g_{3}^{4}  - \frac{24}{900} g_{1}^{4}\nonumber + \frac{6}{2}g_{1}^{4}\\
\delta \beta_{e}^{(2)} (\mu > M_{Rep1} )&=& \frac{264}{100} g_{1}^{4} + \frac{6}{2}g_{1}^{4} 
\end{eqnarray}
\begin{eqnarray}
\delta \beta_{\lambda}^{(2)}(\mu > M_{Rep1} )&=& -\frac{24}{250} g_{1}^{4}\left(12g_{1}^{2} + 20g_{2}^{2} - 
25{\lambda}\right) \nonumber
 - \frac{6}{5}g_{2}^{4} \left(4g_{1}^{2}+20g_{2}^{2}-25{\lambda}\right)  
\end{eqnarray}
A sample unification point is shown in \fig{f:mod9}(a), three copies of weak-isospin triplet vector-like fermions with degenerate mass
of 4.6 TeV and six copies of bottom  like  vector fermion with a mass of 1.6 GeV is considered. The figure shows unification clearly.
The running of $y_t$ and $\lambda$ are also shown. The panel \fig{f:mod9}(b) has the mass distibution in Rep1-Rep2 mass plane. 
\section{Collider Signature of Minimal vector-like fermion models}\label{s:collider-boundstate}
The models listed in \tabl{t:two-ferm} have several exotic states lying close to electroweak scale,
which can be probed at LHC. Models  have exotic lepton like states (uncoloured) mostly in doublet,
triplet and fourplet representation of SU(2). These states are produced at the LHC through Drell-Yan process and typically 
have cross-section of the order 10 fb~\cite{Franceschini:2008pz}(roughly slepton production or exotic lepton production).
These particles decay through Yukawa interaction to lighter SM leptons. In the limit of vanishing Yukawa couplings, these 
particles can manifest as missing energy and disappearing charge track at LHC and limits from monojets and disappearing tracks could
apply to our model.
The LHC at 14 TeV with integrated luminosity 3000 fb$^{-1}$ is only sensitive to mass of order 400 GeV~\cite{Low:2014cba}.
In the following we will concentrate on the strongly interacting exotic sector; 
which appears in all the successful models.
\subsection{Decay Operators}
The models tabulated in the above has exotic fields and some of these fields 
don't have renormalisation level decay operators. These fields are (i) 
$\left(6, 1, \frac13 \right)$, (ii) $\left(6, 1, \frac23 \right)$ and (iii) $\left(8, 1, 0 \right)$.
Now question is whether we can have higher dimensional operators or 
not. Note that if there exists any higher dimensional operator then there must 
be some new fields which got integrated out in some higher scales. Now this 
scale has to be high (close to the GUT scale) as otherwise unification will be 
disturbed. These higher dimensional operators are suppressed as
\begin{equation}
\frac{{\cal O}}{\Lambda^{{\rm dim} ({\cal O})-4}},
\end{equation}
where ${\rm dim} ({\cal O})$ is the dimension of the operator ${\cal O}$.  
Six-dimensional operators are suppressed by square of the GUT scale and thus 
life-time of the particle is expected to be High ($\sim 10^{33}$ years). Thus 
we are focusing only on the five dimensional operators. Any five dimensional 
operator for decay of such particle must have the forms:
\begin{eqnarray}
&&(1) \mbox{ Exotic field} \times \mbox{ a standard model fermion } \times 
\mbox{Higgs }\times \mbox{Higgs}\\
&&(2) \mbox{ Exotic field} \times \mbox{ a standard model fermion} \times 
\mbox{Gauge boson }\times \mbox{Gauge boson},
\end{eqnarray}
where in the place of the Higgs and SM fermions fields one can use their 
conjugate fields. Thus colour charge of the exotic field has to be neutralized 
by SM fermion to form a five dimensional operator involving the Higgs. In the 
SM, there is no such field and hence possibility (1)  is not possible. For the 
second case, colour representation of the exotic times that of the SM fermion 
field must transform as any one of 1, 8, 10 and 27 dimensional representation. 
However we don't have SM field with above representation hence, this  second possibility is also ruled out. 
These exotic fields can form a bound state and in the next subsection we'll 
discuss this in details.

\subsection{Formalism for Bound state}
 
  In this section we investigate the possibility of producing  bound states of the
colour vector-like fermions. The idea of bound state has been  studied, 
in understanding bottom and charm quark through their bound states. 
For the formation of bound state, we assume the new vector-like fermion ($\psi$) is long lived
so that it has time to form a bound state prior to decaying. This condition is easily 
satisfied in our case, as the Yukawa coupling between the new vector-like fermions and 
SM particle is assumed to be negligible. 
The bound state formalism has been studied in~\cite{Kats:2012ym, Kats:2009bv}, where they focus on pair-produced colour particle
Beyond the Standard Model by the observation of diphoton, dijet etc. resonances
arising from QCD bound state. 
 
\begin{figure}
\centering
\subfigure[]{
\includegraphics[height=5.5cm, 
width=7cm]{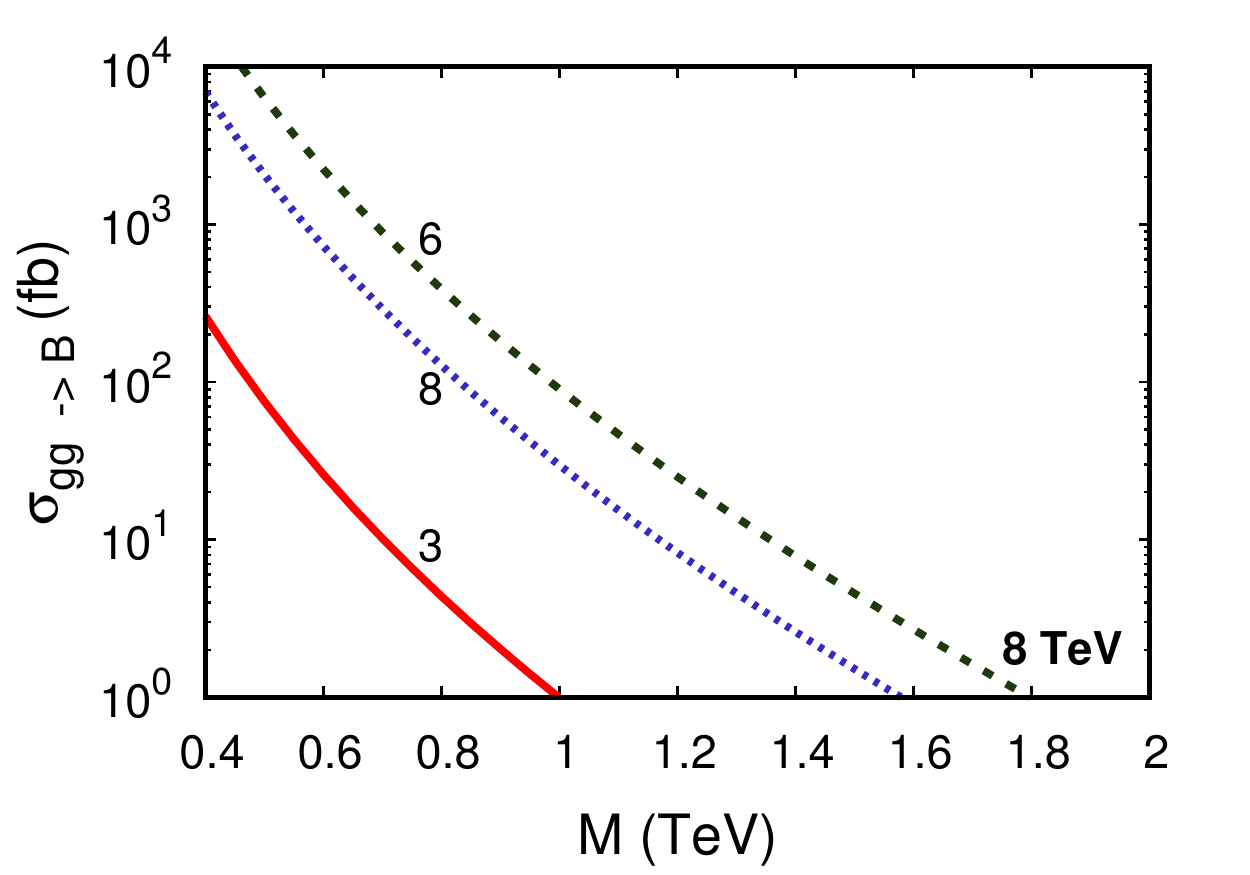}}
\subfigure[]{
\includegraphics[height=5.5cm, 
width=7cm]{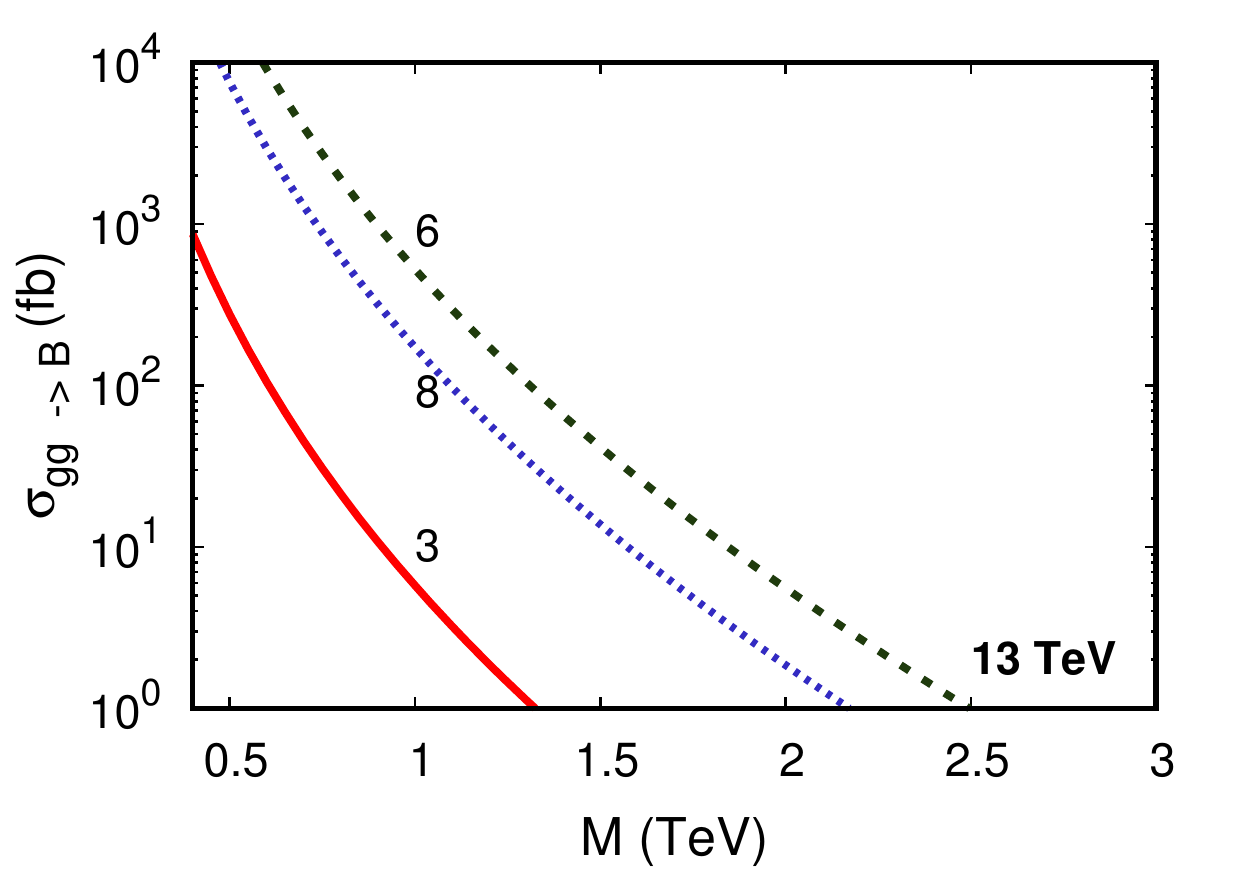}}
\caption{{\sf Cross section of Bound State for Representation $\mathcal{R} = 
1$ and  $J=0$, from constituent particle of Representation $R= 3, 
6, 8$ with respect to mass of bound state. The left fig corespond to $\sqrt s$ = 8 TeV and right at $\sqrt s$ = 13 TeV.}}
\label{fig:prod8and13}
\end{figure}

We assume that the only interaction that contribute to the production of bound state is the Standard Model SU(3) colour gauge interaction.
We estimate the annihilation rates and parton-level cross-section at leading order, along with NLO MSTW parton distribution
functions~\cite{Martin:2009iq}, to compute the LHC signals for $\sqrt{s}$= 8 TeV, 13 TeV and 14 TeV evaluated 
at scale $m_\psi$. The production cross section of colour singlet spin zero bound state from constituent vector-like 
fermion with colour representation 3, 6, 8 are shown in~\fig{fig:prod8and13} and~\fig{fig:14TeVcros}. As pointed out in Ref.~\cite{Younkin:2009zn}, NLO corrections to cross-section can increase the diphoton resonance arising from
stoponium by 25\%. Therefore, large uncertainties are expected in our result of factor of two or so. This still can
allow us to constraints minimal vector-like fermion model. 
 \\
 Further uncertainty in our results arises because of limits extracted from ATLAS and CMS result, which is 
 obtained for a fixed spin and production channel. Signal shape have some dependence on the acceptance, intrinsic width
 and whether a jet is due to parton-level gluon or quark, this adds to some uncertainties.
 \\
 A pair of $\psi\bar{\psi}$ near threshold can form a QCD bound state, 
 which we defined as $\mathcal{O}$. If the decay width of $\mathcal{O}$ is smaller 
 than its respective binding energy, it can be observed as a resonance
 which annihilates to SM particles. For particles ($\psi$) of mass $m_\psi$ $\gg \Lambda_{QCD}$, the Bohr radius of relevant bound state is much smaller than the QCD 
 scale and the velocity of its constituents is non relativistic, we can estimate bound state as modified hydrogenic approximation.
 For a particle $\psi$ in the colour representation $R$, the potential between $\psi$ and $\bar{\psi}$ depends on the colour representation
 $\mathcal{R}$ of the $\psi\bar{\psi}$ pair 
 through the casimirs of $R$ and $\mathcal{R}$ as
\begin{equation}
  V(r)=-C\frac{\bar{\alpha}_s}{r}, \hspace{2cm} C=C(R)-\frac{1}{2}C(\mathcal{R}) 
 \end{equation}
 where $\bar{\alpha}_s$ is defined as the running coupling at the scale of the average distance between the two particle in the corresponding 
 hydrogenic state, which is order of the Bohr radius $a_0=2/(C\bar{\alpha}_s m_\psi)$ (for which we used Ref.~\cite{Pineda:1997hz}).
 The binding energy of the wave 
 functions at the origin for the ground state are given by
 \begin{equation}
  E_b=-\frac{1}{4}C^{2}{\bar{\alpha}_s}^{2}m_\psi, \hspace{1cm} |\psi(0)|^{2}\equiv \frac{1}{4\pi}|R(0)|^{2}=\frac{C^{3}{\bar{\alpha}_s}^{3}m_\psi^{3}}{8\pi}
 \end{equation}
 The quantum number of $\psi$ determines the production as well as the decay modes of bound state particle $\mathcal{O}$.
 The cross-section for the bound state $\mathcal{O}$ to be produced by initial-state partons $x$ and $y$ is given as
 \begin{equation}
 \hat{\sigma}_{xy\rightarrow \mathcal{O}}(\hat{s})=\frac{8\pi}{m_\psi}\frac{ \hat{\sigma}_{xy\rightarrow \psi\bar{\psi}}^{free}(\hat{s})  }{\beta(\hat{s})}
 |\psi(0)|^{2}2 \pi \delta(\hat{s}-M^2)
 \end{equation}
where $M=2m_\psi+E_b$ is the mass of the bound state, $\beta(\hat{s})$ is the velocity of $\psi$ or $\bar{\psi}$ in CM frame. The production cross-section of any narrow 
resonance $\mathcal{O}$ of mass $M$ and spin $J$ from parton $x$ and $y$, and the decay rate 
of bound state to $x$ and $y$, are related by
\begin{equation}\label{e:bound}
  \hat{\sigma}_{xy\rightarrow \mathcal{O}}=\frac{ 2{\pi}(2J+1)d_{\mathcal{O}}(\mathcal{R})} {D_x D_y} \frac{\Gamma_{\mathcal{O}\rightarrow xy}}{M}
  2 \pi \delta(\hat{s}-M^2) \hspace{1cm} (\times \hbox{2 for x=y})
\end{equation}
where $D_{\mathcal{O}}$ denotes the colour representation of particle $\mathcal{O}$.\\

In the next subsection we will strict ourself to study the colour singlet and spin zero (J=0) bound state system. 
Assuming the production cross-section of ${\psi}\bar{\psi}$ is dominated by gluon fusion. The gluon fusion partonic production cross-section of bound state is given by
 \begin{equation}\label{e:gbound}
  \hat{\sigma}_{gg\rightarrow \mathcal{O}}=\frac{{\pi}^{2}}{8} \frac{\Gamma_{\mathcal{O}\rightarrow gg}}{M}
  \delta(\hat{s}-M^2) \hspace{1cm} 
\end{equation}
Depending on the quantum number of $\psi$, bound state $\mathcal{O}$ can decay 
to diphoton, dijet, $Z \gamma$, $ZZ$  and $W^{+}W^{-}$ channels. The production of preceding pair events produced in proton-proton collisions in LHC can be predicted
as $\sigma(pp \rightarrow \mathcal{O}) \times BR(\mathcal{O} \rightarrow X_1 X_2)$.\\
Here we will identify the channels in which the bound state resonance would be most easily measurable and compute the corresponding cross-section as a function of 
the mass, colour representation and charge of the constituent particles.
The promising final states that we analyzed are diphoton and dijet channels.
In the case of SU(2) multiplet the large mass splitting is constrained by
Electroweak precision test, which modifies the oblique parameter T and S~\cite{Lavoura:1992np}, 
hence we have analysed our results in degenerate mass scenario. 
\subsection{Signals}
\subsubsection{$\gamma \gamma$, $ZZ$, $Z\gamma$, $W^{+}W^{-}$ channel}
Any spin half particle can be produced in pairs (in gg collisions ) in an S-wave $J=0$ colour singlet bound state, which can decay as typically 
narrow $\gamma \gamma$, $ZZ$, $Z\gamma$ resonance.
The decay width of the $\gamma \gamma$, $ZZ$, $Z\gamma$ signal due to spin $J=0$ bound state is given as~\cite{PhysRevD.35.3366}
\begin{equation}
 \Gamma(\mathcal{O}^{\mathcal{R}}_{J=0}\rightarrow {\gamma \gamma}) = \frac{Q^{4}C(R)^{3}d_{R}}{2}{\alpha}^{2} {\bar{\alpha}_s}^{3}m_\psi
\end{equation}
\begin{equation}
  \Gamma(\mathcal{O}^{\mathcal{R}}_{J=0}\rightarrow {\gamma Z}) = \frac{Q^{2}C(R)^{3}d_{R}}{\sin^{2} \theta_W \cos^{2} \theta_W }
  (1-R_{Z})v^2{\alpha}^{2} {\bar{\alpha}_s}^{3}m_\psi 
 \end{equation}
 \begin{equation}
  \Gamma(\mathcal{O}^{\mathcal{R}}_{J=0}\rightarrow {Z Z}) = \frac{C(R)^{3}d_{R}}{2 \sin^{4} \theta_W \cos^{4} \theta_W }\frac{\beta^3_{Z}}{(1-2R_{Z})} v^4{\alpha}^{2}
  {\bar{\alpha}_s}^{3}m_\psi 
 \end{equation}\\
 where $v = \frac{1}{2}(T_{3L}+T_{3R}) - Q \sin^{2}\theta_W $, $T_{3L, 3R}$ is the third component of the weak isospin for the 
 left and right handed state of the fermion, $Q$ is the charge of particle, $R_{Z}= M_Z/M$ and 
 $\beta_{Z}= \sqrt{1-4R_{Z}}$.

Model No. 4 of minimal fermion model contains constituent of vector-like fermion (3,2,1/6)
with SU(2) doublet. This can also decay to $W^{+} W^{-}$ channel, which 
is comparable to $\gamma \gamma$ channel. The decay width for $W^{+} W^{-}$
is given as~\cite{PhysRevD.35.3366},
 \begin{equation}
 \Gamma(\mathcal{O}^{\mathcal{R}}_{J=0}\rightarrow {W^{+} W^{-} }) = \frac{3 
\alpha^{2} \beta^{3}_{W}}{16 \sin^{4}\theta_W }\frac{1}{(1-2R_{W})^2}{\bar{\alpha}_s}^{3}m_\psi,
 \end{equation}
where $R_{W}= M_W/M$, 
 $\beta_{W}= \sqrt{1-4R_{W}}$. 
 
The branching fraction of the isoweak singlet fermions which satisfied the gauge 
coupling unification and vacuum stability are tabulated in \tabl{t:branching}.

Model No. 4 with vector-like fermion constituent (3,2,1/6),
can  decay to  $gg$ or $\gamma \gamma$, $Z \gamma$, $ZZ$ and $WW$ channels.
With charge Q=-1/3 the branching fraction at mass $m_\psi=1$  TeV is  
$93.55 \%$, $2.80\times10^{-2}\%$, $0.49\%$, $2.13\%$ and $3.79\%$ respectively and for Q = 2/3 
is $93.49\%$, $ 0.44\%$, $1.31\%$, $0.95\%$
$3.79\%$ respectively. We observed that in a large isoweak SU(2) represenatation the total decay width 
can be larger than its width into gg.

Both ATLAS and CMS  have performed a search of resonant production of photon pairs for scalar particle (J=0).
ATLAS~\cite{ATLAS-CONF-2016-059} analysis is based on data corresponding to an integrated luminosity of 
15.4 fb$^{-1}$ at $\sqrt{s}$=13 recorded in 2015 and 2016. CMS~\cite{Khachatryan:2016yec} data sample correspond to 
luminosity 12.9 fb$^{-1}$ at $\sqrt{s}$=13 in 2016, combined statistically with the previous data of 
2012 and 2015 at $\sqrt{s}$=8 and $\sqrt{s}$=13 respectively, with luminosity of 19.7 and 3.3 fb$^{-1}$. 
\subsubsection{ Dijet channel}

S-wave bound state with spin $J=0$ can be produced via $gg \rightarrow \mathcal{O}$ and 
annihilating mostly to $gg$. For j=1/2 there is also a comparable contribution 
from S-wave $J=1$ colour octet bound states produced via 
$q \bar{q} \rightarrow \mathcal{O}$ and annihilating to $q \bar{q}$, which we will not discuss here. 

The decay width of  $gg$ signal due to spin $J=0$ colour singlet bound  state is,
\begin{equation}
 \Gamma(\mathcal{O}^{\mathcal{R}=1}_{J=0}\rightarrow {gg})= \frac{C(R)^{5}d_{R}}{32}{\alpha}_{s}^{2} {\bar{\alpha}_s}^{3}m_\psi              
\end{equation}

$ (\times \hbox{2 for Complex Representation of constituent fermion})  $

Search for narrow resonances decaying to dijet final states in proton-proton collision has been performed by the ATLAS 
and CMS collaborations using the LHC run data at $\sqrt{s}$ 8 TeV as well as 13 TeV. CMS~\cite{Khachatryan:2016ecr} study has been performed with
integrated luminosity 18.8 fb$^{-1}$ at $\sqrt{s}$ = 8 TeV using a novel technique called data scouting. ATLAS~\cite{Aad:2014aqa} has studied with 
$\sqrt{s}$= 8 TeV using full integrated luminosity of 20.3 fb$^{-1}$ masses upto 4.5 TeV. \\
In run-II, ATLAS~\cite{Aaboud:2017yvp}  with centre-of-mass energy $\sqrt{s}$= 13  has studied the dijet search using the data collected in 2015 and 2016 
with luminosity 3.5 fb$^{-1}$ and 33.5 fb$^{-1}$ respectively and CMS~\cite{CMS-PAS-EXO-16-056} has presented a data with luminosity 36 fb$^{-1}$ 
considering masses above 600 GeV.

\begin{table}
\begin{center}
{\small
\begin{tabular}{|c | c |c|c|c|c|c|c|}
\hline\hline
${\rm Fermion}$ & $\mathcal{O} $   & \multicolumn{1}{c}{}  &\multicolumn{1}{c}{} 
$\hbox{Branching Fraction}\times100  $  & \multicolumn{1}{c}{} & \\	      
\hline
& $\vphantom{\frac{\frac12}{\frac12}}$ & $ 
BR(\mathcal{O}\rightarrow {gg})\times100 $ & $BR(\mathcal{O}\rightarrow 
{\gamma \gamma})\times100$ & $BR(\mathcal{O}\rightarrow {\gamma Z})\times100$ 
& $BR(\mathcal{O}\rightarrow{ZZ})\times100$ \\                                
\hline
\multirow{1}{*}{(6,1,1/3)} & 1 &  $99.99$   & $4.80 \times 10^{-3} $   & $2.79
\times 10^{-3} $  & $\vphantom{\frac{\frac12}{\frac12}}$ $4.98 \times 10^{-4} 
$   \\ \cline{2-6}
\hline

\multirow{1}{*}{(6,1,2/3)} & 1 &  $99.87$   & $7.67 \times 10^{-2} $   & $4.45 
\times 10^{-2} $  & $\vphantom{\frac{\frac12}{\frac12}}$ $7.95 \times 10^{-3} 
$   \\ \cline{2-6}           
\hline

\multirow{1}{*}{(8,1,0)}   & 1 &  $100$     & $-$  & $-$  &  $-$  
    \\ \cline{2-6}
    \hline

\multirow{1}{*}{(3,1,1/3)} & 1 &  $99.95$  & $2.99 \times 10^{-2} $  &  $1.74 
\times 10^{-2} $   & $\vphantom{\frac{\frac12}{\frac12}}$  $3.11 \times 10^{-3} 
$    \\ 
\hline
\multirow{1}{*}{(3,1,2/3)} & 1 &  $99.19$  & $ 0.47 $ &  $0.27$  & 
$\vphantom{\frac{\frac12}{\frac12}}$   $4.94 \times 10^{-2} $   \\ 
\hline                           
\hline
\end{tabular}
}
\end{center}
\caption{{\sf Branching fraction for Bound state of $J=0$, colour 
representation singlet at mass of $m_\psi=1$ TeV}}\label{t:branching}
\end{table}

\begin{figure}
\centering
\subfigure[]{
\includegraphics[height=5.5cm, width=7cm]{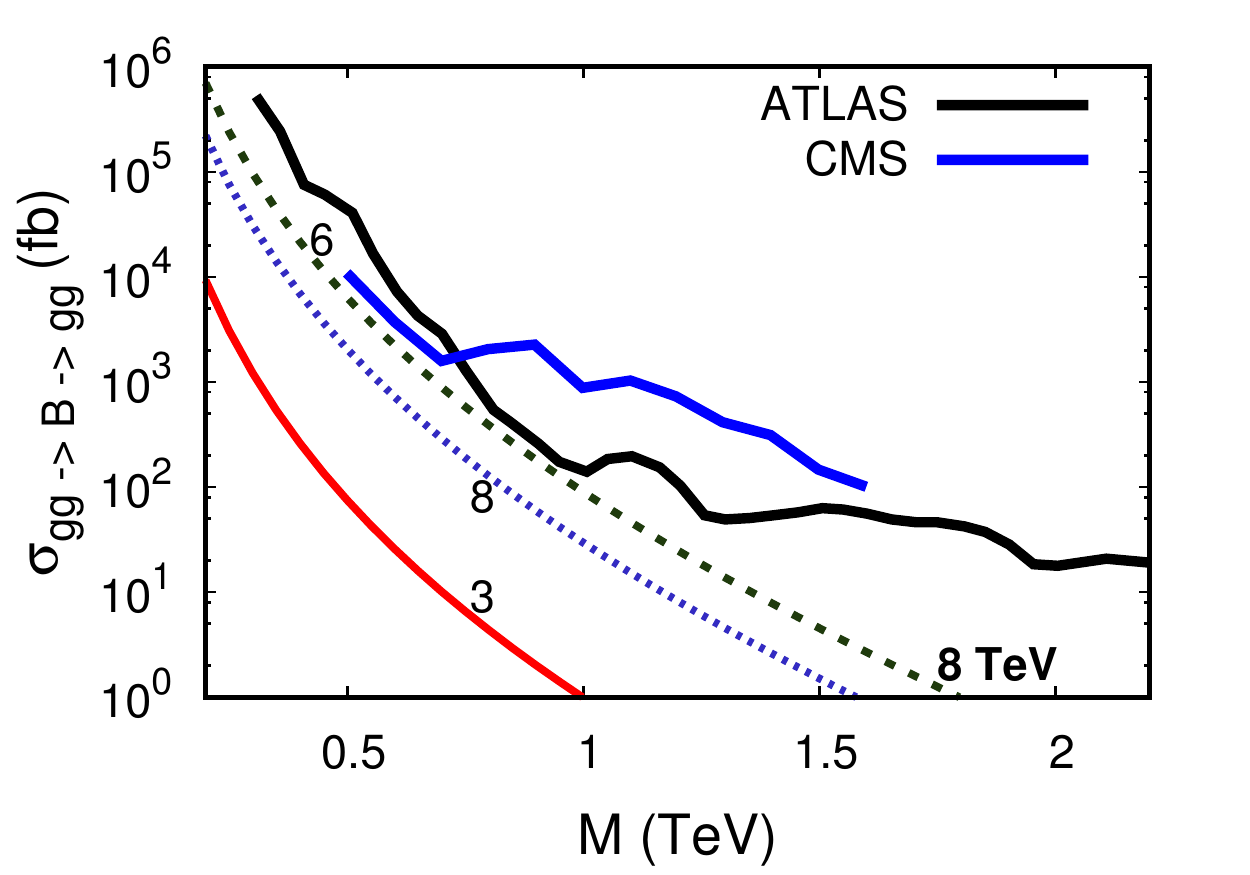}}
\subfigure[]{
\includegraphics[height=5.5cm, 
width=7cm]{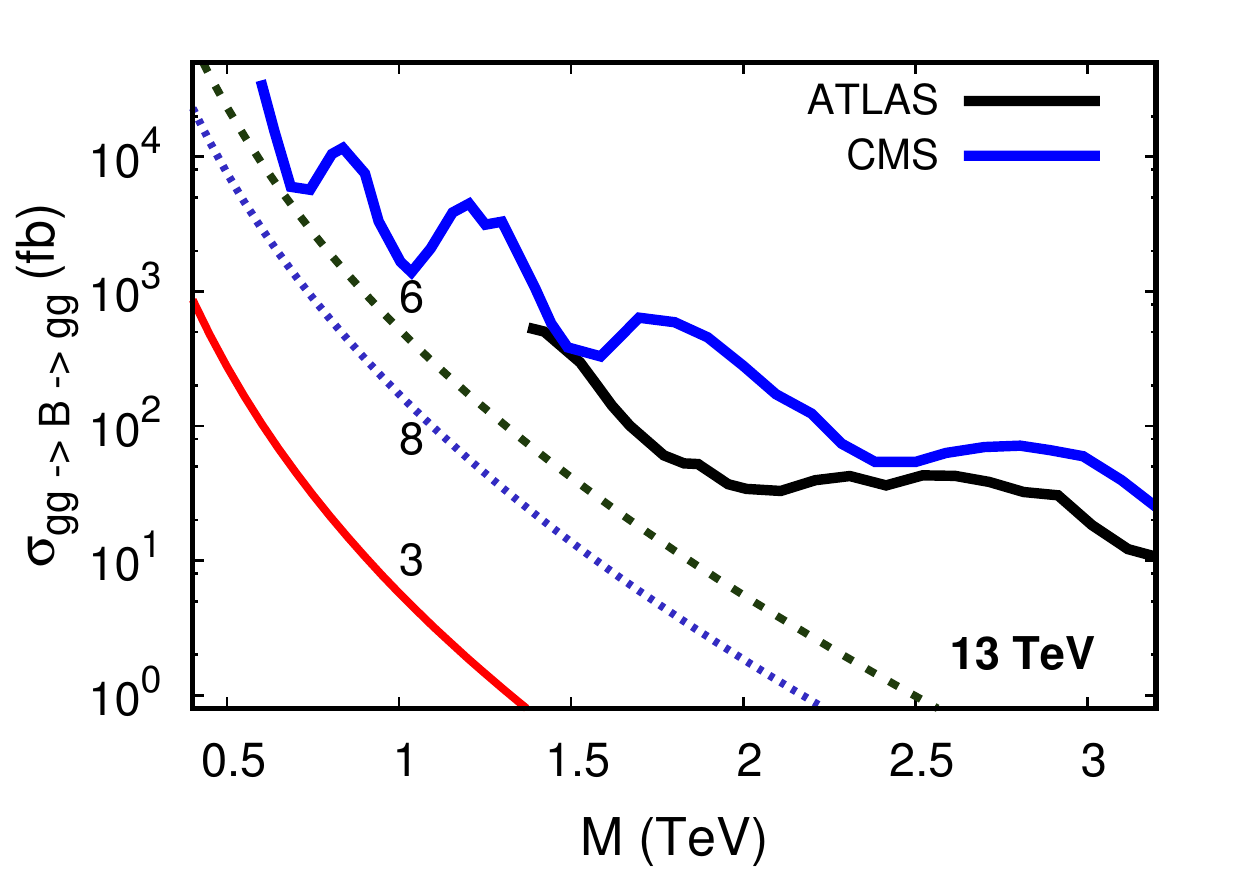}}
\caption{{\sf Cross section of Dijet events at $\sqrt s$ = 8 TeV (left) and 
$\sqrt s$ = 13 TeV (right) for  Bound State of Representation            
$\mathcal{R} = 1$ and  $J=0$, from constituent particle of 
Representation $R= 3, 6, 8$. Limits from ATLAS 8 TeV and 13 TeV are shown in thick 
black and CMS 8 TeV and 13 TeV are shown in thick blue.}}
\label{fig:dijet}
\end{figure}

\subsection{Limits on Signals from CMS and ATLAS}

In next section we examine the constraints on masses of bound state from dijet and diphoton bounds 
considering one copy of constituent vector-like fermions.
We have used the recent limits of ATLAS and CMS for diphoton resonance at centre of energy $\sqrt{s}$=13 TeV from
2015 and as well as 2016 data.
Dijet bounds has been considered for  centre of energy $\sqrt{s}$=8 and 13 TeV from both ATLAS and CMS.

As we have n number of copies of vector-like fermions described in the 
in Section~\ref{s:minmodel}  for two fermions representation, we will give the exclusion limits of 
vector-like fermion particle occurring in different models with n number of copies in  the \tabl{t:bound}.
\begin{table}
\renewcommand{\arraystretch}{1.2}
\begin{center}
{\small
\begin{tabular}{|c |c |c|c|}
\hline\hline
${\rm Model}$ & $ {\rm Representation} $&
$ {\rm Diphoton  (GeV)} $  & ${\rm Dijet  (GeV)}$  \\
\hline                    
{Model1} & Rep2 $\sim$ 1(6,1,1/3) &  $220$   & $-$ 
   \\ \cline{2-4}
\hline
{Model2} & Rep2 $\sim$ 2(8,1,0) &  $-$   & $-$  
   \\ \cline{2-4}
\hline
{Model3} & Rep2 $\sim$ 4(3,1,1/3) &  $150$   & $-$ 
   \\ \cline{2-4}
\hline
\multirow{2}{*}{Model4} & Rep1 $\sim$ 2(3,1,2/3) &  $300$   & $-$  
   \\ \cline{2-4}
                        & Rep2  $\sim$ 2(3,2,1/6)&  $300$   & $-$                  \\ \hline
{Model5} & Rep2 $\sim$ 1(6,1,2/3) &  $390$   & $-$  
   \\ \cline{2-4}
\hline
{Model6} & Rep2 $\sim$ 2(6,1,2/3) &  $450$   & $-$ 
   \\ \cline{2-4}
\hline
\multirow{2}{*}{Model7} & Rep1 $\sim$ 1(3,1,1/3) &  $-$   & $-$ 
   \\ \cline{2-4}
                        & Rep2 $\sim$ 1(3,2,1/6) &  $220$   & $-$                          
    \\ \hline
{Model8} & Rep2 $\sim$ 1(8,1,0) &  $-$   & $-$ 
\\ \cline{2-4}
\hline 
{Model9} & Rep2 $\sim$ 6(3,1,1/3) &  $200$   & $-$ 
\\ \cline{2-4}
\hline
\hline                           
\end{tabular}
}
\end{center}
\caption{{\sf Lower bounds on masses of vector-like fermions ($m_\psi=M/2$)
from dijet and diphoton events.}}\label{t:bound}
\end{table}
 \begin{figure}
 \centering
\includegraphics[width=0.55\linewidth,height=6.0cm]
      {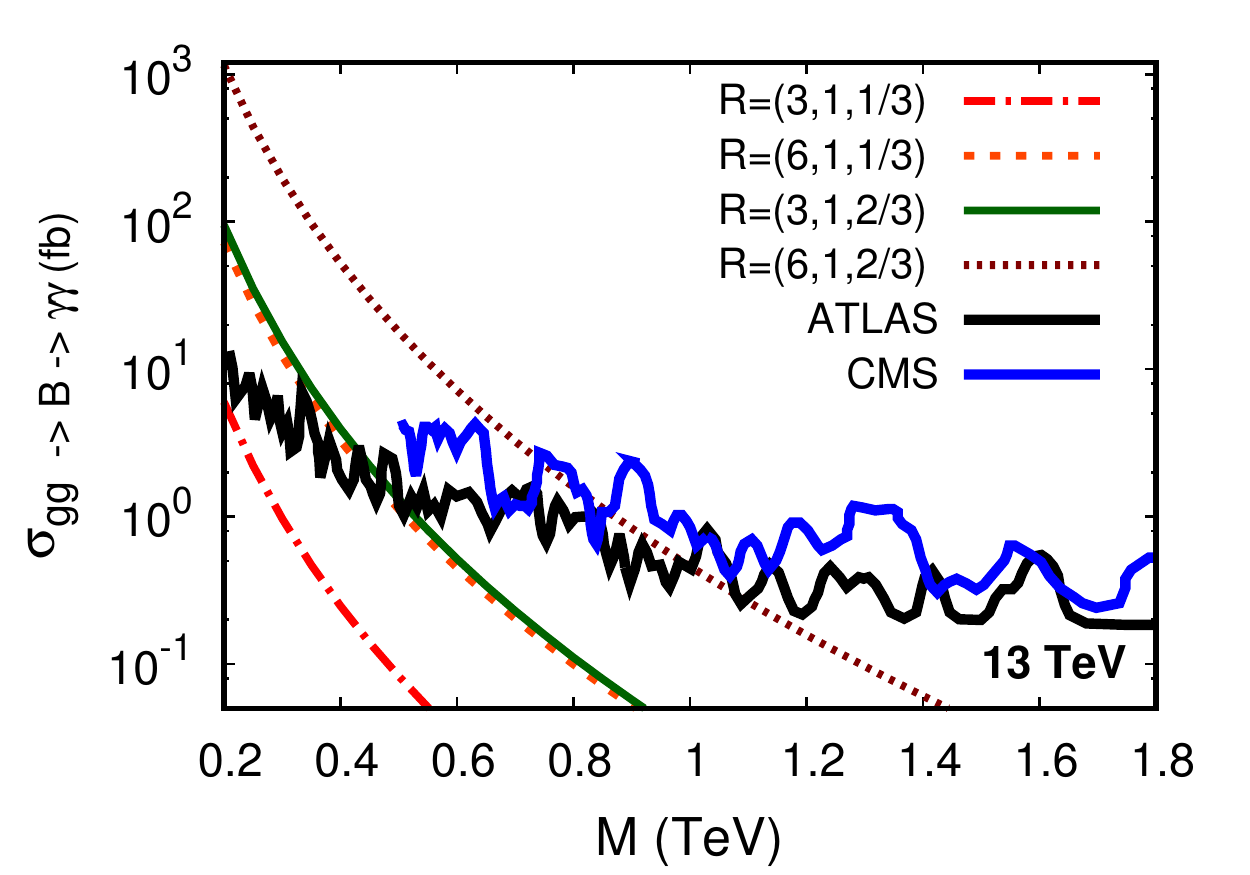}
\caption{{\sf Cross section of diphoton event w.r.t bound state mass at $\sqrt s$ = 13 TeV for Bound 
State of Representation $\mathcal{R} = 1$ and  $J=0$ from constituent particle 
of Color Representation $R= 3, 6$.The red line(dash dot) shows the fermion with $R=3$ and Q=1/3, 
green line(solid) correspond to $R=3$ and Q=2/3, purple line(dotted) shows the fermion with 
$R=6$ and Q=2/3 and orange line(dashed) shows the $R=6$ and 
Q=1/3 fermion. Limits are from ATLAS 13 TeV black line and CMS 13 TeV 
blue line.}}  
\label{fig:diphoton}
\end{figure}
\begin{figure}[h]
\centering
\begin{minipage}[b]{0.48\textwidth}
\includegraphics[height=5.5cm, width=7cm]{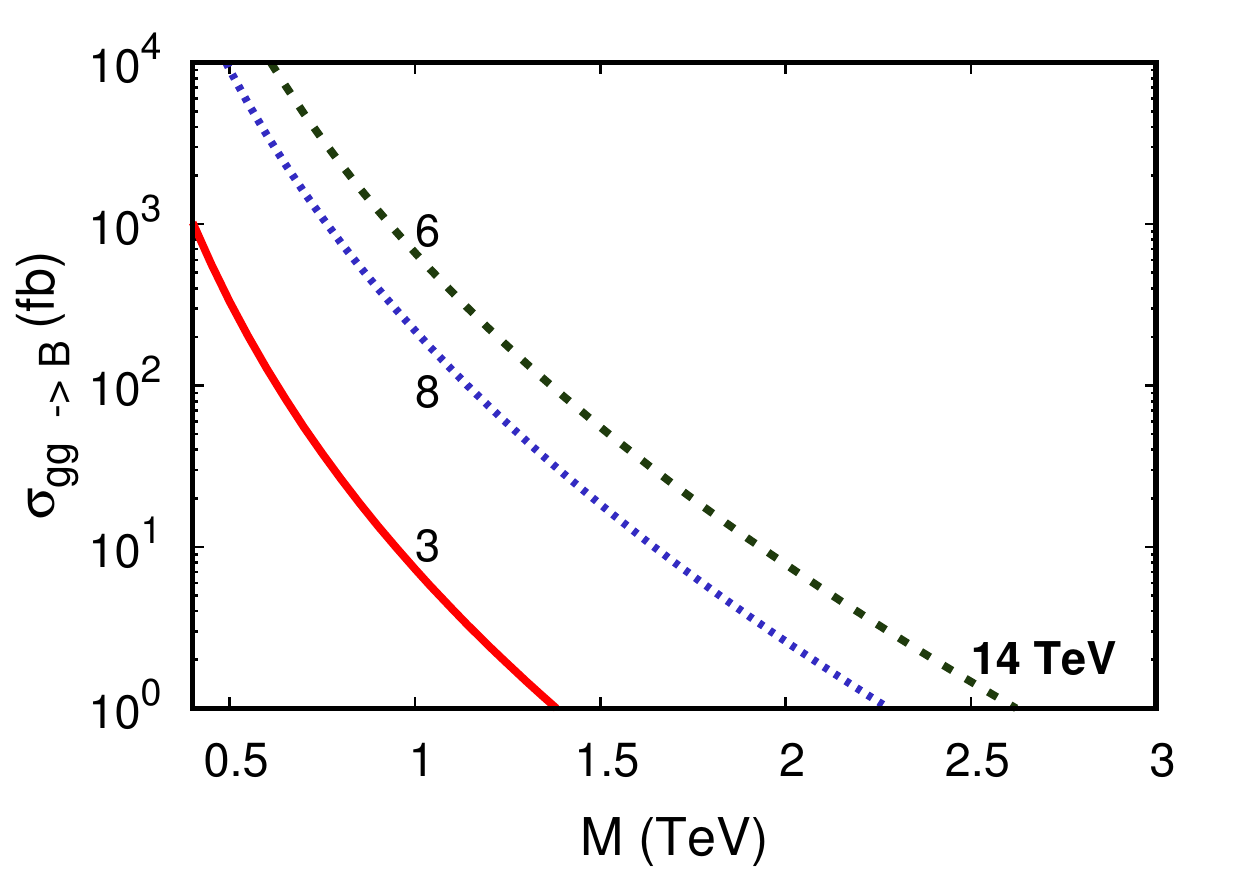}
\\
\caption{\sf Cross section of Bound State w.r.t bound state mass 
at $\sqrt s$ = 14 TeV for 
Representation $\mathcal{R} = 1$ and  $J=0$, from constituent 
particle of Representation $R= 3, 6, 8$.}\label{fig:14TeVcros}
\end{minipage}
\hfill
\begin{minipage}[b]{0.48\textwidth}
\includegraphics[height=5.5cm, width=7cm]{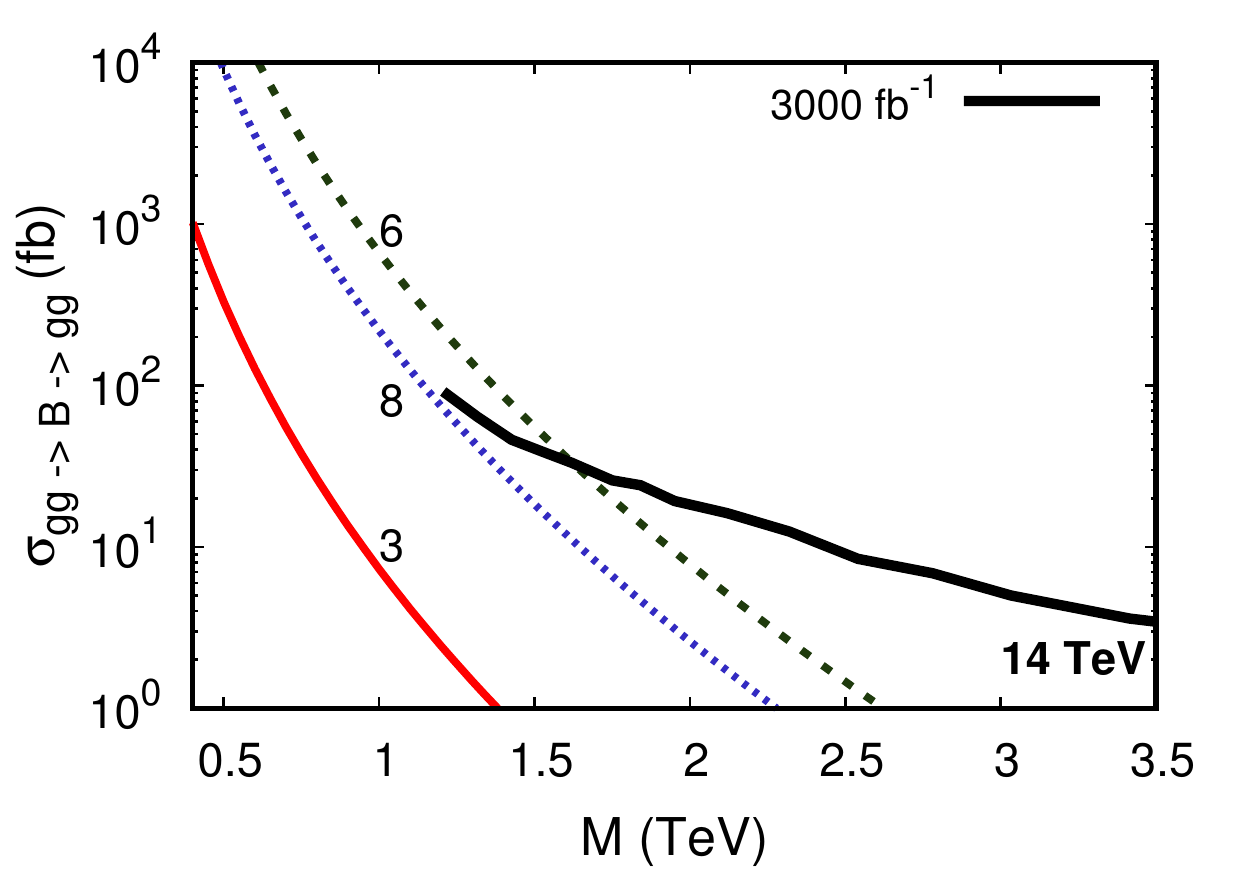}
\caption{{\sf Cross section of Dijet events at $\sqrt s$ = 14 TeV 
for  Bound State for Representation            
$\mathcal{R} = 1$ and  $J=0$, from constituent particle of 
Representation $R= 3, 6, 8$ w.r.t bound state mass. Future limits from 14 TeV at 3000 fb$^{-1}$ is shown in thick 
black line.}}\label{fig:14TeVdijet}
\end{minipage}
\end{figure}
\subsubsection{Dijet Bounds}
 In \fig{fig:dijet}(a)(b) we present the $\sigma(pp \rightarrow \mathcal{O}) \times BR(\mathcal{O} \rightarrow gg)$ as a 
function of the mass of the $\mathcal{O}$ resonance considering one copy of constituent vector-like fermions.
The black line is the upper limit on this cross-section from ATLAS~\cite{Aad:2014aqa} 8 TeV 
and blue line is from CMS~\cite{Khachatryan:2016ecr} 8 TeV data in  \fig{fig:dijet}(a). 
\fig{fig:dijet}(b) shows the  dijet limits from ATLAS(black)~\cite{ATLAS-CONF-2016-069} 13 TeV and CMS(blue)~\cite{Sirunyan:2016iap} 13 TeV data.
We can clearly say that the dijet limits are not strong enough to rule any of the models, if they have only one copy of constituent fermions.

In the \fig{fig:14TeVdijet}, we have plotted (black solid line) the projected limit for 14 TeV LHC at 3000 fb$^{-1}$ for the dijet cross
Section~\cite{Yu:2013wta} . 
Assuming $Z^{'}_{B}$ model, 14 TeV limits on mass of $Z^{'}_{B}$ and coupling between $Z^{'}_{B}$ gauge field 
with quark has been calculated in Ref.~\cite{Yu:2013wta}. Using this limit,
we have calculated 14 TeV projected limit on dijet cross-section. We have found that mass of vector-like fermion with colour 
representation six can be excluded up to 800-900 GeV at the HL-LHC.
\subsubsection{Diphoton Bounds}
The diphoton channel has played a very important role in discovering the Higgs Boson. It can 
be a very important channel to look at BSM physics. 
We present the production of diphoton channel as a function of the resonance 
mass considering one copy of constituent vector-like fermions in \fig{fig:diphoton}. Black line is the upper limit on this 
cross-section from ATLAS~\cite{ATLAS-CONF-2016-059} 13 TeV and blue line is from CMS~\cite{Khachatryan:2016yec} 13 TeV data. 
It can be observed that the upper limits on cross-section can give stringent bound on the masses of vector-like fermions ($m_\psi=M/2$).

There has been searches in $Z\gamma$, $ZZ$ and $WW$ resonances from these bound states.
ATLAS~\cite{Aad:2015ipg} has performed a combination of individual searches in all-leptonic, and all hadronic 
final states to search for heavy bosons decaying to $ZZ$ and $WW$ with integrated luminosity
of 20.3 fb$^{-1}$ at 8 TeV. The sensitivity is weaker than $\gamma \gamma$ channel for ATLAS~\cite{Aad:2015mna} at 8 TeV by around 
1000.
Both CMS~\cite{CMS-PAS-EXO-16-025} and ATLAS~\cite{Aad:2014fha} have performed a resonance decaying to $Z\gamma$ at centre-of-mass energy of 8TeV
at integrated luminosity 20.3 and 19.7 fb$^{-1}$ respectively. Where sensitivity is weaker than diphoton channel is weaker by order 10.

CMS~\cite{Khachatryan:2016odk} has performed a searches in $Z\gamma$ resonance in leptonic channel final decay state 
at centre of mass energies of 8 and 13 TeV. The bounds are weaker than diphoton bounds by 
factor of 200.
ATLAS~\cite{ATLAS-CONF-2016-082} has searched for heavy resonance decaying to $ZZ$ and $ZW$ pair decaying to leptonic and hadronic channels 
at a centre of mass 
energy 13 TeV with total integrated luminosity 13.2 fb$^{-1}$. The sensitivity is still weaker by factor 1000 with respect to diphoton channel. 

\section{Summary and Outlook} 

Unification of gauge couplings is one of the most important signatures of a successful Grand Unified Theory beyond the
electroweak scale.  We look for models with extra vector-like fermions at the weak scale which can lead to successful 
unification of gauge couplings. With two representation, we find a class of nine
models leading to successful unification of gauge couplings. An interesting aspect of these 
is that all of them contain coloured vector-like fermion in the spectrum.
The coloured set of the vector-like fermions can be probed at LHC by looking for
 bound states formed by them and their probable decays. We have already listed the present bounds from LHC for each
 successful model.  The future runs of LHC are sensitive to further mass ranges of these particles. 
Finally, it would be interesting to look for complete GUT models with this particle spectrum. 
\vskip 0.4 true cm 
\textbf{ \Large  Acknowledgments}
\\ \\
The authors thank Tao Han, Rohini Godbole and Xerxes Tata for collaboration in the initial stages 
of the project. We thank Xerxes Tata for comments on the manuscript and a reference.
SKV thanks visits to University of Pittsburgh and University of Hawaii where this project was started.
SKV acknowledges support from  IUSTFF Grant:JC-Physics Beyond Standard Model/23-2010 during 
the visit PITT PACC, Department of physics and Astronomy, 
University of Pittsburgh, USA and Department of Physics and Astronomy, University of Hawaii, Honolulu, USA. 
The work of B.B. is supported by the Department of Science and Technology, Government
of India, under the Grant Agreement number IFA13-PH-75 (INSPIRE Faculty
Award). P.B. thanks Palash B. Pal for giving a clue to calculate 6 dimensional representations of $SU(3)$ generators.
\appendix

\section{Two Representation Case}\label{a:fermdelta}
Here we enlist the models which satisfy gauge coupling unification and positivity of higgs potenatial
for Two fermion representation model, with $\Delta = 3\%$.  
\begin{table}[H]
\centering
$
{\small
\begin{array}{|c|c|c||c|c||c|c|}
\hline\hline
{\rm Mod}&{\rm Rep}\, 1 & M_{Rep1}& {\rm Rep}\, 2 & M_{Rep2}  &  M_{\rm GUT}  & \alpha_{\rm GUT} \\
 {\rm No.}&              & {\rm GeV} &            & {\rm GeV} & \times10^{16} {\rm GeV} &        \\ \hline

 1& 1 \left( 1, 1, 1 \right) &\left( 500-5000\right) & 1 \left( 3, 2, 
\frac{1}{6} \right) &\left(500-5000\right)& \sim 0.15 
& \sim 0.027
\\ \hline
 2& 5 \left( 1, 2, \frac{1}{2} \right) &\left( 250-500\right) & 1 \left( 6, 1, 
\frac{1}{3} \right) &\left(2500-5000\right)& \sim 0.12 
& \sim 0.035
\\ \hline

3& 3 \left( 1, 2, \frac{1}{2} \right) &\left( 250-700\right) & 1 \left( 8, 1, 
0 \right) &\left(1500-5000\right)& \sim 0.14 
& \sim 0.029
\\ \hline

4& 1 \left( 1, 3, 0 \right) &\left( 500-5000\right) & 1 \left( 3, 1, 
\frac{1}{3} \right) &\left(500-5000\right)& \sim 0.11 
& \sim 0.025
\\ \hline

5& 1 \left( 1, 3, 0 \right) &\left( 250-2200\right) & 2 \left( 3, 1, 
\frac{1}{3} \right) &\left(500-5000\right)& \sim 0.13 
& \sim 0.026
\\ \hline

6& 2 \left( 1, 3, 0 \right) &\left( 1300-5000\right) & 3 \left( 3, 1, 
\frac{1}{3} \right) &\left(250-3000\right)& \sim 0.67 
& \sim 0.03
\\ \hline

7& 3 \left( 1, 3, 0 \right) &\left( 3000-5000\right) & 1 \left( 6, 2, 
\frac{5}{6} \right) &\left(250-500\right)& \sim 0.11 
& \sim 0.32
\\ \hline

8& 1 \left( 3, 1, \frac{2}{3} \right) &\left( 250-5000\right) & 1 \left( 3, 2, 
\frac{1}{6} \right) &\left(250-1100\right)& \sim 0.15 
& \sim 0.03
\\ \hline
\hline

\end{array}
}
$
\caption{Model with two vector-like fermions representation satisfying gauge coupling 
unification and vacuum stability condition, with $\Delta = 3\%$.}\label{t:fermdelta}
\end{table}

\section{Three Representation Case}\label{a:three-ferm}

Here we enlist the models which satisfy gauge coupling unification and positivity of higgs potenatial
for three fermion representation model. Unlike Two Representation case, we made a restricted choice that all 
the representations and their copies are degenerate in mass of about 1 TeV, with up to ten copies 
in each representation. 
All of the models have unification scale less than $10^{16}$ GeV, which does not satisfy
with Proton decay constraint. The models are listed below in \tabl{t:three-ferm}
\begin{table}[H]
\centering
$\small
\begin{array}{|c|c|c|c|c|c|}
\hline\hline
{\rm Model No.} & {\rm Rep}\, 1 & {\rm Rep}\, 2 & {\rm Rep}\, 3 & M_{\rm GUT} & \alpha_{\rm GUT} 
\\ 
                 &               &               &               &    \times10^{16} GeV &              
\\                 \hline
1 & 1 \left( 1 , 1 , 1 \right) & 7 \left(  1 , 2 , \frac{1}{2} \right) &  2 \left( 8 , 1 , 0 \right) &0.132 & 0.043 \\
2 & 7 \left( 1 , 1 , 1 \right) & 5 \left(  1 , 3 , 0\right) &  3 \left(8 , 1 , 0 \right) &0.414 & 0.082 \\
3 & 4\left( 1 , 3 , 0 \right) & 1 \left(  3 , 1 , \frac{4}{3}\right) &  2 \left( 8 , 1 , 0 \right) & 0.133 & 0.051 \\
4 & 8 \left( 1 , 2 , \frac{1}{2} \right) & 1 \left(  1 , 3 , 0 \right) & 9 \left( 3 , 1 , \frac{1}{3}\right) & 0.209 & 0.077 \\
5 & 8 \left(1 , 2 , \frac{1}{2} \right) & 4 \left(  3 , 1 , \frac{1}{3} \right) &  1 \left( 8 , 1 , 0 \right) & 0.144 & 0.050 \\
 \hline
\end{array}
$
\caption{Models satisfying three fermion representation of
gauge coupling unification and stable higgs potenatial with degenerate mass of 1TeV. The representation is 
described as $n_i ( R_{SU(3)}, R_ {SU(2)}, R_{U(1)})$, where $n_i$ introduced earlier is the number
of copies of the representation, $R_{G}$ is the representation of the field under the gauge group
G of the SM.}
\label{t:three-ferm}
\end{table}

\section{Four Representation Case}\label{a:four-ferm}

Here we enlist the models which satisfy gauge coupling unification and stable higgs potenatial upto grand 
unified scale for four fermion representation model. Here also we restricted 
representations and their copies are degenerate in mass of about 1 TeV. We have allowed for up to 
ten copies in each model. 
Except one model, all of the models have unification scale less than $10^{16}$ GeV, which does not satisfy
with Proton decay constraint. The models are listed below in \tabl{t:four-ferm}

\begin{table}[H]
\centering
$\small
\begin{array}{|c|c|c|c|c|c|c|}
\hline\hline
{\rm Model No.} & {\rm Rep}\, 1 & {\rm Rep}\, 2 & {\rm Rep}\, 3 & {\rm Rep}\, 4 & M_{\rm GUT} & 
\alpha_{\rm GUT} 
\\ 
                 &               &               &              &   		& \times10^{16} GeV &              
\\                 \hline
 1 & 1 \left(  1 , 1 , 1 \right) &1 \left(  1 , 2 , \frac{3}{2} \right) &1 \left(  1 , 4 , \frac{1}{2} \right) &2 \left(  6 , 1 , \frac{1}{3} \right) &0.837 & 0.14 \\
 2 & 1 \left(  1 , 1 , 1 \right) &4 \left(  1 , 2 , \frac{1}{2} \right) &1 \left(  1 , 3 , 0 \right) &1 \left(  6 , 1 , \frac{1}{3} \right) &0.112 & 0.038 \\
 3 & 1 \left(  1 , 1 , 1 \right) &6 \left(  1 , 3 , 0 \right) &7 \left(  3 , 1 , \frac{1}{3} \right) &4 \left(  3 , 1 , \frac{2}{3} \right) &0.637 & 0.26 \\
 4 & 1 \left(  1 , 1 , 1 \right) &7 \left(  1 , 2 , \frac{1}{2} \right) &2 \left(  1 , 3 , 0 \right) &10 \left(  3 , 1 , \frac{1}{3} \right) &0.317 & 0.11 \\
 5 & 1 \left(  1 , 1 , 1 \right) &8 \left(  1 , 2 , \frac{1}{2} \right) &8 \left(  3 , 1 , \frac{1}{3} \right) &1 \left(  3 , 2 , \frac{1}{6} \right) &0.343 & 0.11 \\
 6 & 2 \left(  1 , 1 , 1 \right) &3 \left(  1 , 3 , 0 \right) &1 \left(  3 , 2 , \frac{5}{6} \right) &2 \left(  8 , 1 , 0 \right) &0.193 & 0.063 \\
 7 & 2 \left(  1 , 1 , 1 \right) &4 \left(  1 , 3 , 0 \right) &2 \left(  3 , 1 , \frac{1}{3} \right) &1 \left(  6 , 1 , \frac{2}{3} \right) &0.123 & 0.051 \\
 8 & 2 \left(  1 , 1 , 1 \right) &4 \left(  1 , 3 , 0 \right) &2 \left(  3 , 1 , \frac{2}{3} \right) &1 \left(  6 , 1 , \frac{1}{3} \right) &0.154 & 0.051 \\
 9 & 2 \left(  1 , 1 , 1 \right) &5 \left(  1 , 2 , \frac{1}{2} \right) &1 \left(  3 , 2 , \frac{1}{6} \right) &1 \left(  6 , 1 , \frac{1}{3} \right) &0.167 & 0.051 \\
 10 & 2 \left(  1 , 1 , 1 \right) &5 \left(  1 , 2 , \frac{1}{2} \right) &1 \left(  1 , 3 , 0 \right) &2 \left(  8 , 1 , 0 \right) &0.137 & 0.044 \\
 11 & 2 \left(  1 , 1 , 1 \right) &5 \left(  1 , 2 , \frac{1}{2} \right) &3 \left(  1 , 3 , 0 \right) &10 \left(  3 , 1 , \frac{1}{3} \right) &0.352 & 0.11 \\
 12 & 2 \left(  1 , 1 , 1 \right) &6 \left(  1 , 3 , 0 \right) &8 \left(  3 , 1 , \frac{1}{3} \right) &3 \left(  3 , 1 , \frac{2}{3} \right) &0.763 & 0.28 \\
 13 & 3 \left(  1 , 1 , 1 \right) &5 \left(  1 , 3 , 0 \right) &3 \left(  3 , 1 , \frac{2}{3} \right) &2 \left(  8 , 1 , 0 \right) &0.274 & 0.080 \\
 14 & 3 \left(  1 , 1 , 1 \right) &6 \left(  1 , 2 , \frac{1}{2} \right) &1 \left(  3 , 2 , \frac{1}{6} \right) &2 \left(  8 , 1 , 0 \right) &0.236 & 0.062 \\
 15 & 4 \left(  1 , 1 , 1 \right) &2 \left(  1 , 3 , 0 \right) &2 \left(  3 , 2 , \frac{1}{6} \right) &1 \left(  6 , 1 , \frac{2}{3} \right) &0.269 & 0.082 \\
 16 & 4 \left(  1 , 1 , 1 \right) &4 \left(  1 , 2 , \frac{1}{2} \right) &2 \left(  3 , 2 , \frac{1}{6} \right) &1 \left(  6 , 1 , \frac{1}{3} \right) &0.358 & 0.082 \\
 17 & 5 \left(  1 , 1 , 1 \right) &1 \left(  1 , 2 , \frac{1}{2} \right) &1 \left(  1 , 4 , \frac{1}{2} \right) &2 \left(  6 , 1 , \frac{1}{3} \right) &1.09 & 0.15 \\
 18 & 5 \left(  1 , 1 , 1 \right) &5 \left(  1 , 2 , \frac{1}{2} \right) &2 \left(  3 , 2 , \frac{1}{6} \right) &2 \left(  8 , 1 , 0 \right) &0.721 & 0.13 \\
 19 & 5 \left(  1 , 1 , 1 \right) &5 \left(  1 , 3 , 0 \right) &4 \left(  3 , 1 , \frac{1}{3} \right) &1 \left(  6 , 1 , \frac{1}{3} \right) &0.300 & 0.081 \\
 20 & 1 \left(  1 , 1 , 1 \right) &1 \left(  1 , 3 , 1 \right) &2 \left(  3 , 2 , \frac{1}{6} \right) &1 \left(  6 , 1 , \frac{2}{3} \right) &0.207 & 0.081 \\
 21 & 1 \left(  1 , 1 , 2 \right) &4 \left(  1 , 2 , \frac{1}{2} \right) &2 \left(  3 , 2 , \frac{1}{6} \right) &1 \left(  6 , 1 , \frac{1}{3} \right) &0.276 & 0.081 \\
 22 & 1 \left(  1 , 1 , 2 \right) &4 \left(  1 , 3 , 0 \right) &2 \left(  3 , 1 , \frac{1}{3} \right) &1 \left(  6 , 1 , \frac{1}{3} \right) &0.157 & 0.051 \\
 23 & 1 \left(  1 , 1 , 2 \right) &6 \left(  1 , 3 , 0 \right) &10 \left(  3 , 1 , \frac{1}{3} \right) &1 \left(  3 , 1 , \frac{2}{3} \right) &0.748 & 0.27 \\
 24 & 1 \left(  1 , 2 , \frac{1}{2} \right) &2 \left(  1 , 3 , 0 \right) &1 \left(  3 , 2 , \frac{5}{6} \right) &1 \left(  6 , 1 , \frac{1}{3} \right) &0.130 & 0.051 \\
 25 & 3 \left(  1 , 2 , \frac{1}{2} \right) &1 \left(  1 , 2 , \frac{3}{2} \right) &2 \left(  3 , 2 , \frac{1}{6} \right) &1 \left(  6 , 1 , \frac{1}{3} \right) &0.266 & 0.081 \\
 26 & 3 \left(  1 , 2 , \frac{1}{2} \right) &4 \left(  1 , 3 , 0 \right) &7 \left(  3 , 1 , \frac{1}{3} \right) &3 \left(  3 , 1 , \frac{2}{3} \right) &0.280 & 0.11 \\
  27 & 4 \left(  1 , 2 , \frac{1}{2} \right) &1 \left(  1 , 3 , 1 \right) &1 \left(  3 , 1 , \frac{1}{3} \right) &2 \left(  8 , 1 , 0 \right) &0.142 & 0.051 \\
  28 & 5 \left(  1 , 2 , \frac{1}{2} \right) &1 \left(  1 , 3 , 0 \right) &1 \left(  3 , 1 , \frac{2}{3} \right) &1 \left(  6 , 1 , \frac{1}{3} \right) &0.112 & 0.043 \\
  29 & 5 \left(  1 , 2 , \frac{1}{2} \right) &1 \left(  1 , 3 , 1 \right) &9 \left(  3 , 1 , \frac{1}{3} \right) &1 \left(  3 , 2 , \frac{1}{6} \right) &0.836 & 0.28 \\
  30 & 5 \left(  1 , 2 , \frac{1}{2} \right) &3 \left(  1 , 3 , 0 \right) &8 \left(  3 , 1 , \frac{1}{3} \right) &2 \left(  3 , 1 , \frac{2}{3} \right) &0.269 & 0.11 \\
  31 & 6 \left(  1 , 2 , \frac{1}{2} \right) &2 \left(  1 , 3 , 0 \right) &8 \left(  3 , 1 , \frac{1}{3} \right) &1 \left(  3 , 1 , \frac{2}{3} \right) &0.200 & 0.077 \\
  32 & 6 \left(  1 , 2 , \frac{1}{2} \right) &4 \left(  3 , 1 , \frac{1}{3} \right) &3 \left(  3 , 1 , \frac{2}{3} \right) &2 \left(  3 , 2 , \frac{1}{6} \right) &0.922 & 0.30 \\
   33 & 8 \left(  1 , 2 , \frac{1}{2} \right) &7 \left(  3 , 1 , \frac{1}{3} \right) &1 \left(  3 , 1 , \frac{2}{3} \right) &1 \left(  3 , 2 , \frac{1}{6} \right) &0.319 & 0.11 \\
   34 & 1 \left(  1 , 2 , \frac{3}{2} \right) &1 \left(  1 , 3 , 1 \right) &2 \left(  3 , 2 , \frac{1}{6} \right) &2 \left(  8 , 1 , 0 \right) &0.570 & 0.13 \\
   35 & 1 \left(  1 , 3 , 0 \right) &2 \left(  1 , 3 , 1 \right) &3 \left(  3 , 1 , \frac{1}{3} \right) &2 \left(  8 , 1 , 0 \right) &0.239 & 0.080 \\
   36 & 1 \left(  1 , 2 , \frac{3}{2} \right) &4 \left(  1 , 3 , 0 \right) &3 \left(  3 , 1 , \frac{1}{3} \right) &1 \left(  6 , 1 , \frac{1}{3} \right) &0.185 & 0.062 \\
   37 & 3 \left(  1 , 3 , 0 \right) &1 \left(  3 , 1 , \frac{2}{3} \right) &1 \left(  3 , 2 , \frac{5}{6} \right) &1 \left(  6 , 1 , \frac{1}{3} \right) &0.156 & 0.062 \\
   38 & 3 \left(  1 , 3 , 0 \right) &1 \left(  3 , 1 , \frac{4}{3} \right) &1 \left(  3 , 2 , \frac{1}{6} \right) &1 \left(  6 , 1 , \frac{1}{3} \right) &0.157 & 0.062 \\
   39 & 4 \left(  1 , 3 , 0 \right) &1 \left(  1 , 3 , 1 \right) &9 \left(  3 , 1 , \frac{1}{3} \right) &2 \left(  3 , 1 , \frac{2}{3} \right) &0.681 & 0.27 \\
   40 & 5 \left(  1 , 3 , 0 \right) &1 \left(  3 , 1 , \frac{1}{3} \right) &5 \left(  3 , 1 , \frac{2}{3} \right) &1 \left(  8 , 1 , 0 \right) &0.223 & 0.079 \\
   41 & 5 \left(  1 , 3 , 0 \right) &5 \left(  3 , 1 , \frac{1}{3} \right) &1 \left(  3 , 1 , \frac{4}{3} \right) &1 \left(  8 , 1 , 0 \right) &0.188 & 0.078 \\
\hline
 \end{array}
 $
\caption{Models satisfying four fermion representation of
gauge coupling unification and stable higgs potenatial with degenerate mass of 1 TeV.  The representation is 
described as $n_i ( R_{SU(3)}, R_ {SU(2)}, R_{U(1)})$, where $n_i$ introduced earlier is the number
of copies of the representation, $R_{G}$ is the representation of the field under the gauge group
G of the SM.}
\label{t:four-ferm}
\end{table}
\section{Representations and Dynkin indices}\label{a:RepDyn}
We considered all the $SU(3)\times SU(2)\times U(1)$ representations coming 
from $SU(5)$ representations upto dimension 75. In \tabl{t:rep-list},
we listed those forty representations \cite{Slansky:1981yr} with their contribution to beta function 
(i.e. Dynkin index) considering them as scalar fields. One can 
straight-forwardly 
derive corresponding Dynkin indices if the fileds are vector-like fermion just by 
multiplying the tabulated value with 2 if the representation is real and by 
multiplying with 4 if the considered representation is complex.
\begin{table}[H]
	{\footnotesize
		\begin{tabular}{|c|c|c|c|c|c|c|c|}
			\hline\hline
			\mbox{S.No.} & \mbox{SM Rep} & \mbox{Source} &
			\mbox{Dynkin Indices} & {\rm S.No.} & 
			\mbox{SM Rep} & \mbox{Source} & \mbox{Dynkin Indices}\\
			\hline\hline
			1 & $\left(1, 1, 1 \right)$ & 10 & $\left(0, 0, -\frac{1}{5} \right)$& 21 & 
			$\left(3, 2, \frac{7}{6} \right)$& $\overline{45}$, $\overline{50}$ & 
			$\left(-\frac{1}{3}, -\frac{1}{2}, 
			-\frac{49}{30} \right)$\\ \hline
			2 & $\left(1, 1, 2 \right)$ & $\overline{50}$ & $\left(0, 0, -\frac{4}{5} 
			\right)$& 22 & 
			$\left(3, 3, -\frac{1}{3} \right)$& 45, 70 & $\left(-\frac{1}{2}, -2, 
			-\frac{1}{5} 
			\right)$\\ \hline
			3 & $\left(1, 1, 3 \right)$ &  & $\left(0, 0, -\frac{9}{5} \right)$& 23 & 
			$\left(3, 3, \frac{2}{3} \right)$& $\overline{35}$, $\overline{40}$ & 
			$\left(-\frac{1}{2}, -2, -\frac{4}{5} 
			\right)$\\ \hline
			4 & $\left(1, 1, 4 \right)$ &  & $\left(0, 0, -\frac{16}{5} \right)$& 24 & 
			$\left(\bar 3, 3, \frac{4}{3} \right)$& 70 & $\left(-\frac{1}{2}, -2, 
			-\frac{16}{5} 
			\right)$\\ \hline
			5 & $\left(1, 1, 5 \right)$ &  & $\left(0, 0, -5 \right)$& 25 & $\left(3, 
			4, \frac{7}{6} \right)$& $\overline{70'}$ & $\left(-\frac{2}{3}, -5, 
			-\frac{49}{15} \right)$
			\\ \hline
			6 & $\left(1, 2, \frac{1}{2} \right)$ & 5, 45, 70 & $\left(0, -\frac{1}{6}, 
			-\frac{1}{10} \right)$& 26 & $\left(6, 1, \frac{1}{3} \right)$& $\overline{45}$ 
			& $\left(
			-\frac{5}{6}, 0, -\frac{2}{15} \right)$\\ \hline
			7 & $\left(1, 2, -\frac{3}{2} \right)$ & 40 & $\left(0, -\frac{1}{6}, 
			-\frac{9}{10} \right)$& 27 & $\left(6, 1, -\frac{2}{3} \right)$& 15 & $\left(
			-\frac{5}{6}, 0, -\frac{8}{15} \right)$\\ \hline
			8 & $\left(1, 3, 0 \right)$ & 24 & $\left(0, -\frac{2}{3}, 0 \right)$& 28 
			& $\left(6, 1, \frac{4}{3} \right)$& 50 & $\left(-\frac{5}{6}, 0, 
			-\frac{32}{15} \right)$\\ \hline
			9 & $\left(1, 3, 1 \right)$ & 15 & $\left(0, -\frac{2}{3}, -\frac{3}{5} 
			\right)$& 29 & $\left(\bar 6, 2, \frac{1}{6} \right)$& 35, 40 & 
			$\left(-\frac{5}{3}, -1 
			, -\frac{1}{15} \right)$\\ \hline
			10 & $\left(1, 4, \frac{1}{2} \right)$ & 70 & $\left(0, -\frac{5}{3}, 
			-\frac{1}{5} \right)$& 30 & $\left(6, 2, \frac{5}{6} \right)$& 75 & $\left(
			-\frac{5}{3}, -1, -\frac{5}{3} \right)$\\ \hline
			11 & $\left(1, 4, -\frac{3}{2} \right)$ & 35 & $\left(0, -\frac{5}{3}, 
			-\frac{9}{5} \right)$& 31 & $\left(6, 2, -\frac{7}{6} \right)$& 70 & $\left(
			-\frac{5}{3}, -1, -\frac{49}{15} \right)$\\ \hline
			12 & $\left(1, 5, -2 \right)$ & $70'$ & $\left(0, -\frac{10}{3}, -4 \right)$& 
			32 
			& $\left(6, 3, \frac{1}{3} \right)$& $\overline{50}, \overline{70'}$ & 
			$\left(-\frac{5}{2}, -4, 
			-\frac{2}{5} \right)$\\ \hline
			13 & $\left(1, 5, 1 \right)$ &  & $\left(0, -\frac{10}{3}, -1 \right)$& 33 
			& $\left(8, 1, 0 \right)$& 24 & $\left(-1, 0, 0 \right)$\\ \hline
			14 & $\left(1, 5, 0 \right)$ &  & $\left(0, -\frac{10}{3}, 0 \right)$& 34 
			& $\left(8, 1, 1 \right)$& 40 & $\left(-1, 0, -\frac{8}{5} \right)$\\ \hline
			15 & $\left(3, 1, -\frac{1}{3} \right)$ & 5, 45, 50, 70 & 
			$\left(-\frac{1}{6}, 0, -\frac{1}{15} \right)$& 35 & $\left(8, 2, \frac{1}{2} 
			\right)$& 45, 50, 70 & $\left(-2, 
			-\frac{4}{3}, -\frac{4}{5} \right)$\\ \hline
			16 & $\left(\bar 3, 1, -\frac{2}{3} \right)$ & 10, 40 & 
			$\left(-\frac{1}{6} 
			, 0, -\frac{4}{15} \right)$& 36 & $\left(8, 3, 0 \right)$& 75 & $\left(-3, 
			-\frac{16}{3}, 0 \right)$\\ \hline
			17 & $\left(\bar3, 1, \frac{4}{3} \right)$ & 45 & $\left(-\frac{1}{6}, 0, 
			-\frac{16}{15} \right)$& 37 & $\left(\overline{10}, 1, 1 \right)$& 35 & 
			$\left(-\frac{5}{2} 
			, 0, -2 \right)$\\ \hline
			18 & $\left(3, 1, \frac{5}{3} \right)$ & 75 & $\left(-\frac{1}{6}, 0, 
			-\frac{5}{3} \right)$& 38 & $\left(\overline{10}, 2, \frac{1}{2} \right)$& 
			$70'$ & $\left(-5, 
			-\frac{5}{3}, -1 \right)$\\ \hline
			19 & $\left(3, 2, \frac{1}{6} \right)$ & 10, 15, 40 & $\left(-\frac{1}{3}, 
			-\frac{1}{2}, -\frac{1}{30} \right)$& 39 & $\left(15, 1, -\frac{1}{3} \right)$
			& 70 & $\left(-\frac{10}{3}, 0, -\frac{1}{3} \right)$\\ \hline
			20 & $\left(3, 2, -\frac{5}{6} \right)$ & 24, 75 & $\left(-\frac{1}{3}, 
			-\frac{1}{2}, -\frac{5}{6} \right)$& 40 & $\left(15, 1, \frac{4}{3} \right)$& 
			$70'$ & $\left(-\frac{10}{3}, 0, -\frac{16}{3} \right)$\\ \hline
		\end{tabular}
	}
	\caption{{\sf Representation of fields considered in this paper. In the column 
			entitled with ``SM Rep'' we put incomplete multiplets of $SU(5)$ and the 
			entries inside the brackets are $SU(3), SU(2)$ and $U(1)$ representations 
			respectively. In the column with title we'd written the $SU(5)$ representations 
			from which those representations are coming. Dynkin indices are calculated 
			assuming the fields are scalar fields. Note that we had considered up the SU(5) 
			representation of dimension 75. There are some extra representations as well.}}
	\label{t:rep-list}
\end{table}
\section{Mixing between SM particle with vector-like fermion}\label{a:mixing-SM}

In this section we will assume that the new vector-like fermions interact with 
the SM fermions via Yukawa interactions. SM contains $l= (1,2,-1/2)$
$e_R=(1,1,-1)$, $q=(3,2,1/6)$and $d_R=(3,1,-1/3)$, $u_R=(3,1,2/3)$ and Higgs doublet,
$H=(1,2,1/2)$
. It can be easily be understood that,
among the vector-like fermions considered in this work, new vector-like
fermions coupling to the SM ones with renormalisable 
couplings can only appear in top and bottom partner gauge-covariant multiplets, 
and in lepton and neutrino partner with 
definite $SU(3)_C \times SU(2)_L \times U(1)_Y$ quantum numbers, which has 
been studied in \cite{Cacciapaglia:2015ixa, delAguila:2008pw,
Bizot:2015zaa,delAguila:2000rc, Cacciapaglia:2010vn, Gopalakrishna:2011ef, 
Aguilar-Saavedra:2013qpa, Fujikawa:1994we} and some of them tabulated in \tabl{t:mixing}.
\begin{table}
\centering
\renewcommand{\arraystretch}{1.3}
\begin{tabular}{|c|c|}
\hline
 Vector-Like Fermion &  Couples to\\
\hline\hline
$E  (1,1,-1)$ &  $l,e_R$\\
\hline
$L (1,2,-\frac12)$ & $l,e_R$\\
\hline
$\Lambda  (1,2,-\frac32)$ &  $e_R$\\
\hline
$\Delta (1,3,-1)$ & $l$\\
\hline
$\Sigma (1,3,0)$ & $l$\\
\hline
$T (3,1,+\frac23)$ &  $q,u_R$\\
\hline
$B  (3,1,-\frac13)$ &  $q,d_R$\\
\hline
$X_T (3,2,+\frac76)$ &  $u_R$\\
\hline
$Q  (3,2,+\frac16)$ &  $q,d_R,u_R$\\
\hline
$Y_B (3,2,-\frac56)$ &   $d_R$\\
\hline
$X_Q (3,3,+\frac23)$ &   $q$\\
\hline
$Y_Q (3,3,-\frac13)$ &  $q$\\
\hline
\end{tabular}
\caption{Vector-like fermions, that provide a consistent extension of the SM and modify the Higgs boson couplings 
\cite{Bizot:2015zaa}.} 
\label{t:mixing}
\end{table}
Here we will briefly overview the leading order constraints coming from EW precision tests,
direct searches at colliders and Higgs physics. It is reasonable to assume that, only 
third family of SM fermions have sizable contribution from new vector-like fermions.
\subsection{Vector like quarks}
Due to mixing of the SM top and bottom quark with vector-like fermions partners, the
resulting physical up and down type quark mass eigenstates $u^0,c^0,t^0,T^0$ and $d^0,s^0,b^0,B^0$ may
contain non-zero $T$ and $B$ components, leading to a deviation in their couplings to Z and W$\pm$ 
bosons.  
In this case, the relation between weak and mass eigenstates for up quark 
can be parameterized as two $2\times2$ matrices $V_{L,R}^U$,
 \begin{equation}
\left(\! \begin{array}{c} t^0_{L,R} \\ T^0_{L,R} \end{array} \!\right) =
\left(\! \begin{array}{cc} \cos \theta_{L,R}^u & -\sin \theta_{L,R}^u \\ \sin \theta_{L,R}^u & \cos \theta_{L,R}^u \end{array}
\!\right)
\left(\! \begin{array}{c} t_{L,R} \\ T_{L,R} \end{array} \!\right) \,.
\label{ec:mixu}
\end{equation}
Similar unitary matrices can be written for down sector.
The mixing angles in the left and right sectors are not independent, but have a relation 
(see also~\cite{Dawson:2012di,Fajfer:2013wca,Atre:2011ae})
\begin{eqnarray}
\tan \theta_R^q & = & \frac{m_q}{m_Q} \tan \theta_L^q \quad \text{(singlets, triplets)} \,, \notag \\
\tan \theta_L^q & = & \frac{m_q}{m_Q} \tan \theta_R^q \quad \text{(doublets)} \,, 
\label{ec:rel-angle1}
\end{eqnarray}
where $m_q$ and $m_Q$ are the mass of SM fermion and vector-like fermion respectively.\\ 
This mixing gives new contributions to the oblique parameters S and T \cite{Peskin:1990zt}, which is precisely 
measured at LEP and SLC. The contributions to S, T in models with T, B 
singlets and (T B) doublets are studied in \cite{Lavoura:1992np, Aguilar-Saavedra:2013qpa, Bizot:2015zaa, Angelescu:2015kga}, 
which would give a constraints in mixing parameters between SM and their 
vector-like fermions partners. For singlet B quark, the constraints from $R_b$ is strong, which gives 
upper bound on mixing $\sin \theta_{L}^d$ to be 0.04. For singlet T quark upper bound of $\sin \theta_{L}^u$
is 0.15 to 0.10 for mass range 600 GeV to 2 TeV respectively, from S and T parameter.
For (T B) doublet, the constraints from EW precision gives upper bound on $\sin \theta_{R}^d$ to be 0.06 and, 
$\sin \theta_{R}^u$ between 0.13 to 0.09 for mass range 600 GeV to 2 TeV respectively, 
considering the splitting between $M_B$ and $M_T$ of 2 GeV.\\
\paragraph{Direct Searches}
A full model of vector-like Quark decaying to SM particles and search strategies 
to discover at LHC has been studied in Ref.~\cite{AguilarSaavedra:2009es, Contino:2008hi, Aguilar-Saavedra:2013qpa} and Ref. within.
The singlet $T$ Quark decays as, 
\begin{equation}
T \to W^+ b \,,\quad \quad T \to Zt \,,\quad \quad T \to Ht \,.
\end{equation}
The singlet  $B$ quark decays are
\begin{equation}
B \to W^- t \,,\quad \quad B \to Zb \,,\quad \quad B \to Hb \,.
\end{equation}
 $T B$ doublet assuming that they couple to the third generation, are the same as for singlets,
\begin{align}
& T \to W^+ b \,,\quad \quad T \to Zt \,,\quad \quad T \to Ht \,, \notag \\
& T \to W^- t \,,\quad \quad B \to Zb \,,\quad \quad T \to Hb \,,
\end{align}
We would summaries the mass constraints coming from direct searches of VLQ at the LHC.


For Integrated luminosity of 19.5 fb$ ^{-1}$ at $\sqrt{s} = 8$ TeV
CMS \cite{Chatrchyan:2013uxa} experiment at the Large Hadron Collider 
searched for the $T$ quark decaying into three different final states, bW, tZ, and tH. 
The search is carried out using events with at least one isolated lepton. The 
lower limits are set on the T quark mass at 95$\%$ confidence level between 687 and 782 GeV 
for all possible values of the branching fractions into the three different final 
states assuming strong production.

A search in CMS \cite{Khachatryan:2015oba} is performed in five exclusive channels: a single-
lepton channel, a multilepton channel, two all-hadronic channels optimized either
for the bW or the tH decay, and one channel in which the Higgs boson decays into
two photons. A statistical combination of these results is
performed and lower limits on the T quark mass are set. Depending on the branch-
ing fractions, lower mass limits between 720 and 920 GeV at 95 $\%$ confidence level are
found.
A search similar to Top like vector quark,  heavy B quark  vec-
tor couplings to W, Z, and H bosons, is carried out by CMS experiment~\cite{Khachatryan:2015gza}. 
The B quark is assumed to be pair 
produced and to decay in one of three ways: to tW, bZ, or bH. The search is carried
out in final states with one, two, and more than two charged leptons, as well as 
in fully hadronic final states.Each of the channels in the exclusive final-state 
topologies is designed to be sensitive to specific combinations of the 
B quark-antiquark pair decays. A statistical combination of these results gives 
lower limits on the B quark mass between 740 GeV and 900 GeV with 95 $\%$ confidence level, 
depending on the values of the branching fractions of the B quark to tW, bZ, and bH.

ATLAS has also searched for exotic quark, heavy X quark with $Q=5/3$ decaying to tW gives a lower bound of
mass 840 GeV \cite{Aad:2015mba} with 95$\%$ C.L. . Quark Y with chagre $Q=-4/3$ decaying to Wb gives 
lower bound of mass 770 GeV \cite{Aad:2015kqa} with 95$\%$ C.L.. The experimental searches assume pair production via strong 
interactions and prominent decays in the indicated channels.

\subsection{Vector like leptons}
In this section we discuss new colourless fermions.
Weak iso-triplet with zero hyper-charge vector-like 
fermion can couple to left $l$ handed SM fermions and higgs as:
\begin{equation}
 {\cal L}_{\Sigma} = - \sqrt{2}Y_{\Sigma} \overline{\Sigma}{l}\tilde{H}  - \frac 12 {\rm Tr}\left(
 \overline{\Sigma} M_\Sigma \Sigma^c \right) + h.c. ~,
\label{sigma}
\end{equation}
where the matrix notation of $\Sigma$ is as follows
\begin{equation}
\Sigma_\equiv \sqrt2 \Sigma^a \tau^a =  \left(\begin{array}{cc} 
\frac{1}{\sqrt 2}\Sigma^0 & -\Sigma^+ \\
\Sigma^- & -\frac{1}{\sqrt 2}\Sigma^0 
\end{array}\right) ~
\end{equation}
The contribution of $\Sigma$ to the EW precision parameters is vanishingly small \cite{delAguila:2008pw}, since the mixing angle 
are suppressed by $\sim m_{\nu}/M_{\Sigma}$ and the loop induced mass splitting between the 
$M_{\Sigma^{\pm}}-M_{\Sigma^{0}}$ = $ 164 -165$ GeV \cite{Ibe:2012sx}.
In the limit $Y_\Sigma \ll M_\Sigma /v$ we can realize it as a type III seesaw model \cite{Foot:1988aq} with 
neutrino mass $m_\nu = Y_\Sigma^2v^2/M_\Sigma$.
\\
In the limit $Y_\Sigma \rightarrow 0$, this can be realized as a wino like dark matter \cite{Cirelli:2005uq}.
\\
SM fermions can also couple to four different possible vector-like leptons, a weak singlet E, a weak doublet L or
$\Lambda$, a weak triplet $\Delta$. The effect of these vector-like leptons on modification on the 
Higgs decays, anomalous magnetic moment to the muon and lepton flavour violation decays  
are studied in Refs. \cite{Falkowski:2013jya,Altmannshofer:2013zba,
Kumar:2015tna,Chen:2016lsr,Fujikawa:1994we, Ishiwata:2015cga}.   
\paragraph{Direct search}
The limits on $M$ strongly depend on the SM generation that couples to the
heavy leptons. The limits on doublet L, couplings only to the third generation is  $M_L$ $ > $ 270 GeV and coupling 
with $e$ and $\mu$ gives bound of  $M_L$ $ > $ 450 GeV, Ref.~\cite{Falkowski:2013jya},
while the LEP limit remains more constraining in the case of the singlet
E, $M_{E} > $ 100 GeV. For the exotic doublet $\Lambda$ with a doubly-charged component, Ref.~\cite{Altmannshofer:2013zba} 
reports $M_{\Lambda} > $ 320 GeV.

\section{ Earlier Scan of models By Tom Rizzo }\label{a:tom-rizzo}
In this section we update the work done in Ref. \cite{Rizzo:1991tc}. They studied the grand unified theories in context of
additional degree of freedom at electroweak scale. S (F) indicates that the quantum numbers following it refer to a 
complex scalar (vector like fermion) representation. $N_A$ ($N_B$) is the number of fields of type A (B)
in the scenario.

\begin{eqnarray}
{\small
	\begin{array}{|c|c|c|c|c|c|c|c|c|c|c|c|c|}
	\hline
	N_A & & SU(3) & SU(2) &U(1) & N_B & & SU(3) & SU(2) &U(1) & M_{\rm GUT} & 
	\alpha_{\rm GUT} & Status\\
	\hline
	1 & S & 8 & 1 & \frac{2}{3} & 1 & S & 3 & 3 & 1 & 5.15393\times 10^{14} & 
	0.0310221 & {\rm No} \\ \hline
	2 & S & 3 & 2 & \frac{1}{6} & 2 & S & 1 & 2 & \frac{1}{2} & 5.07162\times 
	10^{14} & 0.026024 & {\rm Yes} \\ \hline
	2 & S & 6 & 2 & \frac{1}{2} & 2 & S & 1 & 3 & \frac{2}{3} & 4.07143\times 
	10^{16} & 0.0353412 & {\rm Yes} \\ \hline
	2 & S & 6 & 2 & \frac{1}{6} & 2 & S & 1 & 3 & 0 & 8.54256\times 10^{20} & 
	0.0326849 & {\rm No} \\ \hline
	1 & F & 3 & 2 & \frac{1}{6} & 1 & F & 1 & 1 & 1 & 5.07162\times 10^{14} & 
	0.0283188 & {\rm Yes} \\ \hline
	1 & F & 3 & 2 & \frac{1}{2} & 1 & S & 8 & 2 & \frac{1}{6} & 1.29764\times 
	10^{17} & 0.034587 & {\rm Yes} \\ \hline
	1 & F & 3 & 2 & \frac{1}{6} & 1 & S & 1 & 1 & 2 & 5.07162\times 10^{14} & 
	0.0283188 & {\rm Yes} \\ \hline
	1 & F & 3 & 2 & \frac{1}{6} & 2 & S & 3 & 1 & 1 & 1.69262\times 10^{15} & 
	0.0292299 & {\rm Yes} \\ \hline
	1 & F & 3 & 1 & 0 & 2 & S & 1 & 3 & \frac{2}{3} & 5.02121\times 10^{14} & 
	0.0264338 & {\rm Yes} \\ \hline
	1 & F & 3 & 1 & \frac{1}{3} & 2 & S & 1 & 3 & 1 & 1.7518\times 10^{14} & 
	0.0275409 & {\rm Yes} \\ \hline
	1 & S & 8 & 1 & \frac{2}{3} & 1 & S & 3 & 1 & \frac{5}{3} & 1.91539\times 
	10^{22} & 0.0440168 & {\rm No} \\ \hline
	1 & S & 8 & 2 & \frac{1}{6} & 1 & S & 1 & 3 & 1 & 2.10093\times 10^{16} & 
	0.0285893 & {\rm Yes} \\ \hline
	1 & F & 3 & 2 & \frac{1}{6} & 1 & F & 3 & 1 & 0 & 4.07855\times 10^{16} & 
	0.0276959 & {\rm Yes} \\ \hline
	1 & F & 1 & 2 & \frac{1}{6} & 1 & F & 8 & 2 & \frac{1}{6} & 8.51879\times 
	10^{48} & -0.0774188 & {\rm No} \\ \hline
	1 & F & 3 & 2 & \frac{1}{6} & 2 & S & 3 & 1 & \frac{2}{3} & 1.90667\times 
	10^{15} & 0.0281254 & {\rm Yes} \\ \hline
	2 & F & 1 & 2 & \frac{1}{2} & 1 & S & 8 & 2 & \frac{1}{2} & 5.07162\times 
	10^{14} & 0.0310574 & {\rm Yes} \\ \hline
	2 & F & 1 & 2 & \frac{1}{6} & 1 & S & 6 & 1 & 0 & 1.22375\times 10^{15} & 
	0.0258567 & {\rm Yes} \\ \hline
	2 & F & 1 & 2 & \frac{1}{6} & 1 & S & 6 & 1 & \frac{1}{3} & 6.47456\times 
	10^{14} & 0.0259887 & {\rm Yes} \\ \hline
	2 & F & 1 & 1 & 0 & 1 & S & 3 & 3 & \frac{2}{3} & 3.63426\times 10^{14} & 
	0.0272945 & {\rm Yes} \\ \hline
	2 & F & 3 & 1 & \frac{2}{3} & 1 & S & 6 & 3 & 1 & 1.2987\times 10^{16} & 
	0.0690751 & {\rm No} \\ \hline
	2 & F & 3 & 2 & \frac{1}{2} & 1 & S & 8 & 1 & 0 & 8.1903\times 10^{15} & 
	0.0390761 & {\rm Yes} \\ \hline
	2 & F & 3 & 2 & \frac{1}{2} & 1 & S & 8 & 1 & \frac{1}{3} & 3.12815\times 
	10^{15} & 0.0388947 & {\rm Yes} \\ \hline
	2 & F & 1 & 2 & \frac{1}{6} & 2 & S & 6 & 2 & \frac{1}{2} & 1.29764\times 
	10^{17} & 0.034587 & {\rm Yes} \\ \hline
	2 & F & 3 & 2 & \frac{1}{6} & 2 & S & 6 & 2 & \frac{5}{6} & 6.99517\times 
	10^{17} & 0.0713552 & {\rm Yes} \\ \hline
	\end{array}
}
\end{eqnarray}
\section{Two loop Beta Function}\label{a:2lbeta}
For Standard Model, in Yukawa sector the beta function are \cite{Jones:1981we, Arason:1991ic, Luo:2002ey}
\begin{equation}
 \frac{dY_{u,d,e}}{dt}=Y_{u,d,e}\frac{1}{16\pi^2}{\beta}^{(1)}_{u,d,e}+
 \frac{1}{(16\pi^2)^2}{\beta}^{(2)}_{u,d,e}
\end{equation}
where one loop contribution are given as
\begin{eqnarray}
 \beta_{u}^{(1)}&=&\frac{3}{2}(Y_{u}^{\dag}Y_{u}-Y_{d}^{\dag}Y_{d})+Y_{2}(S)-\Big(\frac{17}{20}g_{1}^{2}
 +\frac{9}{4}g_{2}^{2}+8g_{3}^{2}\Big)\\
 \beta_{d}^{(1)}&=&\frac{3}{2}(Y_{d}^{\dag}Y_{d}-Y_{u}^{\dag}Y_{u})+Y_{2}(S)-\Big(\frac{1}{4}g_{1}^{2}
 +\frac{9}{4}g_{2}^{2}+8g_{3}^{2}\Big)\\
 \beta_{e}^{(1)}&=&\frac{3}{2}Y_{e}^{\dag}Y_{e}+Y_{2}(S)-\frac{9}{4}(g_{1}^{2}+g_{2}^{2})
\end{eqnarray}
with
\begin{equation}
 Y_{2}(S)=Tr(3Y_{u}^{\dag}Y_{u}+3Y_{d}^{\dag}Y_{d}+Y_{e}^{\dag}Y_{e})
\end{equation}
the two-loop contribution are given as
\begin{eqnarray}
 \beta_{u}^{(2)}&=& \frac{3}{2}(Y_{u}^{\dag}Y_{u})^{2} - Y_{u}^{\dag}Y_{u}Y_{d}^{\dag}Y_{d}
  - \frac{1}{4} Y_{d}^{\dag}Y_{d}Y_{u}^{\dag}Y_{u} + \frac{11}{4}(Y_{d}^{\dag}Y_{d})^{2}
  + Y_{2}(S) \Big( \frac{5}{4}Y_{d}^{\dag}Y_{d} - \frac{9}{4}Y_{u}^{\dag}Y_{u}\Big)      
   \nonumber \\
 &-&\chi_{4}(S) + \frac{3}{2}\lambda^{2} - 6\lambda Y_{u}^{\dag}Y_{u} 
 +\Big(\frac{223}{80}g_{1}^{2} +\frac{135}{16}g_{2}^{2}+16g_{3}^{2}\Big) Y_{u}^{\dag}Y_{u}
  \nonumber \\
 &-&\Big(\frac{43}{80}g_{1}^{2} -\frac{9}{16}g_{2}^{2}+16g_{3}^{2}\Big)Y_{d}^{\dag}Y_{d} 
 + \frac{5}{2}Y_{4}(S)
 + \left(
 \frac{9}{200}+\frac{29}{45}n_g\right)g_{1}^{4}
  \nonumber \\
 &-&\frac{9}{20}g_{1}^{2}g_{2}^{2}  + \frac{19}{15}g_{1}^{2}g_{3}^{2} - 
 \left(
 \frac{35}{4}-n_g\right)g_{2}^{4}+ 9g_{2}^{2}g_{3}^{2}-\left(
 \frac{404}{3}-\frac{80}{9}n_g\right)g_{3}^{4}
\end{eqnarray}
\begin{eqnarray}
 \beta_{d}^{(2)}&=& \frac{3}{2}(Y_{d}^{\dag}Y_{d})^{2} - Y_{d}^{\dag}Y_{d}Y_{u}^{\dag}Y_{u}
  - \frac{1}{4} Y_{u}^{\dag}Y_{u}Y_{d}^{\dag}Y_{d} + \frac{11}{4}(Y_{u}^{\dag}Y_{u})^{2}
  + Y_{2}(S) \Big( \frac{5}{4}Y_{u}^{\dag}Y_{u} - \frac{9}{4}Y_{d}^{\dag}Y_{d} \Big)      
   \nonumber \\
 &-&\chi_{4}(S) + \frac{3}{2}\lambda^{2} - 2\lambda 3Y_{d}^{\dag}Y_{d} 
 +\Big(\frac{187}{80}g_{1}^{2} +\frac{135}{16}g_{2}^{2}+16g_{3}^{2}\Big)Y_{d}^{\dag}Y_{d}
  \nonumber \\
 &-& \Big(\frac{79}{80}g_{1}^{2} -\frac{9}{16}g_{2}^{2}+16g_{3}^{2}\Big)Y_{u}^{\dag}Y_{u} +
  \frac{5}{2}Y_{4}(S)
 - \left(
 \frac{29}{200}+\frac{1}{45}n_g\right)g_{1}^{4}
  \nonumber \\
 &-&\frac{27}{20}g_{1}^{2}g_{2}^{2}  + \frac{31}{15}g_{1}^{2}g_{3}^{2} - 
 \left(
 \frac{35}{4}-n_g\right)g_{2}^{4}
 9g_{2}^{2}g_{3}^{2}-\left(
 \frac{404}{3}-\frac{80}{9}n_g\right)g_{3}^{4}
\end{eqnarray}
\begin{eqnarray}
 \beta_{e}^{(2)}&=& \frac{3}{2}(Y_{e}^{\dag}Y_{e})^{2} 
  - \frac{9}{4}Y_{2}(S) Y_{e}^{\dag}Y_{e} - \chi_{4}(S) + \frac{3}{2}\lambda^{2} - 
  6\lambda Y_{e}^{\dag}Y_{e} + \Big(\frac{387}{80}g_{1}^{2} +\frac{135}{15}g_{2}^{2}\Big)Y_{e}^{\dag}Y_{e}
  \nonumber \\
 &+& \frac{5}{2}Y_{4}(S)+
\left(
 \frac{51}{200}+\frac{11}{5}n_g\right)g_{1}^{4} + \frac{27}{20}g_{1}^{2}g_{2}^{2} - \left(
 \frac{35}{4}-n_g\right)g_{2}^{4})
\end{eqnarray}
with
\begin{equation}
 Y_{4}(S)= \Big(\frac{17}{20}g_{1}^{2} +\frac{9}{4}g_{2}^{2}+8g_{3}^{2}\Big)Tr[Y_{u}^{\dag}Y_{u}]
 + \Big(\frac{1}{4}g_{1}^{2} +\frac{9}{4}g_{2}^{2}+8g_{3}^{2}\Big)Tr[Y_{d}^{\dag}Y_{d}]
 + \frac{3}{4}(g_{1}^{2} +g_{2}^{2})Tr[Y_{e}^{\dag}Y_{e}]
\end{equation}
and
\begin{equation}
 \chi_{4}(S) = \frac{9}{4}\Big(3(Y_{u}^{\dag}Y_{u})^{2} + 3(Y_{d}^{\dag}Y_{d})^{2} + (Y_{e}^{\dag}Y_{e})^{2}
 - \frac{2}{3}Y_{u}^{\dag}Y_{u}Y_{d}^{\dag}Y_{d}\Big)
\end{equation}
In Higgs sector we present $\beta$ functions for the quartic coupling:
\begin{equation}
 \frac{d\lambda}{dt}=\frac{1}{16\pi^{2}}\beta^{(1)}_{\lambda}+\frac{1}{(16\pi^{2})^{2}}\beta^{(2)}_{\lambda}
\end{equation}
where the one loop contribution is given as,
\begin{eqnarray}
\beta_{\lambda}^{(1)}=12\lambda^{2}-\left(\frac{9}{5}g_{1}^{2}+9g_{2}^{2}  \right)\lambda
 + \frac{9}{4}\left(\frac{3}{5}g_{1}^{4} +\frac{2}{5}g_{2}^{2}g_{2}^{2}+g_{2}^{4}\right)+4Y_{2}(S)\lambda
 -4H(S),
\end{eqnarray}
with 
\begin{equation}
 H(S)=Tr(3(Y_{u}^{\dag}Y_{u})^{2}+3(Y_{d}^{\dag}Y_{d})^{2}+(Y_{e}^{\dag}Y_{e})^{2})
\end{equation}
and the two loop contribution is given as:
\begin{eqnarray}
\beta_{\lambda}^{(2)}&=&-78\lambda^{3}+18\left(\frac{3}{5}g_{1}^{2}+3g_{2}^{2}\right)\lambda^{2}
-\left[\left(\frac{313}{8}-10n_{g}\right)g_{2}^{4}-\frac{117}{20}g_{1}^{2}g_{2}^{2}-\left(
\frac{687}{200}+2n_g\right)g_{1}^{4}\right]\lambda 
\nonumber \\
&+&\left(\frac{497}{8}-8n_{g}\right)g_{2}^{3}-\frac{3}{5}\left(\frac{97}{24}+\frac{8}{3}n_{g}\right)g_{1}^{2}g_{2}^{4}
-\frac{9}{25}\left(\frac{239}{24}+\frac{40}{9}n_{g}\right)g_{1}^{4}g_{2}^{2}-
\frac{27}{125}\left(\frac{59}{24}+\frac{40}{9}n_{g}\right)g_{1}^{6}  
\nonumber \\
&-&64g_{3}^{2}Tr((Y_{u}^{\dag}Y_{u})^{2}+(Y_{d}^{\dag}Y_{d})^{2})-
\frac{8}{5}g_{1}^{2}Tr(2(Y_{u}^{\dag}Y_{u})^{2}-(Y_{d}^{\dag}Y_{d})^{2}+3(Y_{e}^{\dag}Y_{e})^{2})-
\frac{3}{2}g_{2}^{4}Y_{2}(S)  
\nonumber \\
&+&10\lambda\left[\left(\frac{17}{20}g_{1}^{2} +\frac{9}{4}g_{2}^{2}+8g_{3}^{2}\right)Tr(Y_{u}^{\dag}Y_{u})
+ \left(\frac{1}{4}g_{1}^{2} +\frac{9}{4}g_{2}^{2}+8g_{3}^{2}\right)Tr(Y_{d}^{\dag}Y_{d})
+ \frac{3}{4} \left( g_{1}^{2}+g_{2}^{2}\right) Tr(Y_{e}^{\dag}Y_{e}) \right]
\nonumber \\
&+& \frac{3}{5}g_{1}^{2}\left[\left(-\frac{57}{10}g_{1}^{2} +21g_{2}^{2}\right)Tr(Y_{u}^{\dag}Y_{u})
+ \left(\frac{3}{2}g_{1}^{2} +9g_{2}^{2}\right)Tr(Y_{d}^{\dag}Y_{d})
+  \left(-\frac{15}{2}g_{1}^{2} +11g_{2}^{2}\right)Tr(Y_{e}^{\dag}Y_{e}) \right]
\nonumber \\
&-& 24 \lambda^{2} Y_{2}(S)-\lambda H(S)-42\lambda Tr(Y_{u}^{\dag}Y_{u}Y_{d}^{\dag}Y_{d})+
20Tr(3(Y_{u}^{\dag}Y_{u})^{3}+3(Y_{d}^{\dag}Y_{d})^{3}+(Y_{e}^{\dag}Y_{e})^{3})
\nonumber \\
&-& 12 Tr \{ Y_{u}^{\dag}Y_{u}(Y_{u}^{\dag}Y_{u}+Y_{d}^{\dag}Y_{d})Y_{d}^{\dag}Y_{d} \}
\end{eqnarray}
where $n_{g}$ is the number of generation of fermions in SM.
\providecommand{\href}[2]{#2}\begingroup\raggedright
\endgroup
\end{document}